\documentclass[prd,aps,a4paper,superscriptaddress,twocolumn,nofootinbib]{revtex4}
\usepackage{graphicx}
\usepackage{color}
\usepackage{dcolumn}
\usepackage{bm}
\usepackage{slashed}
\usepackage{amsmath}
\usepackage{latexsym}
\usepackage{amssymb}
\usepackage{mathrsfs}
\usepackage{amsfonts}
\usepackage{url}
\usepackage{graphicx}

\newcommand{\be}{\begin{equation}}
\newcommand{\ee}{\end{equation}}

\allowdisplaybreaks
\begin{document}
	
\title{Using horizon shadows to distinguish a black hole and a white hole}
	
\author{Chengyu Bi}
\email[ChengYu Bi:~]{bichengyu@nefu.edu.cn}
\affiliation{College of Science, Northeast Forestry University, Harbin 150040, China}
\author{Zhoujian Cao\footnote{corresponding author}}
\email[Zhoujian Cao:~]{zjcao@bnu.edu.cn}
\affiliation{School of Physics and Astronomy, Beijing Normal University, Beijing 100875, China}
\affiliation{School of Fundamental Physics and Mathematical Sciences, Hangzhou Institute for Advanced Study, University of Chinese Academy of Sciences, %No.1 Xiangshan Branch,
Hangzhou 310024, China}

\begin{abstract}
Within theoretical frameworks such as loop quantum gravity, black holes may evolve into white holes through a quantum bounce. This paper uses general relativistic ray-tracing techniques to calculate the ray-traced imaging of accretion disks from the previous cosmic stage during the Kerr black hole and post-bounce Kerr white hole phases. Calculations show that the black hole image presents a crescent emission ring and a central shadow. In contrast, after radiation from the previous universe penetrates the rotating white hole, eccentric and asymmetric nested intensity ring structures form in the synthetic image due to frame-dragging and lensing effects. We analyze the influence of spin parameters, observation inclinations, and accretion disk geometric configurations on the distribution of this nested ring structure using synthetic images and intensity profiles. Building upon this, we introduce polarized ray-tracing calculations for radiation across evolutionary stages. This process results in the polarization image features after the polarization vector is subjected to the gravitational field and spacetime spin dragging during the photon propagation through the white hole horizon and internal spacetime. The spatial rotation patterns and concentric interference fringes in the white hole polarization images exhibit a distinct inter-ring polarization discontinuity. This phenomenon differs from the polarization behavior of black holes. The intensity ring structures and polarization inter-ring discontinuity features provide multi-band and polarimetric interferometry baselines to overcome morphological observational degeneracies. This provides theoretical guidance for future very-long-baseline interferometry (VLBI) to distinguish black holes and white holes.
\end{abstract}

\maketitle

\section{Introduction}

The singularity problem in general relativity prompts physicists to explore quantum gravity effects. In principle, if one can accurately measure the metric of the spacetime, we may find out the deviation from general relativity (GR) \cite{6sws-hfj7}. However, such accuracy requirement is not reachable in short terms. In related theories, the quantum bounce evolution mechanism from black holes to white holes receives attention \cite{PhysRevD.92.104020,Bianchi:2018mml,Simpson:2018tsi,PhysRevLett.130.101501}. Recently, many regular black hole models are proposed to describe this process \cite{PhysRevLett.96.031103,PhysRevD.94.104056,2023IJTP...62..202L,10.1088.1572-9494.ae662c}. If this physical process really exists, some black holes in our universe should be results of white holes \cite{2025ChPhC..49b5109L} instead of being formed by stellar collapse, which is currently believed by astronomers or being formed by curvature fluctuation in the early universe which is referred as primordial black holes \cite{PhysRevLett.116.201301,2026NCimR..49..225C,2026arXiv260623846S}. Then it is interesting to ask whether it is possible to distinguish black holes formed from white holes from those formed through normal mechanisms. If specific observations can rule in or rule out white holes, valuable guidelines may be provided to quantum gravity research.

In the current paper, we would like to investigate the possibility of using radio astronomical observations such as the Event Horizon Telescope to looking for black holes resulting from white holes. There are already several works related to this question \cite{2025ChPhC..49b5109L,PhysRevD.108.104004,PhysRevD.109.064083}. These works only considered spinless black holes and white holes. The white hole images can be mimicked by proper disk configuration which makes white holes indistinguishable from black holes \cite{2025ChPhC..49b5109L}.

Compact objects in astrophysics generally possess angular momentum. Imaging calculations based on rotating Kerr or Kerr-Newman (KN) spacetimes align with realistic physical scenarios. The current paper first calculates the ray-traced appearance of the accretion disk radiation left over from the previous universe penetrating the white hole interior to reach the current observer after a rotating Kerr system undergoes a quantum bounce. By comparing the synthetic imaging and intensity profiles of Kerr black holes and white holes under the same parameters, we analyze their basic image differences. The study examines the nested ring structure features of the white hole intensity.

Considering realistic astrophysical environments, accreting matter usually exists around the post-bounce object in the current universe. Therefore, we further introduce a foreground accretion disk model in the current universe to explore the physical obscuration effect of the foreground fluid on the transmitted signal inside the white hole in a dual accretion disk system. Furthermore, because the total intensity is susceptible to optical thickness and local accretion flow physical uncertainties, separating the complex fluid radiation background is difficult.

Besides the intensity information, polarization observations provide electromagnetic vector information reflecting spacetime geometric characteristics. Due to differences in spacetime topology and photon propagation history, the polarization evolution processes of black holes and white holes have objective differences. Based on this, we calculate the radiation polarization images after the photons experience the previous evolutionary stage. By analyzing the mapping pattern of polarization phase angle and electric vector position angle (EVPA), we study the action mechanism of frame-dragging effects on white hole polarization images. We detail the polarization discontinuity behavior between adjacent concentric rings. We comprehensively analyze the effects of spin parameters, observation inclinations, and accretion disk geometric configurations on the nested intensity rings and polarization discontinuity features to provide a practical numerical reference for future very-long-baseline interferometry to compare and distinguish related celestial bodies.

The arrangement of the rest of the current paper is as follows. Based on the null rays and time like curves analysis, we set up the observers and accretion disks in the next section. Then we investigate the related black hole images and white hole images in Sec.~\ref{sec3}. Following the intensity images investigation, we study the polarization properties in Sec.~\ref{sec4}. We interestingly find out the characteristic nested ring behavior and the interring discontinuity polarization structures of white hole images. Finally we conclude the paper in the last section and some discussion is also presented there.

Throughout the paper we use geometric units where $c=G=1$.
\section{Observers, null rays and accretion disk models}
\begin{figure}[htbp]
    \centering
    \includegraphics[width=0.48\textwidth]{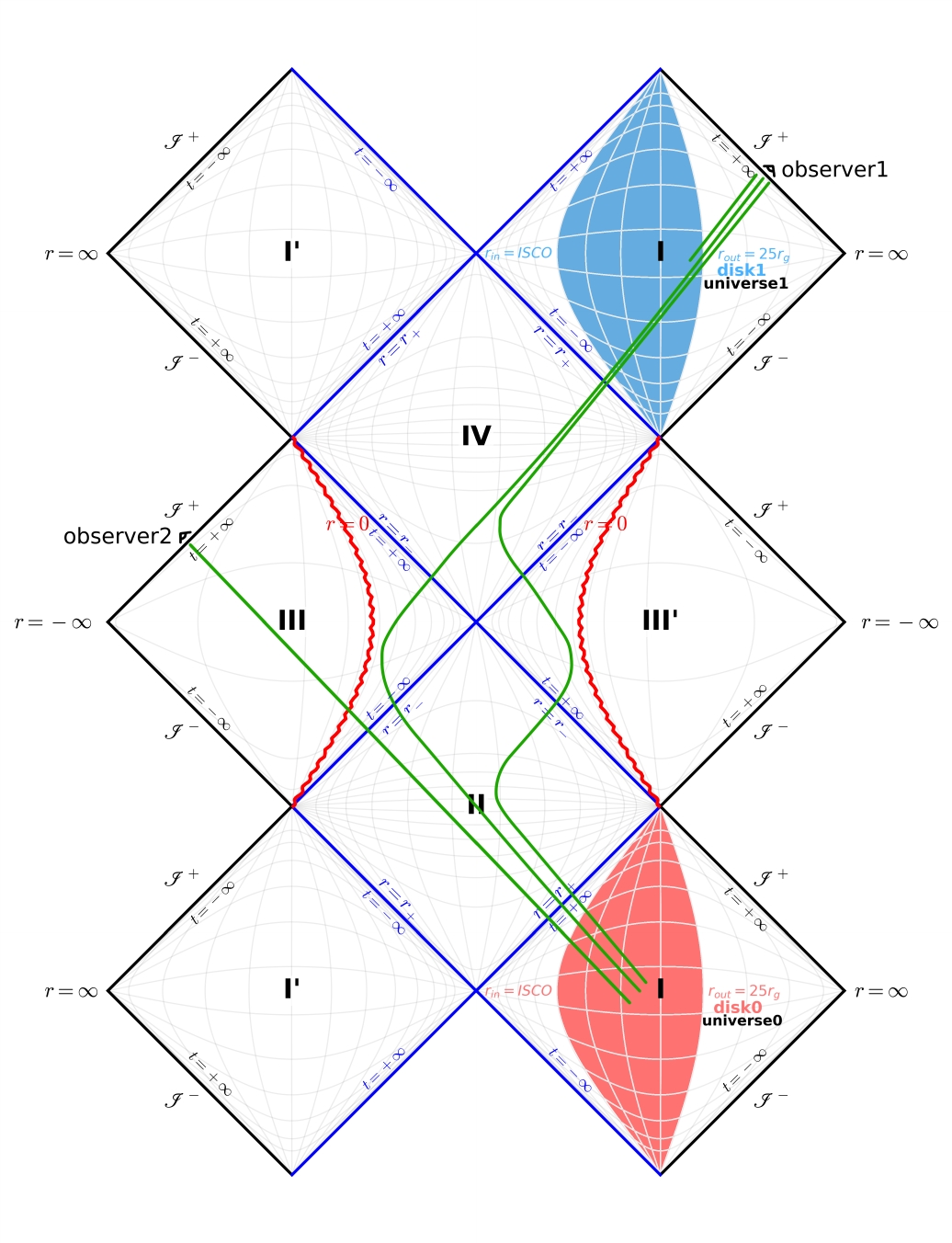}
    \caption{A portion of the maximally extended Penrose diagram for KN spacetime. The structure for Kerr spacetime is similar. The observer1 sitting in the usual asymptotically flat region may detect the radiation from the accretion disk1 locating in the current universe (Universe 1) and the radiation from the accretion disk0 in the previous cosmic stage (Universe 0) penetrates the internal region and reaches the asymptotic observer in the current universe (Universe 1). The observer 2 sitting in the region III can only detect the radiation from the accretion disk0 in universe0.}
    \label{fig:penrose}
\end{figure}
\subsection{Possible accretion disks in maximally extended Kerr-Newman spacetime}
The Kerr-Newman (KN) spacetime geometry is determined by the mass $M$, the spin $a$, and the charge $Q$. Its maximally extended Penrose diagram is illustrated in Fig.~\ref{fig:penrose}. From this figure we can see there are two types of asymptotically flat regions marked with I (I') and III (III'). In I (I') regions, the Boyer-Lindquist coordinate $r$ goes to $+\infty$ for asymptotically null infinity, while in III (III') regions, the $r$ goes to $-\infty$. In the following we will show that the phenomenological behavior is quite different for region I and III. Consequently accretion disks may only appear in regions I and I'.

For an asymptotically flat region, within the Boyer-Lindquist coordinate system we consider the behavior of object initially moving along $\left(\frac{\partial}{\partial t}\right)^a$. The geodesic equation leads
\begin{equation}
    \frac{d^2 r}{d\tau^2} \approx -\Gamma^r_{tt} = \frac{1}{2} g^{rr} \partial_r g_{tt}\approx-\frac{M}{r^2},
\end{equation}
where $\tau$ is the proper time of the object. The above equation is nothing but the Newton limit for general relativity and it indicates the inverse square law of gravitational interaction. This equation can also be understood as the gravitational force lets the object move along the direction of decreasing $r$. For regions I and I' where $r$ asymptotically goes to $+\infty$, this behaves as an attraction force. In contrast for regions III and III' where $r$ asymptotically goes to $-\infty$, it behaves as a repulsion force. Direct calculation for ADM mass and Bondi mass for the region III can result a negative mass \cite{2018A&A...620A..92F}. Actually the negative sign of the mass will affect the multipoles accordingly \cite{RevModPhys.52.299,10.1063.1.1666501,Gursel1983,10.1063.1.1665427}. This fact makes the formation of a matter disk in regions III and III' impossible.

Based on the time-like geodesic equation with the Boyer-Lindquist coordinate, we have \cite{2019MNRAS.488.2722G,PhysRevD.106.084048,LIU2025139616}
\begin{align}
&\Sigma^2\left(\frac{dr}{d\tau}\right)^2=R(r),\\
&R(r)\equiv[(r^2+a^2)E-aL]^2-\Delta[\mathcal{K}+r^2+(L-aE)^2],\nonumber\\
&\Delta\equiv r^2-2Mr+a^2+Q^2,\nonumber\\
&\Sigma\equiv r^2+a^2\cos^2\theta,\nonumber\\
&\mathcal{K}\equiv p_\theta^2+(1-E^2)\cos^2\theta a^2+L^2\cot^2\theta,\nonumber
\end{align}
where $p_\mu\equiv g_{\mu\nu}\frac{dx^\nu}{d\tau}$, and $E$, $L$ and $\mathcal{K}$ are conserved energy, angular momentum and Carter constant. The event horizon and Cauchy horizon are located at the two roots $r_+$ and $r_-$ of $\Delta \equiv r^2 - 2Mr + a^2 + Q^2 = 0$, respectively.

Following the analysis of \cite{PhysRevD.105.024075} we find that there are possible parameters $E$, $L$, $\mathcal{K}$, $M$, $a$ and $Q$ to get two roots $r_+<r_1<r_2<+\infty$ of $R(r)=0$ to make $R(r_1<r<r_2)>0$, but it is not possible to get two roots $-\infty<r_1<r_2<r_-$ of $R(r)=0$ satisfying $R(r_1<r<r_2)>0$. Situation $R(r_1<r<r_2)>0$ with $r_+<r_1<r_2<+\infty$ corresponds to the usual bound state, or to say the elliptic orbits in Newtonian limit. And this type of orbits corresponds to the accretion disk formation. Regarding the bound state $R(r_1<r<r_2)>0$ with $r_1<r_-$ and $r_2>r_+$, although part of the orbit happens in spacetime region III, the staying time in region III is short. The orbit will go through the ladder shown in Fig.~\ref{fig:penrose}. The object will move quickly from one region III to the near region IV, then the region I, then the region II, then the region III but the one sitting in the latter time of the aforementioned region III, and so on. This fact means even in strong gravitational region of the region III, it is not possible to have an accretion disk.

In principle, observers can locate at any of the asymptotically flat regions including region III. But as analyzed above, the observers locating at the asymptotically flat region of III should see celestial bodies escaping from a region which corresponds to the central `KN black hole'. To date, there are no reports about the observation of this kind of phenomenon..

For reference, we call the one where observer1 locates as universe 1 as shown in Fig.~\ref{fig:penrose}. We call the one located in the past of universe 1 "universe 0" as shown Fig.~\ref{fig:penrose}. Note that there is an asymptotic flat region dual to the universe 0 which sits at the left of the universe 0 in Fig.~\ref{fig:penrose}. But later we will show that the null rays reaching the observers must come from universe 1 or universe 0. Apparently the white hole image is made of the rays coming from the accretion disk sitting in the universe 0 and the black hole image is made of the rays coming from the accretion disk sitting in the universe 1.

Although there is no evidence of their existence, we denote the observer sitting in region III as observer2. Such observers may detect the radiation from the accretion disk0 in universe0. In the current paper we will also give theoretical prediction on the shadows detected by observer2.
\subsection{Null ray properties in maximally extended Kerr-Newman spacetime}
Again with the Boyer-Lindquist coordinate system $(t, r, \theta, \phi)$, given the three conserved quantities including energy $E$, axial angular momentum $L$, and the Carter constant $\mathcal{K}$, the null geodesic equation for the polar angle $\theta$ and radial coordinate $r$ with respect to the affine parameter $\lambda$ can be written as \cite{PhysRevD.106.084048}
\begin{align}
    &\Sigma^2 \left(\frac{d\theta}{d\lambda}\right)^2 = \Theta(\theta) = \mathcal{K} + a^2E^2\cos^2\theta - L^2\cot^2\theta \label{eq:theta},\\
    &\Sigma^2 \left(\frac{dr}{d\lambda}\right)^2 = R(r) = P(r)^2-\Delta[\mathcal{K}+(L-aE)^2] \label{eq:radial},\\
    &\Sigma \equiv r^2 + a^2 \cos^2\theta,\\
    &E \equiv -p_t,\\
    &L \equiv p_\phi,\\
    &\mathcal{K} \equiv p_\theta^2 -a^2E^2\cos^2\theta + L^2\cot^2\theta,\label{eq1}\\
    &P(r)\equiv (r^2+a^2)E - aL,
\end{align}
where $p_\mu\equiv g_{\mu\nu}\frac{dx^\nu}{d\lambda}$.

We firstly investigate the turning points in $r$ direction which correspond to $R(r)=0$. In the regions between the two horizons ($r_- < r < r_+$) like region II and IV in Fig.~\ref{fig:penrose}, since $\Delta = (r-r_+)(r-r_-) < 0$ and $\mathcal{K}\geq0$ from (\ref{eq1}), $R(r)$ must be positive in region II and IV. Similar but simpler than the time-like geodesic situation discussed in the above subsection \cite{PhysRevD.105.024075}, the roots of the quartic polynomial $R(r)$ distribute as follows. If there is at least one root, it must locate in region III $-\infty<r_1<r_-$. If there are more than two roots, two of them locate in III and others locate in region I.

As shown in Fig.~\ref{fig:penrose}, the rays reaching the observer may come from the disk1. Otherwise the rays come from the white hole (region IV) which penetrate the horizon $r=r_+$. The $R(r)$ for these rays must be positive for all $r_+<r<\infty$. In order to determine these rays come from region III or III', we consider \cite{PhysRevD.106.084048}
\begin{align}
\Sigma \frac{dt}{d\lambda} &= \frac{r^2+a^2}{\Delta}P(r)+a(L-aE\sin^2\theta).
\end{align}
When $r$ approaches to $r_-$ where $\Delta=0$,
\begin{align}
\frac{dt}{d\lambda}\approx P(r_-)/(\Delta\Sigma).
\end{align}
We can see when $r$ approaches to $r_-$, $t\rightarrow{\rm sgn}(P(r_-))\cdot{\rm sgn}(\Delta \lambda)\infty$ where ${\rm sgn}(\Delta \lambda)$ means the sign of $\Delta \lambda$ from $\Delta>0$ side to horizon. Near $r_-$, $\Delta>0$ in region IV, so $\Delta\lambda<0$. This means the rays with positive $P(r_-)$ come from region III', while the rays with negative $P(r_-)$ come from region III. These rays may or may not admit $R(r)=0$ in region $-\infty<r<r_-$. If no, these rays must come from the past null infinity of III or III' because there is no accretion disks in region III and III'. If yes, the corresponding rays will penetrate the horizon $r=r_-$ and these rays come from region II.

Again since $R(r)>0$ in the whole region II ($r_- < r < r_+$), these rays must penetrate the horizon $r=r_+$. Similar to the situation when $r$ approaches to $r_-$, when $r$ approaches to $r_+$, $t\rightarrow{\rm sgn}(P(r_+))\cdot{\rm sgn}(\Delta \lambda)\infty$. Since these rays pass the interface between region I and region IV, where $t=-\infty$ and $\Delta \lambda>0$, we are sure $P(r_+)<0$ when $r$ approaches to $r_+$. For the interface between region I (or I') and region II, $\Delta \lambda<0$. This means these rays come from region I.

Besides the radiation from disk1 and disk0, the observer1 can also detect radiations from the past null infinity of region I, of region III and III', and of region I. In the current paper we only consider the radiation from disks.

For the rays observed by observer2, if they penetrate the horizon $r=r_-$, they must come from region II. Then these rays must come from region lower I or I'. In the following we will show that these rays can not come from region I'. For rays penetrating both the interface between II and III and the interface between II and I', we have $P(r_-)<0$ and $P(r_+)>0$. That is to say there must be $r_-<r_0<r_+$ with $P(r_0)=0$. Noting $P(r)$ depends on $r$ in the form $r^2$, we are sure $P(-r_0)=0$. Since $-r_0<0$, $-r_0$ must locate in region III. We consider
\begin{align}
&R(-r_0)=P(-r_0)^2-\Delta(r_0)[\mathcal{K}+(L-aE)^2]\nonumber\\
&=-[r_0^2 + 2Mr_0 + a^2 + Q^2][\mathcal{K}+(L-aE)^2]<0.
\end{align}
That is to say the rays penetrating both the interface between II and III and the interface between II and I' can not reach the infinity of region III and consequently can not be detected by the observer2. So observer2 can only detect the radiations from disk0.

Besides the radiation from disk0, the observer2 can also detect radiations from the past null infinity of region I and region III. In the current paper we do not consider the radiation from past null infinity.

\subsection{Accretion disk models}
\begin{figure}[htbp]
    \centering
\begin{tabular}{ccc}
\includegraphics[width=0.14\textwidth]{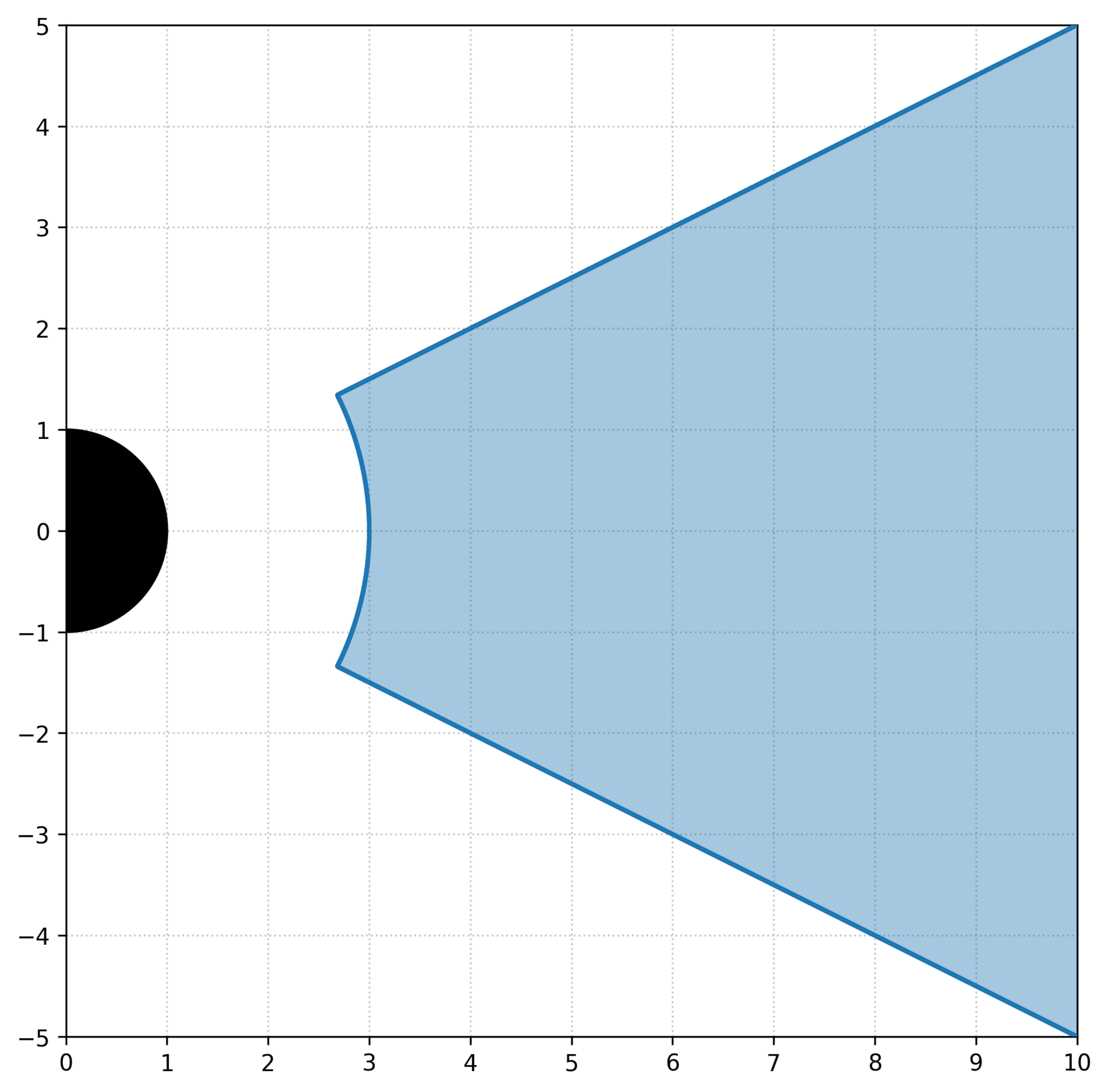}&
\includegraphics[width=0.14\textwidth]{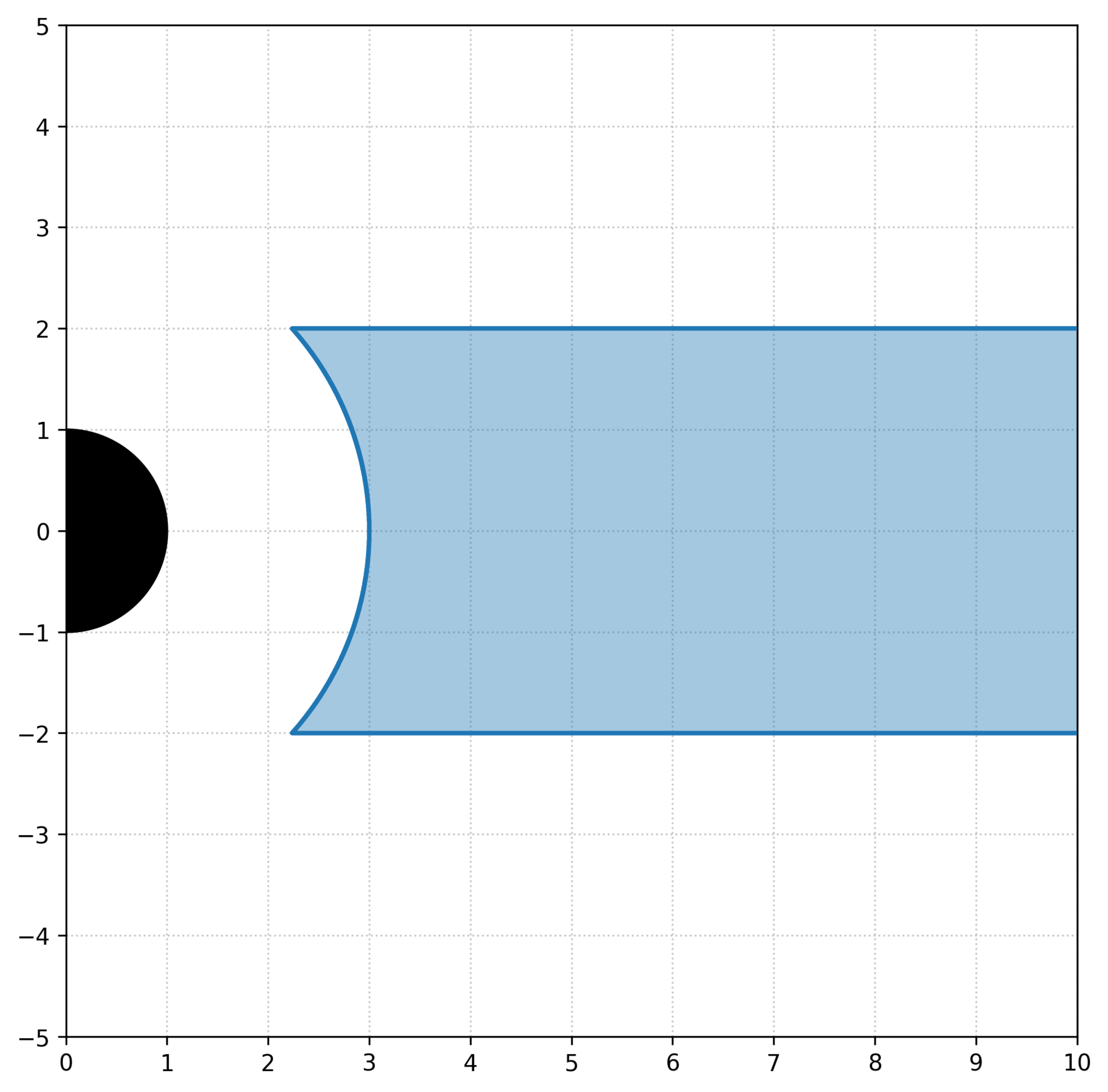}&
\includegraphics[width=0.14\textwidth]{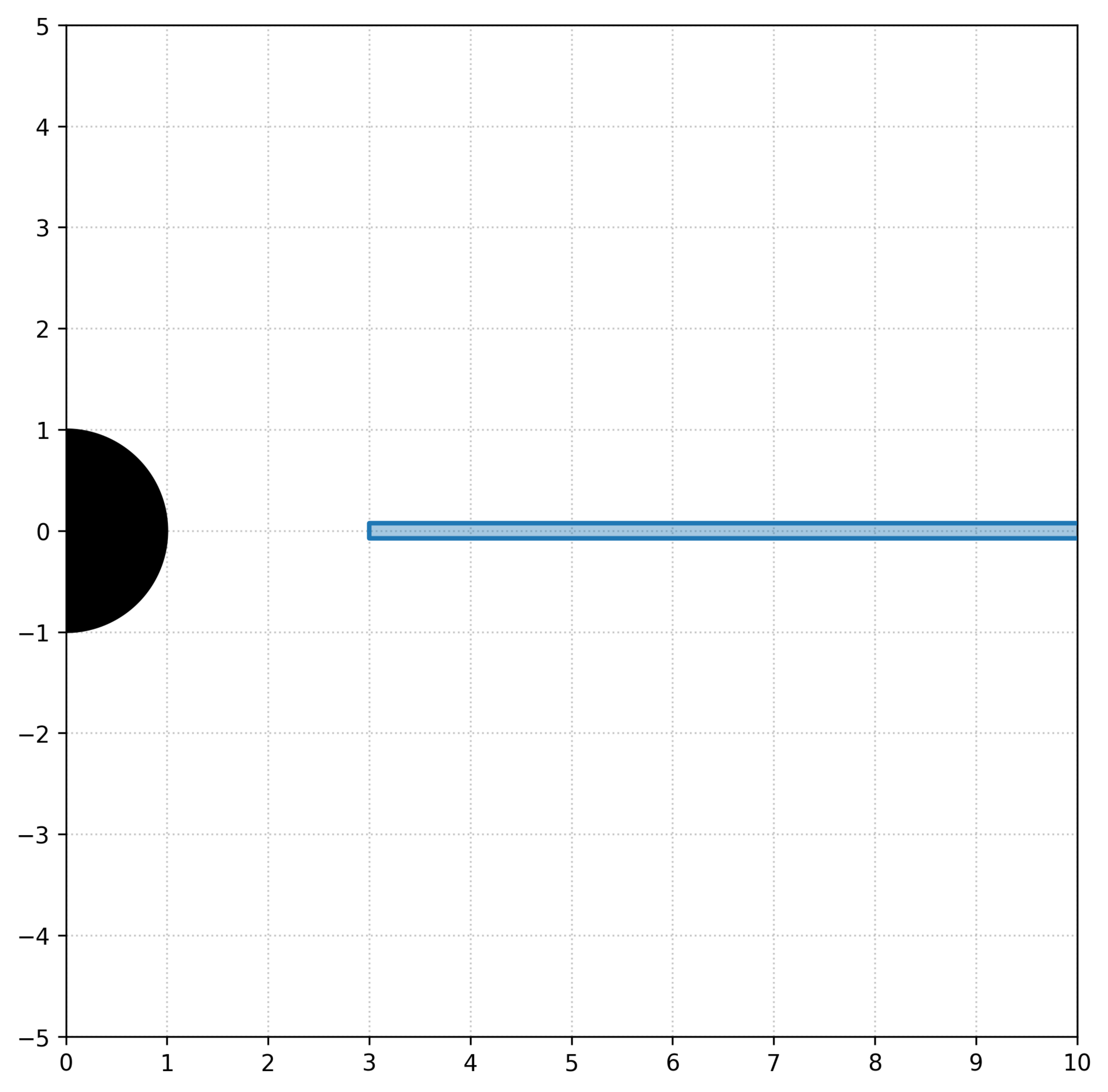}
\end{tabular}
    \caption{Schematic diagram of geometric cross-sections of accretion disks simulated by ray tracing. From left to right, the subplots respectively correspond to the thin disk model (the vertical distribution range is $0.075R_g$), the thick disk model (the vertical distribution range is $2R_g$) and the hopper disk model. Here the gravitational radius is defined as $R_g\equiv2M$. The unit of the axis of the plots are $R_g$.}
    \label{fig:disk_models}
\end{figure}

In the current paper, we consider both optically thin and thick disk models. The optically thin model means photons only accumulate radiation without self-absorption attenuation when penetrating this region. To verify whether the image distribution depends on specific geometric parameters of the radiation source, three different accretion disk models are investigated (Fig.~\ref{fig:disk_models}). In the thin disk model, radiation is concentrated near the equatorial plane. The thick disk model has a broader vertical distribution range. The hopper model simulates an accretion environment with outflow channels. Different shapes of accretion flows alter the radiation accumulation rate of photons on the internal path.

\begin{figure}[htbp]
    \centering
    \includegraphics[width=0.48\textwidth]{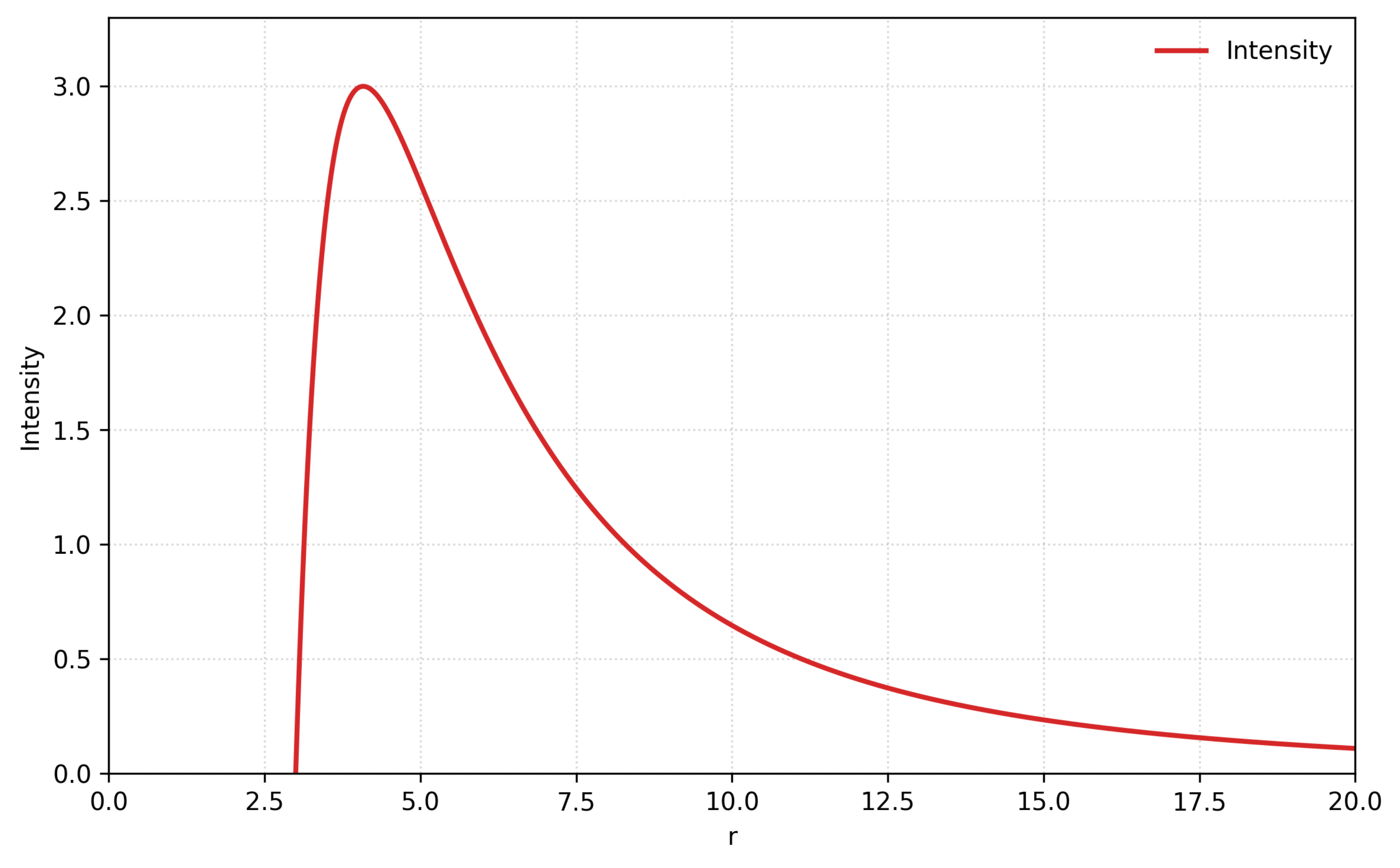}
    \caption{Radial local emission intensity curve of the accretion disk calculated based on the Page-Thorne thin disk model. The inner edge radius is set as $R_{\text{in}} = 3.0 R_g$ in this figure.}
    \label{fig:intensity_curve}
\end{figure}

Regarding local radiation characteristics, we assume that the radial local emission intensity of the accretion disk follows the classic Page-Thorne thin disk model radiation distribution. The variation of its local temperature $T$ with radial coordinate $r$ is proportional to:
\begin{equation}
    T \propto \left[ \frac{1}{r^3} \left( 1 - \sqrt{\frac{R_{\text{in}}}{r}} \right) \right]^{0.25}
\end{equation}
where $R_{\text{in}}$ is the inner edge truncation radius of the accretion disk. The local emission intensity calculated based on this temperature profile presents a distribution that first rises and then decays with the radius (the distribution curve with $R_{\text{in}} = 3.0 R_g$ is shown in Fig.~\ref{fig:intensity_curve}). In ray-tracing calculations, the observation image intensity depends not only on this local distribution but also includes frequency shifts during the covariant radiative transfer process. Let the emission energy of the photon in the fluid rest frame of the accretion disk be $E_{\text{em}} = -(p_\mu u^\mu)_{\text{em}}$, where $u^\mu_{\text{em}}$ is the four-velocity of the disk fluid. The energy received by the observer is $E_{\text{obs}} = -(p_\mu u^\mu)_{\text{obs}}$. According to Liouville's theorem, the observed specific intensity $I_{\nu_{\text{obs}}}$ and the local emitted specific intensity $I_{\nu_{\text{em}}}$ satisfy $I_{\nu_{\text{obs}}} = g^4 I_{\nu_{\text{em}}}$, where the redshift factor is $g = E_{\text{obs}} / E_{\text{em}}$. This calculation introduces the Doppler effect produced by fluid motion and the gravitational frequency shift accumulated by photons traversing spacetime regions to determine the physical intensity of each pixel in the final image.

In order to calculate the polarization evolution when radiation propagates in a gravitational field, we calculate the polarization electric vector position angle (EVPA) through general relativistic ray tracing. In the fluid rest frame of the accretion disk, we assume the disk surface has a helical ordered magnetic field $B^\mu$ composed of radial and toroidal components. Its configuration is described by the magnetic pitch angle $\chi$. In the subsequent calculations and analysis, this magnetic pitch angle is set to $-0.7$ radians. The initial linear polarization vector $f_\mu$ of the emitted photon is orthogonal to the plasma four-velocity $u^\mu$, the photon four-momentum $p^\mu$, and the local magnetic field. Its covariant components are given by the Levi-Civita totally antisymmetric tensor $\epsilon_{\mu\nu\rho\sigma}$
\begin{equation}
    f_\mu = \mathcal{N} \epsilon_{\mu\nu\rho\sigma} u^\nu p^\rho B^\sigma
\end{equation}
where $\mathcal{N}$ is a normalization factor used to ensure the polarization vector satisfies $f^\mu f_\mu = 1$ and the orthogonality condition $f^\mu p_\mu = 0$.

In curved spacetime, the polarization vector satisfies the parallel transport equation $\nabla_p f^\mu = 0$ along null geodesics. To avoid numerical errors that direct integration of vector fields along trajectories may introduce, we calculate the complex Walker-Penrose constant $\mathcal{K}_{\text{WP}}$ which is conserved along geodesics due to the symmetries of Kerr spacetime \cite{PhysRevD.101.084020,PhysRevD.104.044060,Li:2026lca}
\begin{equation}
    \mathcal{K}_{\text{WP}} = K_{\text{re}} + i K_{\text{im}} = (k_{\mu\nu} + i h_{\mu\nu}) p^\mu f^\nu
\end{equation}
where $k_{\mu\nu}$ is the Killing tensor and $h_{\mu\nu}$ is its conjugate tensor. After calculating $\mathcal{K}_{\text{WP}}$ at the radiation emission point, this constant remains invariant along the photon propagation path.

To calculate the final observation results on the screen, the polarization vector of the photon needs to be projected into the two-dimensional coordinate system of the observer plane. Let the observer four-velocity be $U^\mu$, and establish a purely spatial orthonormal basis $(X^\mu, Y^\mu)$ within their local reference frame. Subsequently, through the Gram-Schmidt orthogonalization method, the projection components of the observer velocity $U^\mu$ and the relative spatial motion direction of the photon $D^\mu = p^\mu - (p_\alpha U^\alpha)U^\mu$ are removed. This constructs the horizontal and vertical orthogonal basis of the observation screen and calculates their corresponding reference Walker-Penrose constants $\mathcal{K}_X$ and $\mathcal{K}_Y$, respectively.

The electromagnetic wave amplitude components $(\alpha, \beta)$ arriving at the screen are derived by solving a linear system of equations. By integrating the fluid thermal radiation on the photon trajectory, the Stokes parameters $Q$ and $U$ of the image pixels are obtained:
\begin{align}
    \Delta Q &= (\alpha^2 - \beta^2) g^4 I_{\nu_{\text{em}}}, \\
    \Delta U &= (2\alpha\beta) g^4 I_{\nu_{\text{em}}}.
\end{align}
From this, the electric vector position angle $\text{EVPA} = \frac{1}{2}\arctan(U/Q)$ affected by gravitational lensing and frame-dragging is calculated. This provides the data foundation for subsequent analysis of the inter-ring discontinuity features of white hole polarization.

\section{Image of rotating white holes}\label{sec3}

Image of rotating black holes,which corresponds to the observation by the observer1 of the radiations from disk1 has been extensively studied. In the current paper we focus on the observation of radiations from disk0. Such observations corresponds to null rays passing through inner horizon. We consequently call it image of rotating white holes.

To visually display the photon ring structure and shadow details more clearly, the colormap of all two-dimensional synthetic images in this paper applies a logarithmic scale to the intensity. The horizontal and vertical intensity profiles on the side of the figures maintain the original linear numerical values of the radiation intensity.

\begin{figure}[htbp]
    \centering
    \includegraphics[width=0.48\textwidth]{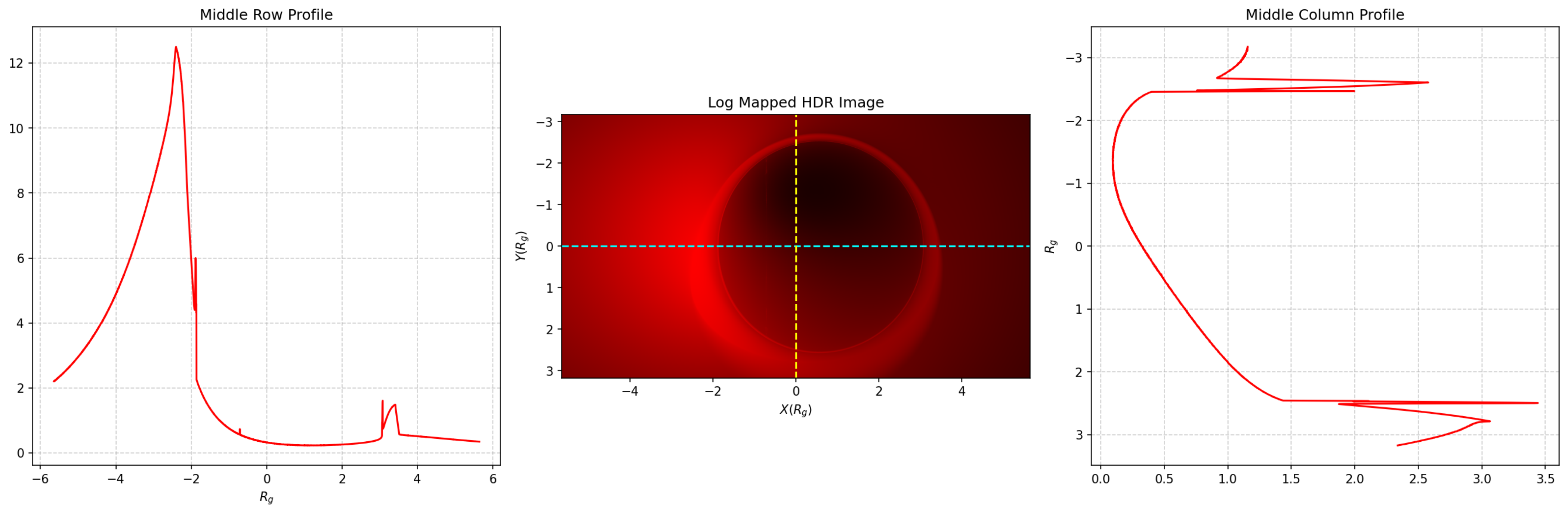} \\
    \vspace{0.2cm}
    \includegraphics[width=0.48\textwidth]{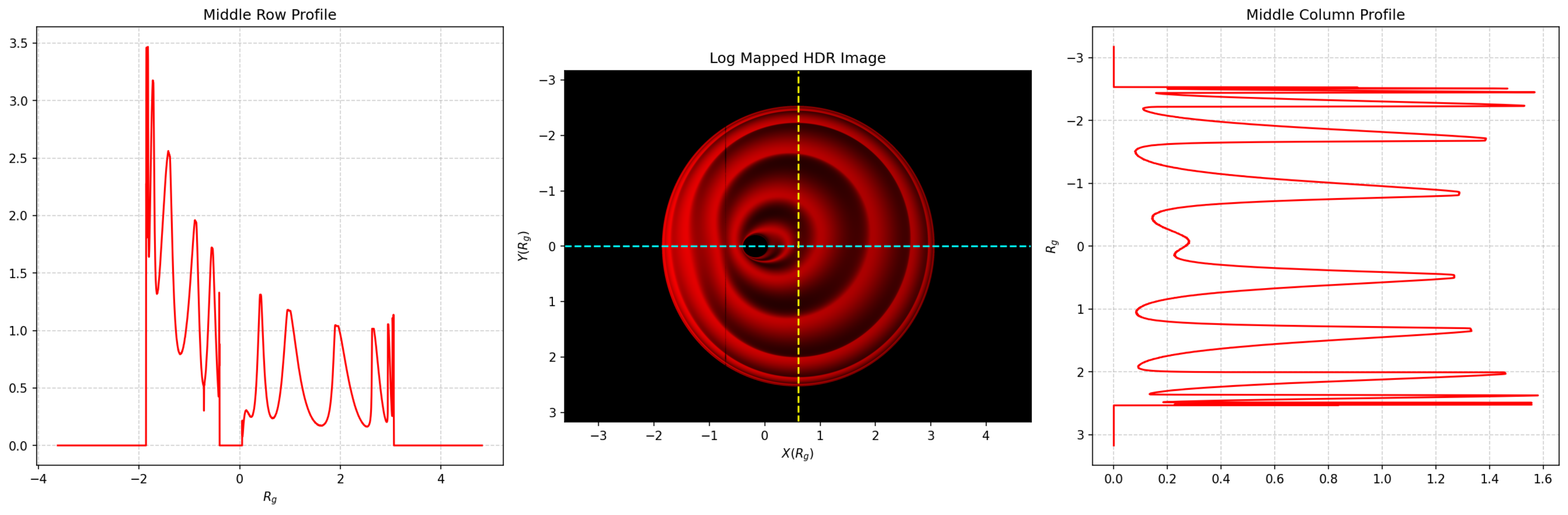} \\
    \vspace{0.2cm}
    \includegraphics[width=0.48\textwidth]{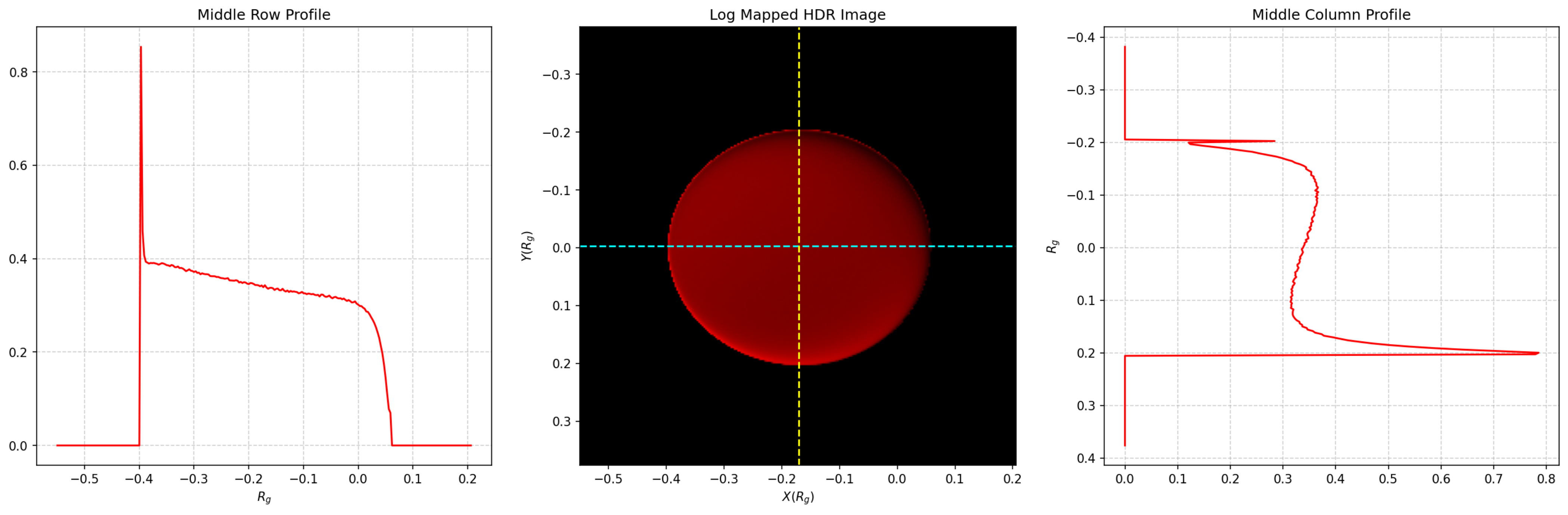}
    \caption{Images comparison between the Kerr black hole (top row) and the Kerr white hole (middle row and bottom row) with parameters spin $a=0.8$ and inclination $\Phi=45^\circ$. The thick disk model is used here. The middle column shows apparent morphology, and the left and right show horizontal and vertical intensity profiles. The top row and the middle row are for observer1 shown in Fig.~\ref{fig:penrose}, and the bottom row is for the observer2. Note that the size for these three images is different.}
    \label{fig:BH_vs_WH}
\end{figure}

Firstly we investigate the white hole image for observer1 shown in Fig.~\ref{fig:penrose}. In the top row of Fig.~\ref{fig:BH_vs_WH}, the Kerr black hole presents a crescent emission ring and Doppler beaming effects, with a black hole shadow in the center. Its intensity profile exhibits a single main peak at the edge, and the intensity in the shadow region is near zero.

In contrast, in the Kerr white hole image shown as the bottom row of Fig.~\ref{fig:BH_vs_WH}, the radiation from the previous universe penetrates the white hole interior. Photon distributions exist in the original shadow region, presenting a nested ring structure shifted to one side. In the horizontal intensity profile as shown in the left and right columns of Fig.~\ref{fig:BH_vs_WH}, the prograde side shows a series of tightly clustered intensity peaks due to frame-dragging. The retrograde side corresponds to a series of sparse secondary intensity peaks. Compared to the low-intensity shadow at the center of the black hole, the eccentric nested rings and multiple intensity oscillation peaks constitute the primary structural feature of the white hole image in the intensity distribution.

From Fig.~\ref{fig:BH_vs_WH} we can see a characteristic feature of white holes, the nested ring structure, which can not be mimicked by complicated disks. This behavior is very different to the nonrotating white holes \cite{2025ChPhC..49b5109L}. In the following we will show this feature is very robust against the variation of system parameters.

Secondly we investigate the white hole image for observer2 shown in Fig.~\ref{fig:penrose}. We find that the image appears as a disk which is very similar to the image of a non-rotating white hole \cite{2025ChPhC..49b5109L}. This kind of behavior is difficult from distinguish with images of black holes when the accretion disk is complicated.

\begin{figure}
    \centering
\begin{tabular}{ccc}
\includegraphics[width=0.15\textwidth]{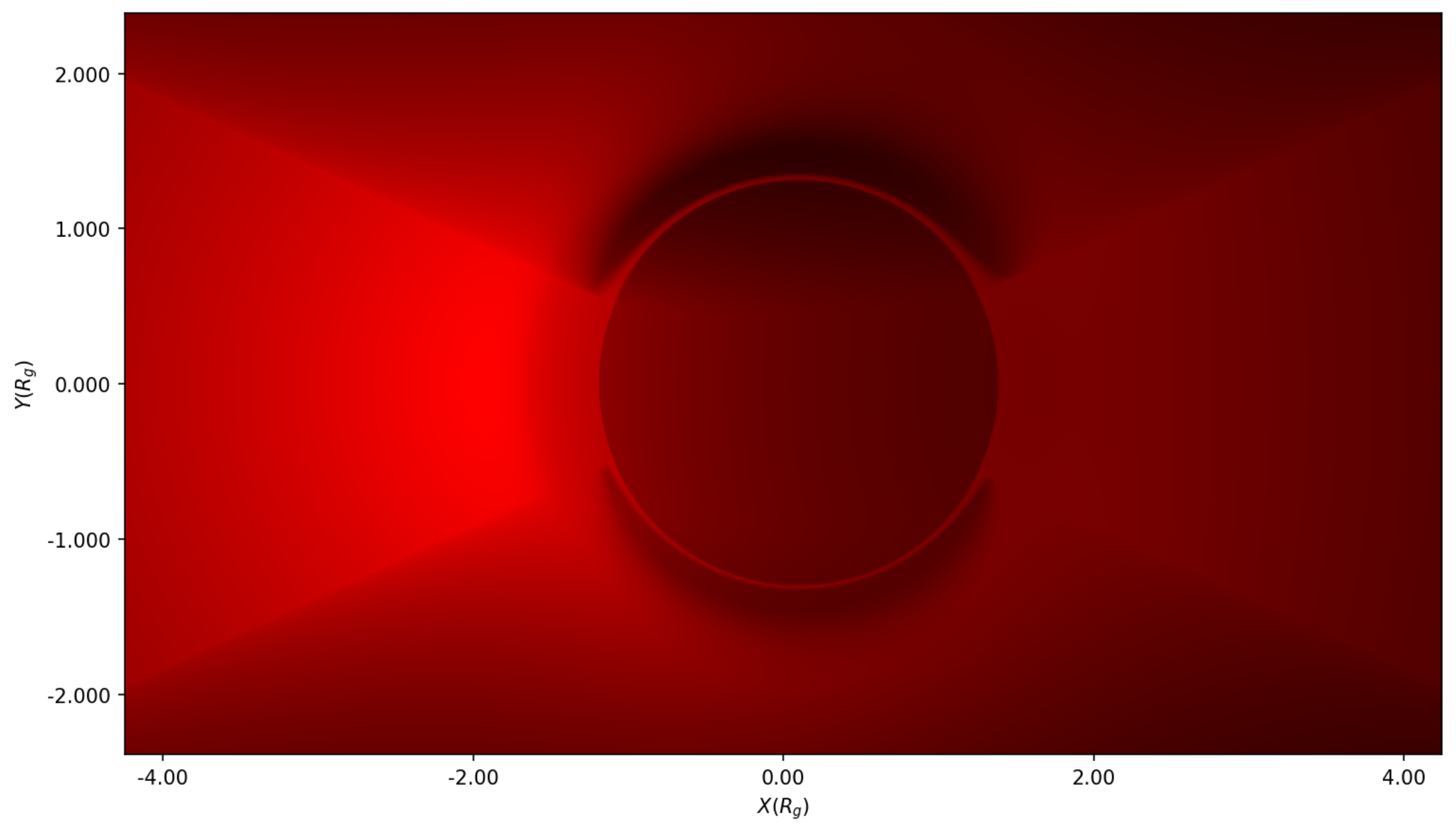}&
\includegraphics[width=0.15\textwidth]{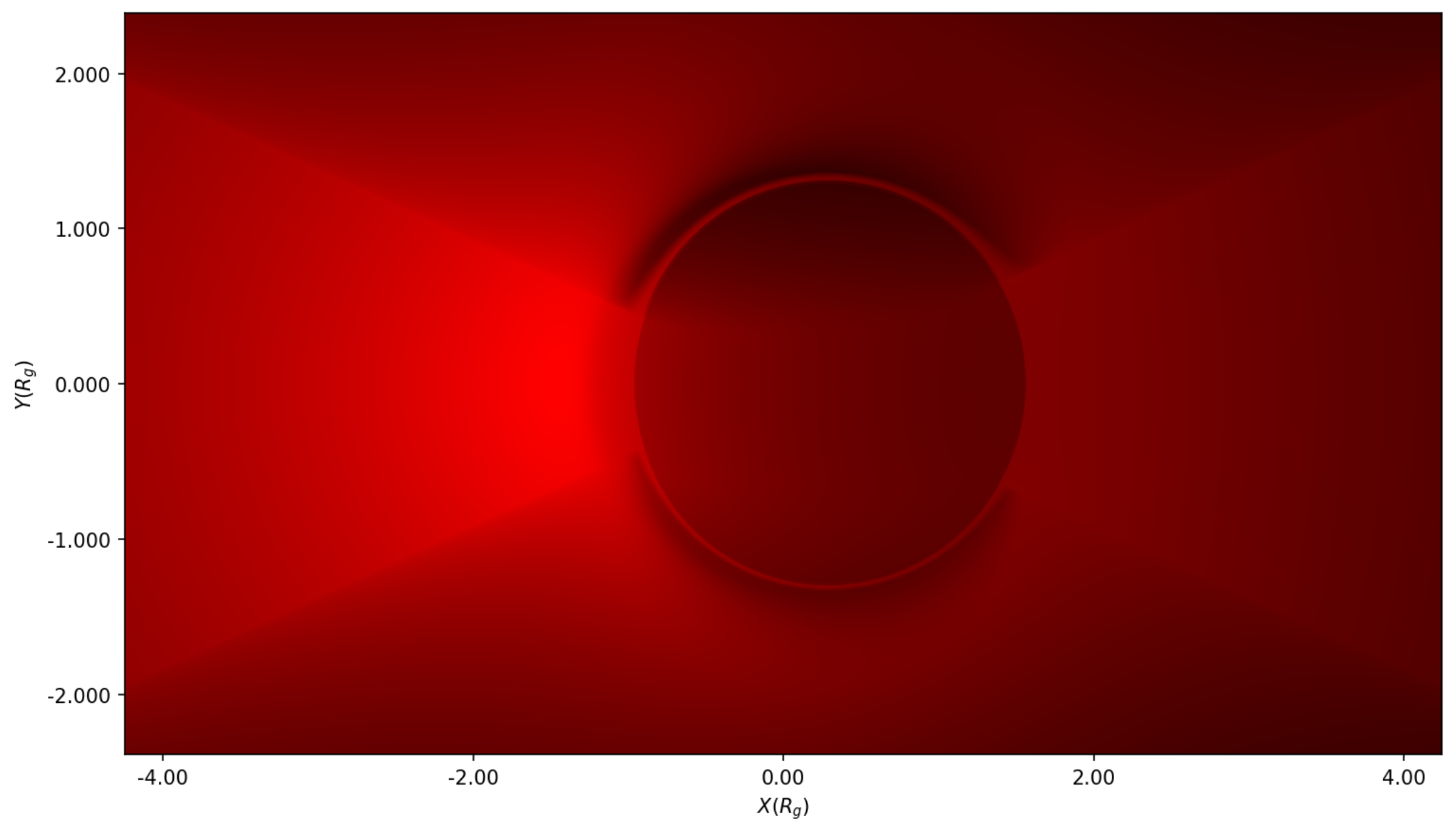}&
\includegraphics[width=0.15\textwidth]{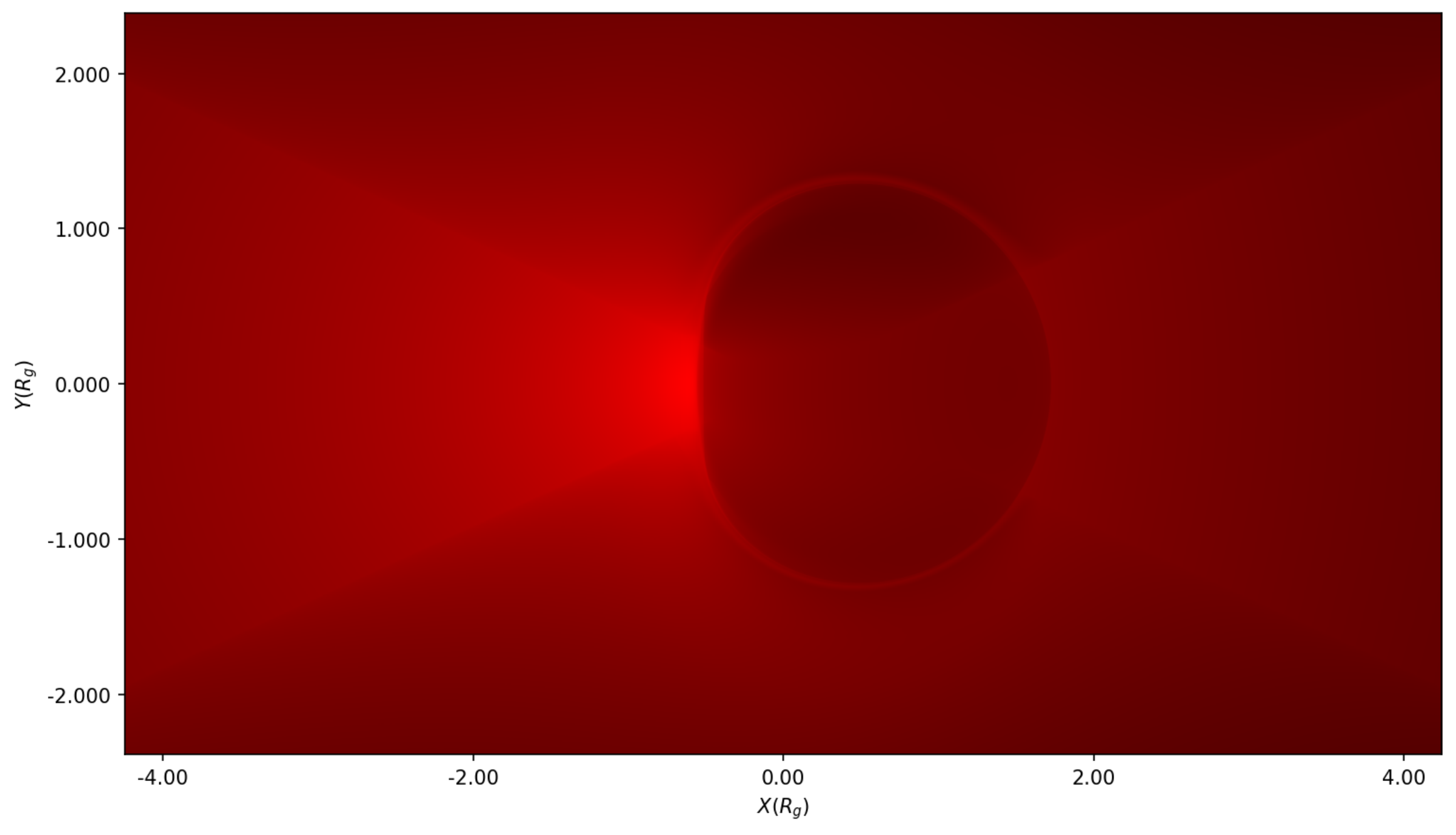}\\
\includegraphics[width=0.15\textwidth]{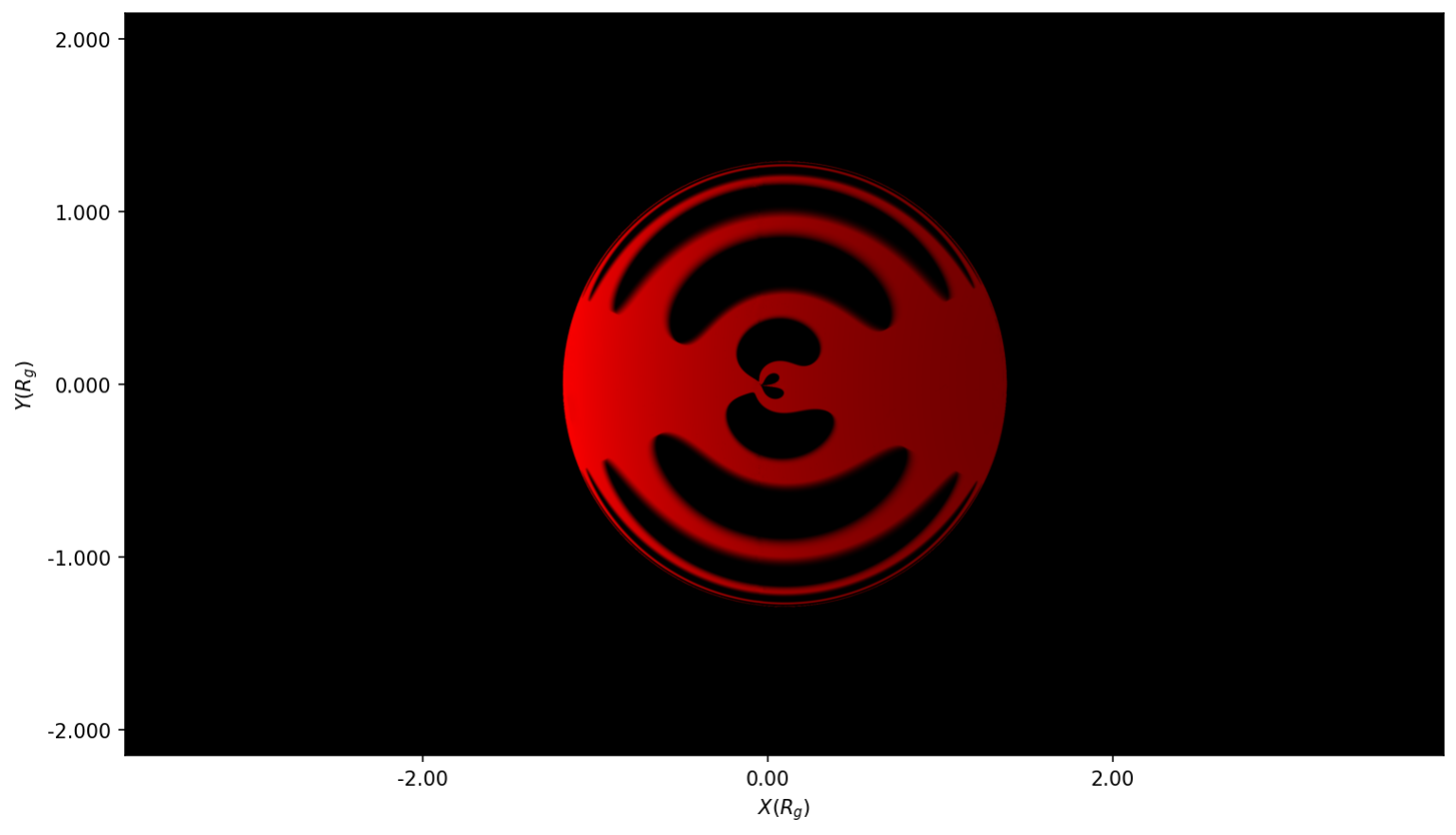}&
\includegraphics[width=0.15\textwidth]{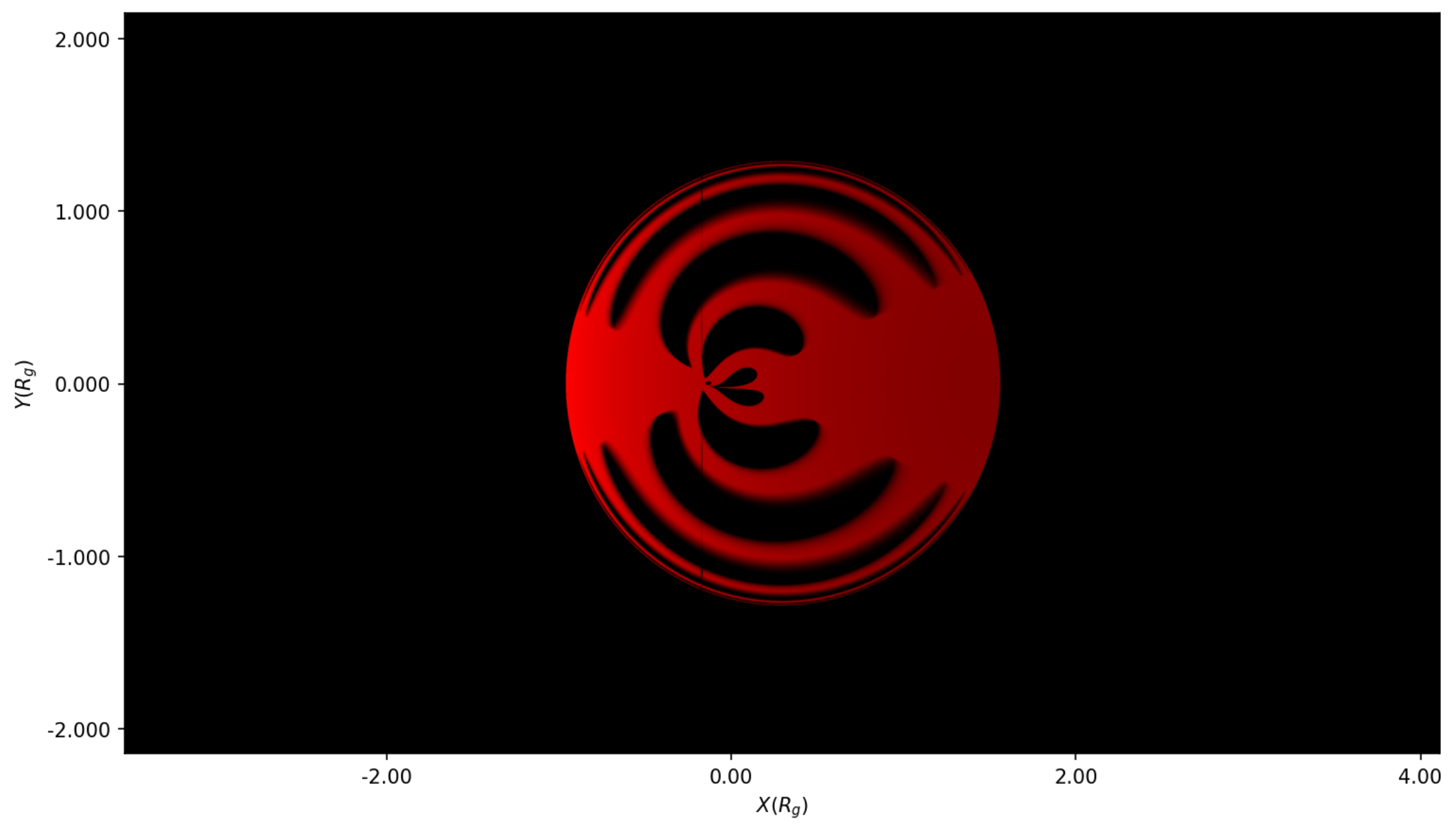}&
\includegraphics[width=0.15\textwidth]{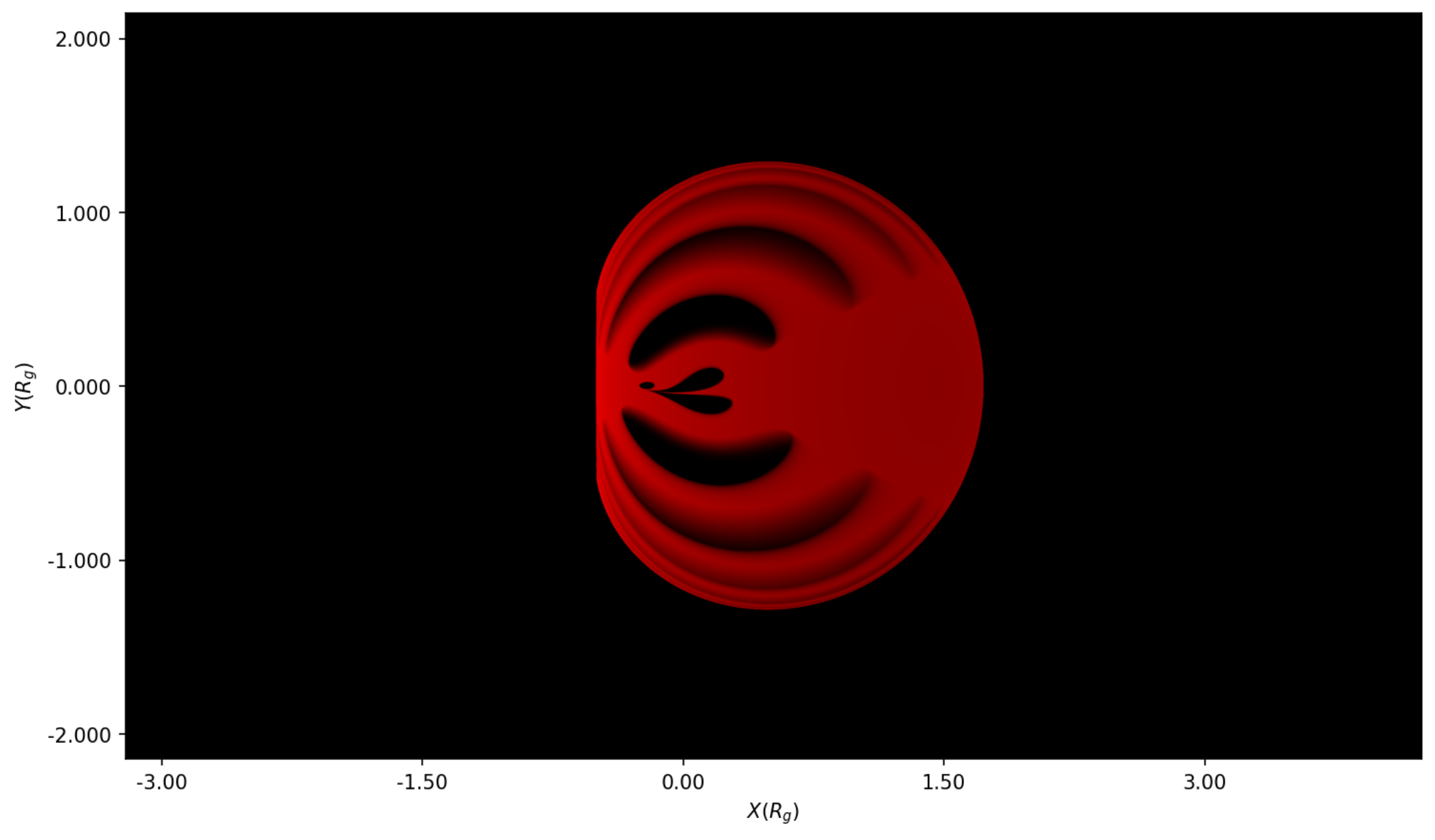}\\
\includegraphics[width=0.15\textwidth]{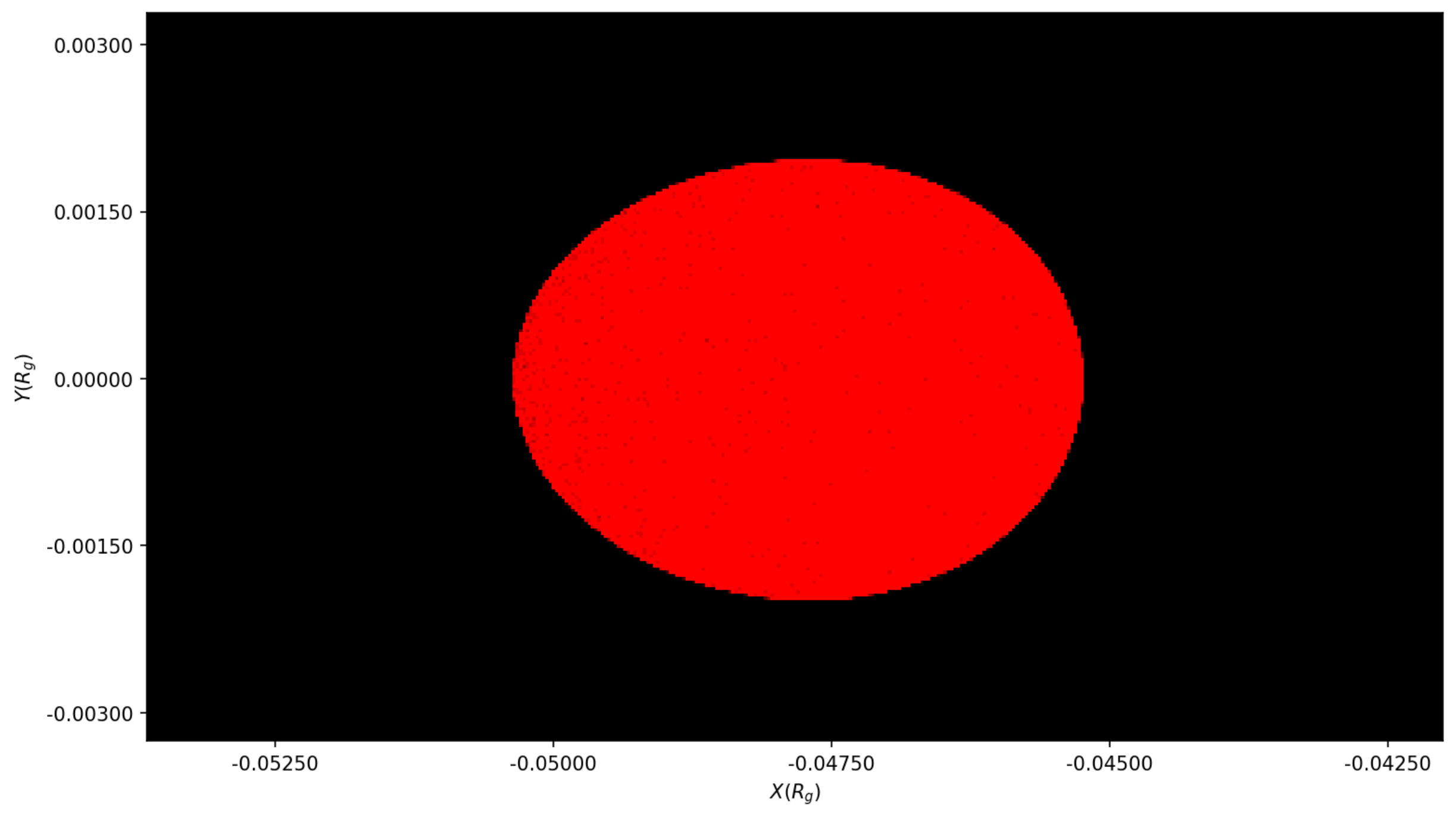}&
\includegraphics[width=0.15\textwidth]{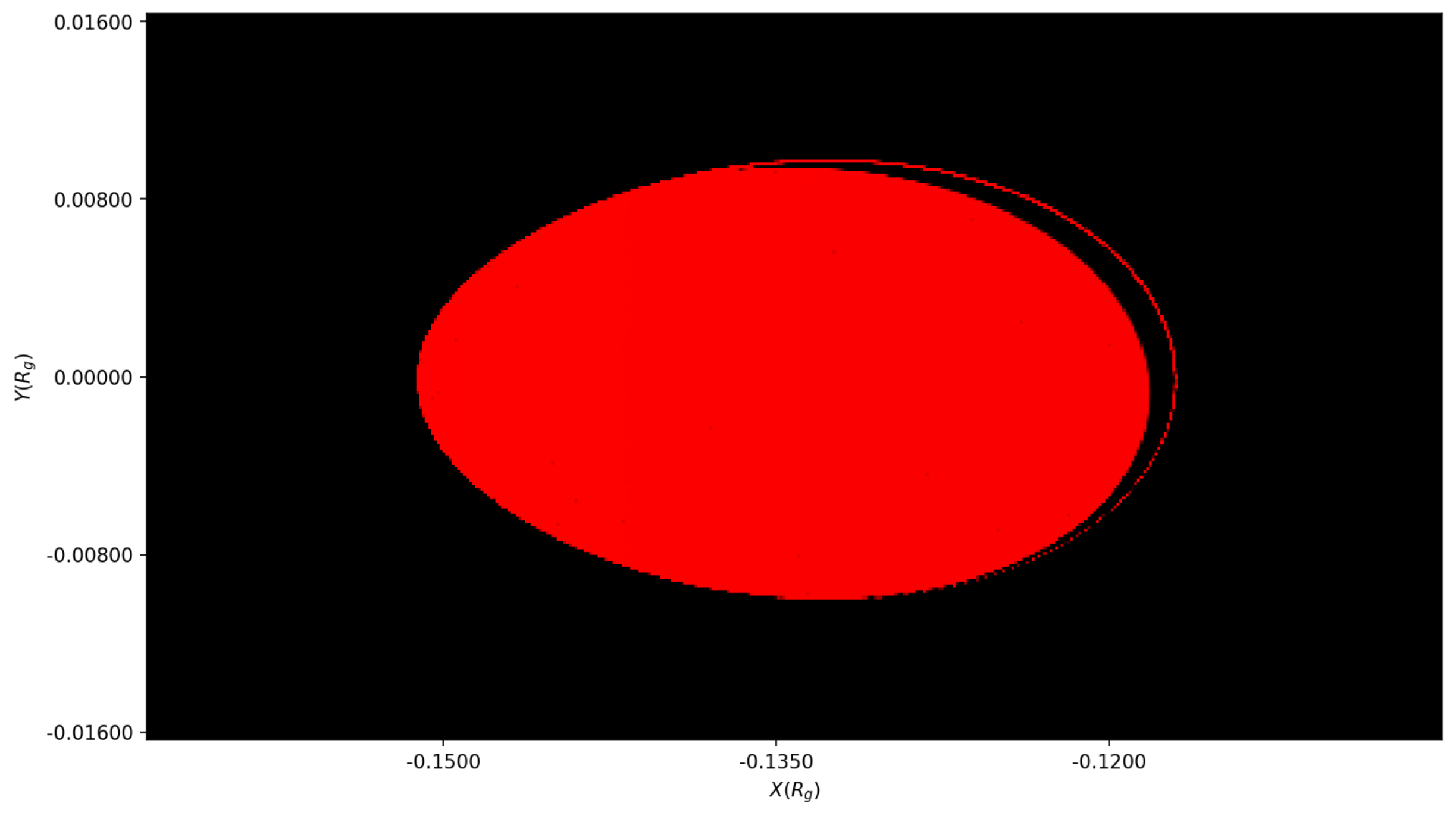}&
\includegraphics[width=0.15\textwidth]{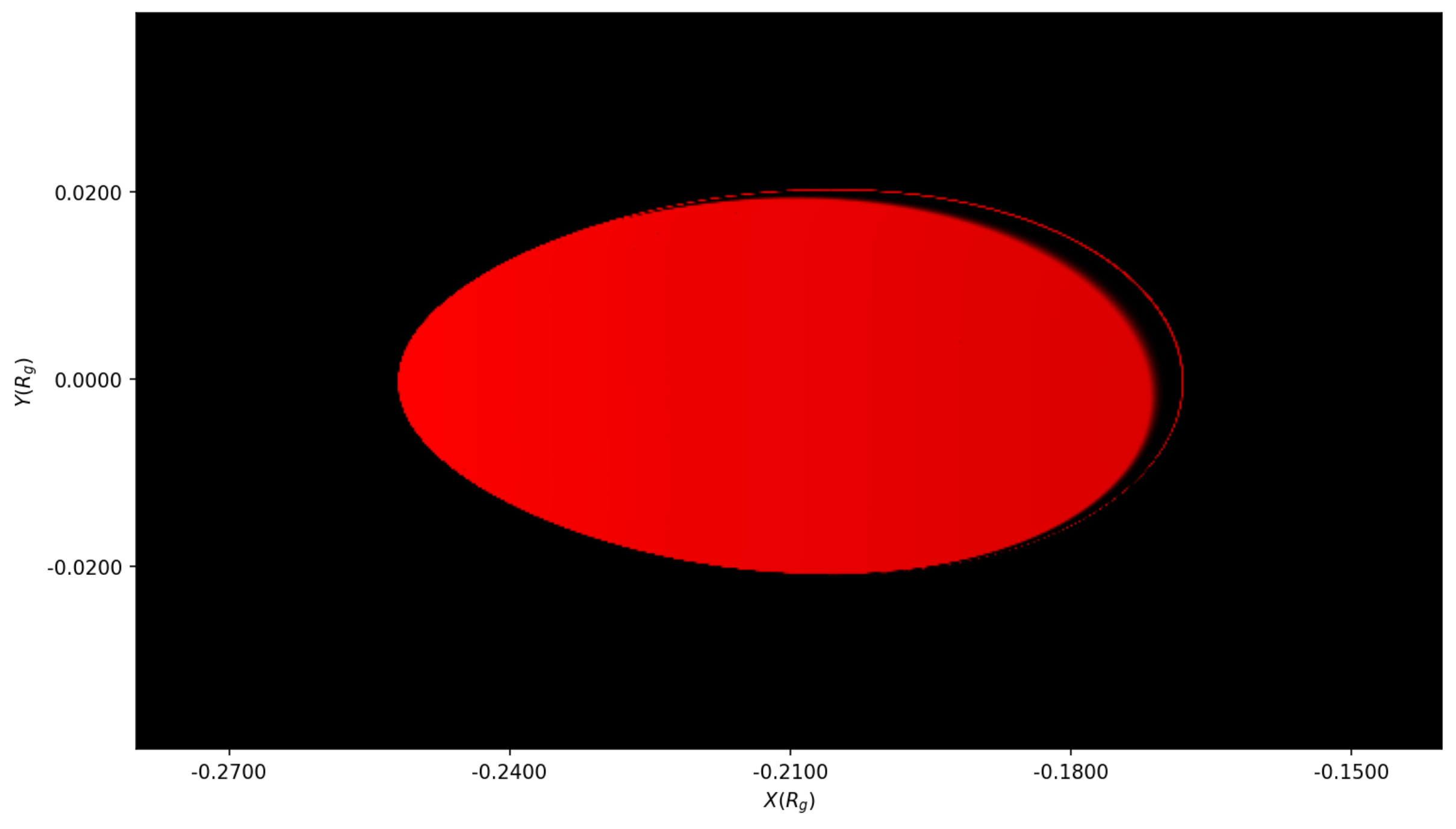}
\end{tabular}
    \caption{Dependence of black hole/white hole images on spin. Left column $a=0.2$, middle column $a=0.6$, right column $a=1$. This figure corresponds to Fig.~\ref{fig:BH_vs_WH} but with different spin parameters.}
    \label{fig:spin_effect}
\end{figure}

We check the effect of spin parameter on the nested ring feature of white hole image. We compare spin parameters $a=0.2$ and $a=0.8$ with inclination $\Phi=80^\circ$ in Fig.~\ref{fig:spin_effect}. Here we use hopper disk model. In the lower spin $a=0.2$ case, the offset of the inner ring of the white hole is small, and the intensity distribution on both sides is similar. After the spin increases to $a=0.8$, the frame-dragging effect of spacetime strengthens, and the photon orbits present an asymmetric distribution. The emission ring on one side of the image contracts inward. Calculation results show that the increase in the spin parameter amplifies the asymmetry of the transverse intensity distribution.

\begin{figure}
\begin{tabular}{ccc}
\includegraphics[width=0.15\textwidth]{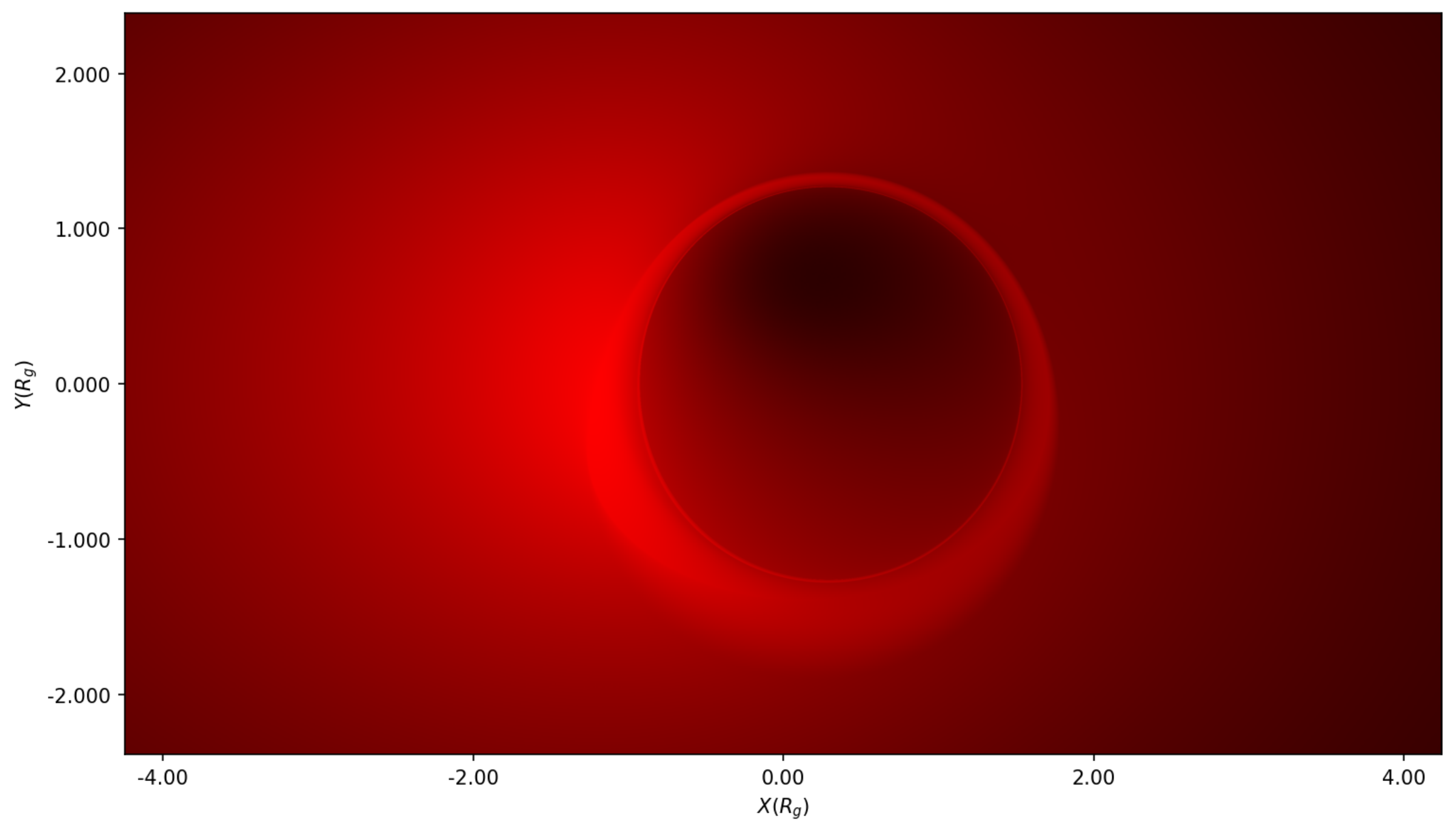}&
\includegraphics[width=0.15\textwidth]{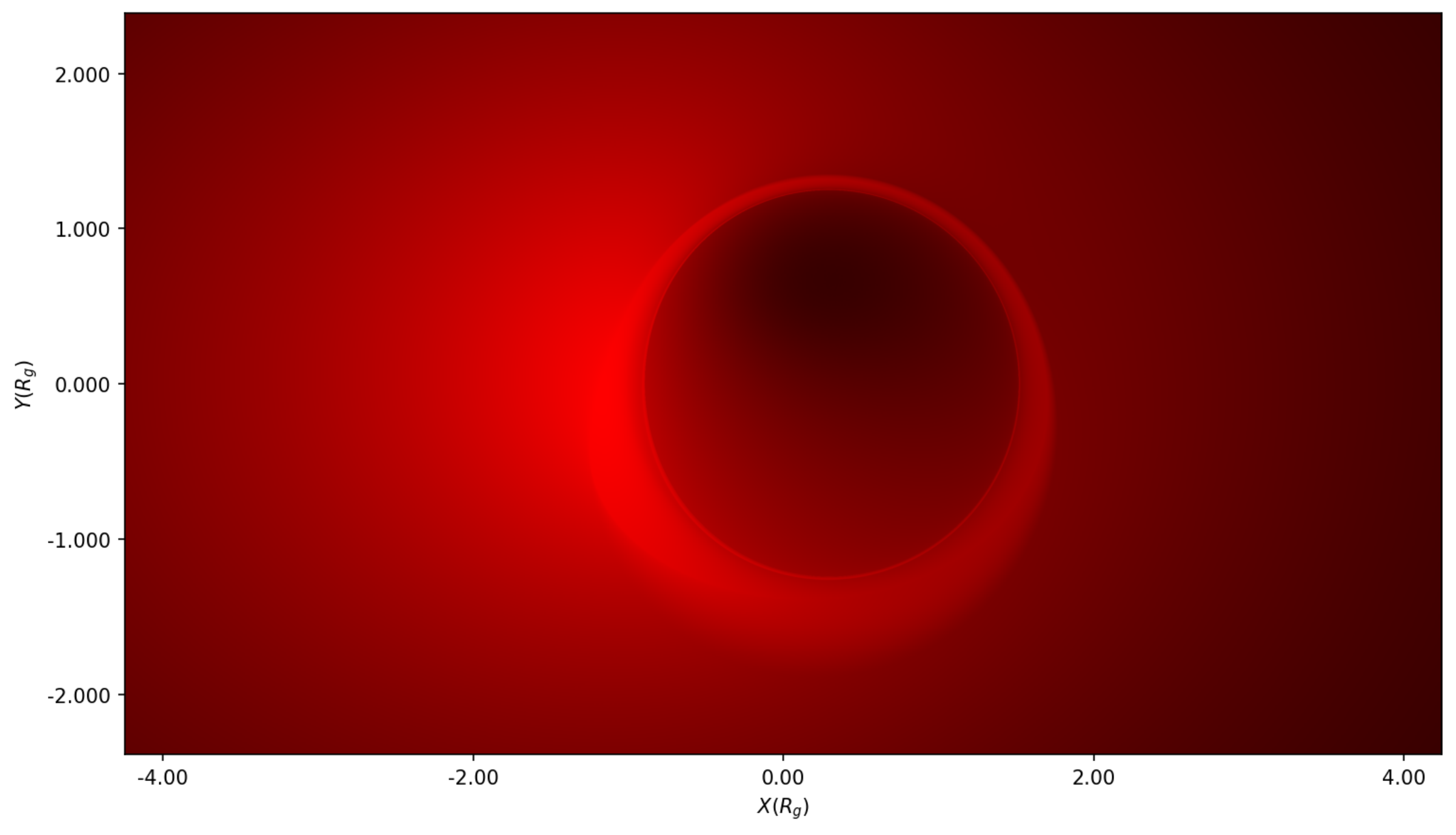}&
\includegraphics[width=0.15\textwidth]{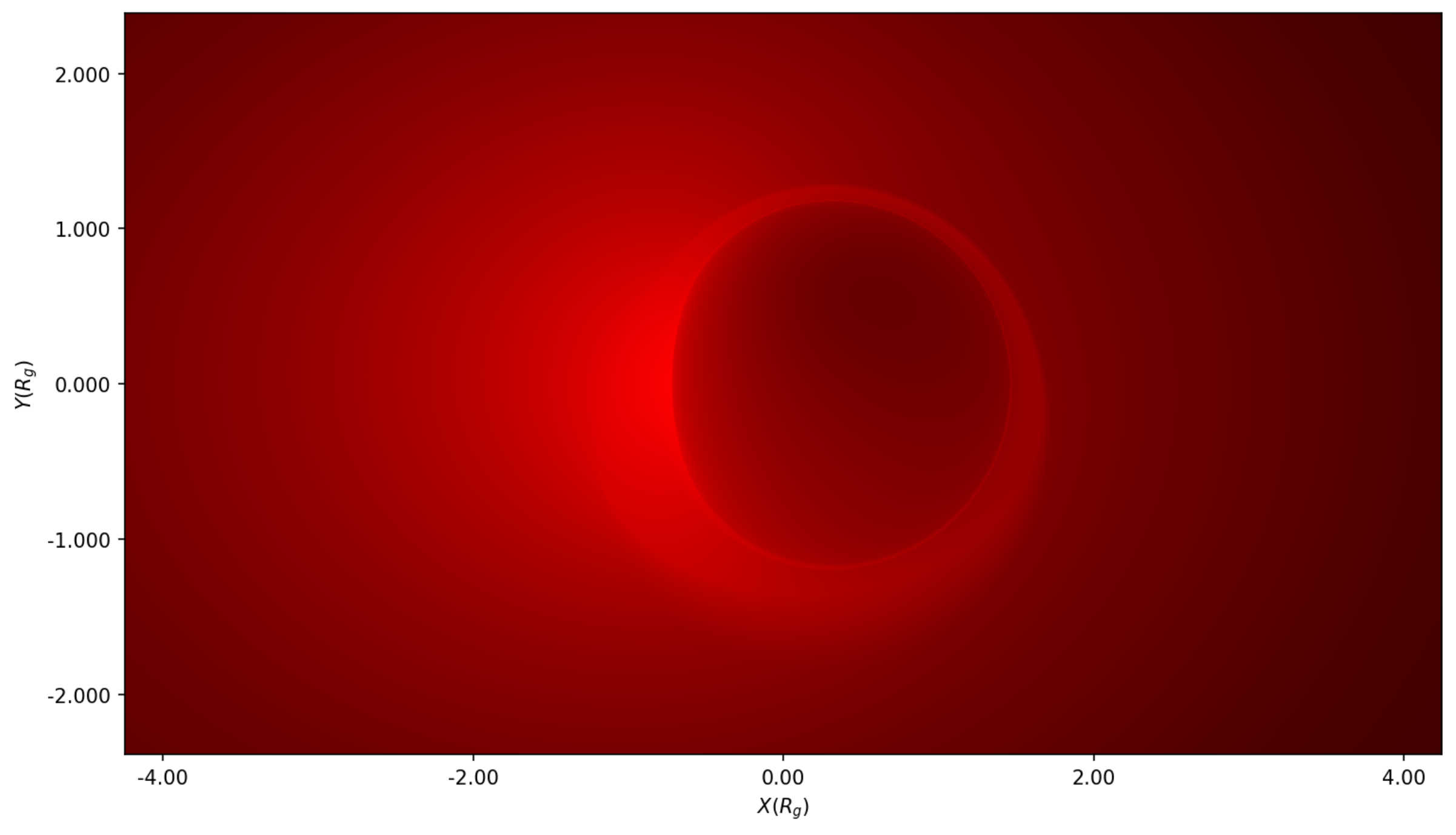}\\
\includegraphics[width=0.15\textwidth]{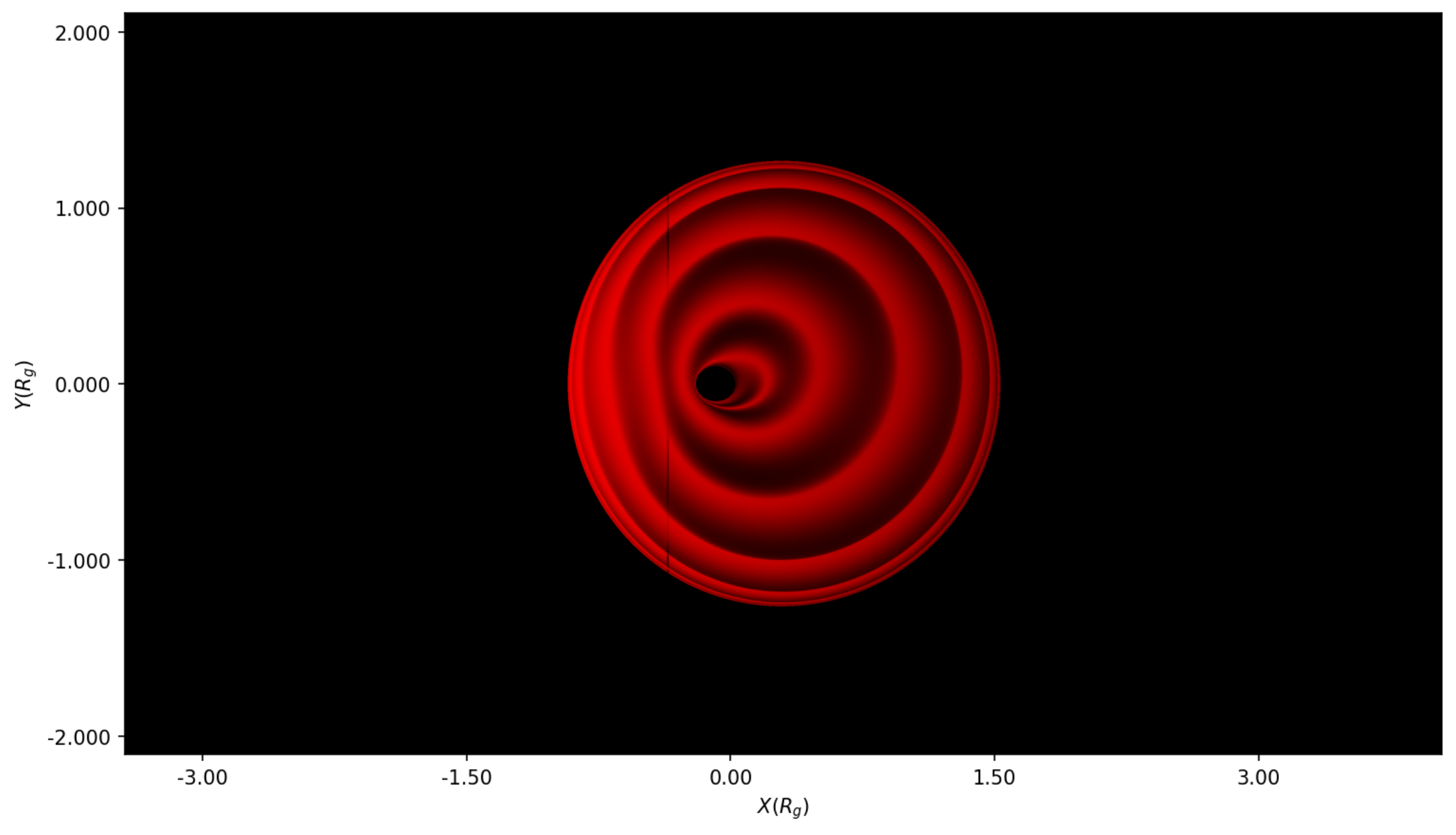}&
\includegraphics[width=0.15\textwidth]{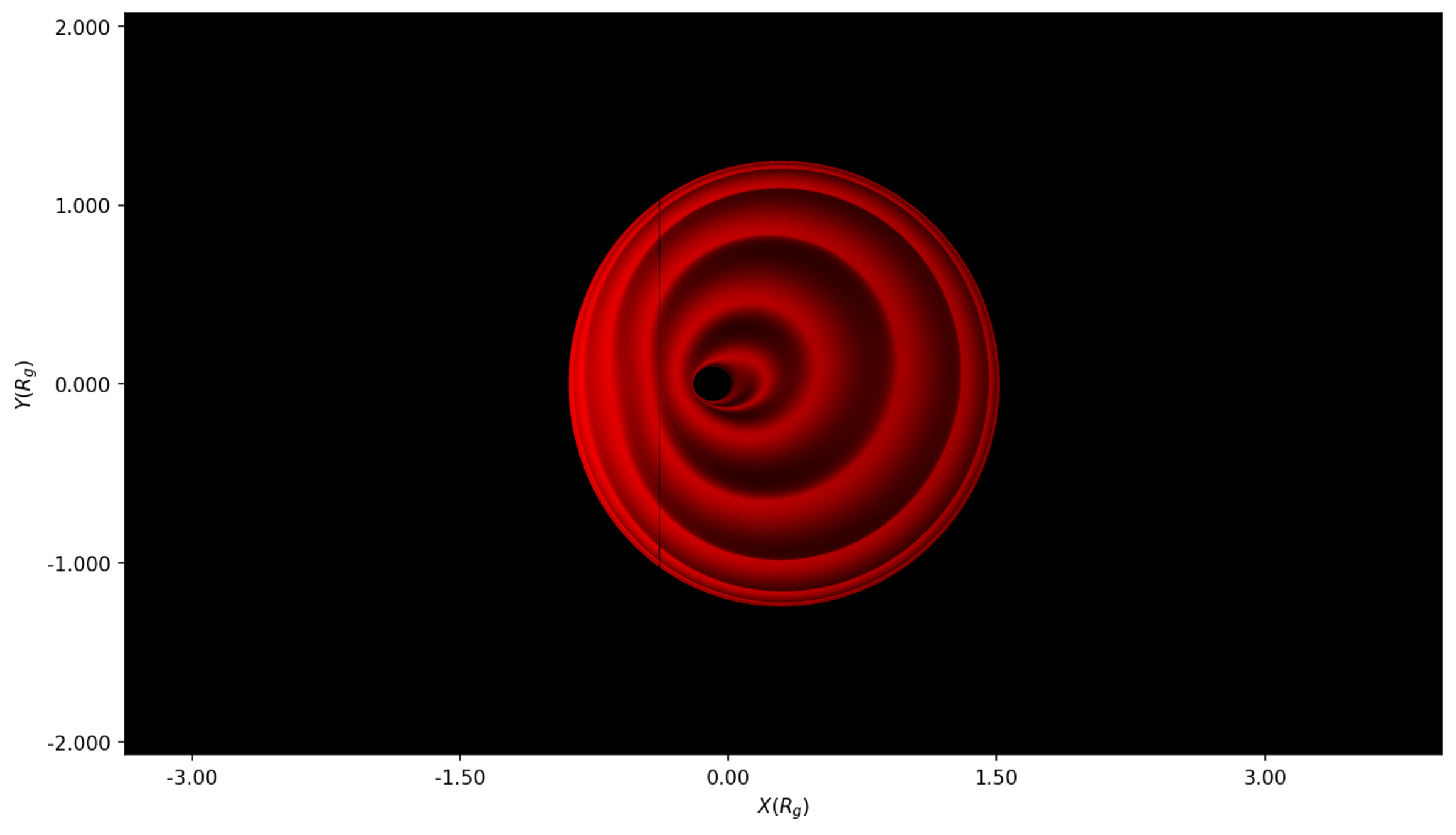}&
\includegraphics[width=0.15\textwidth]{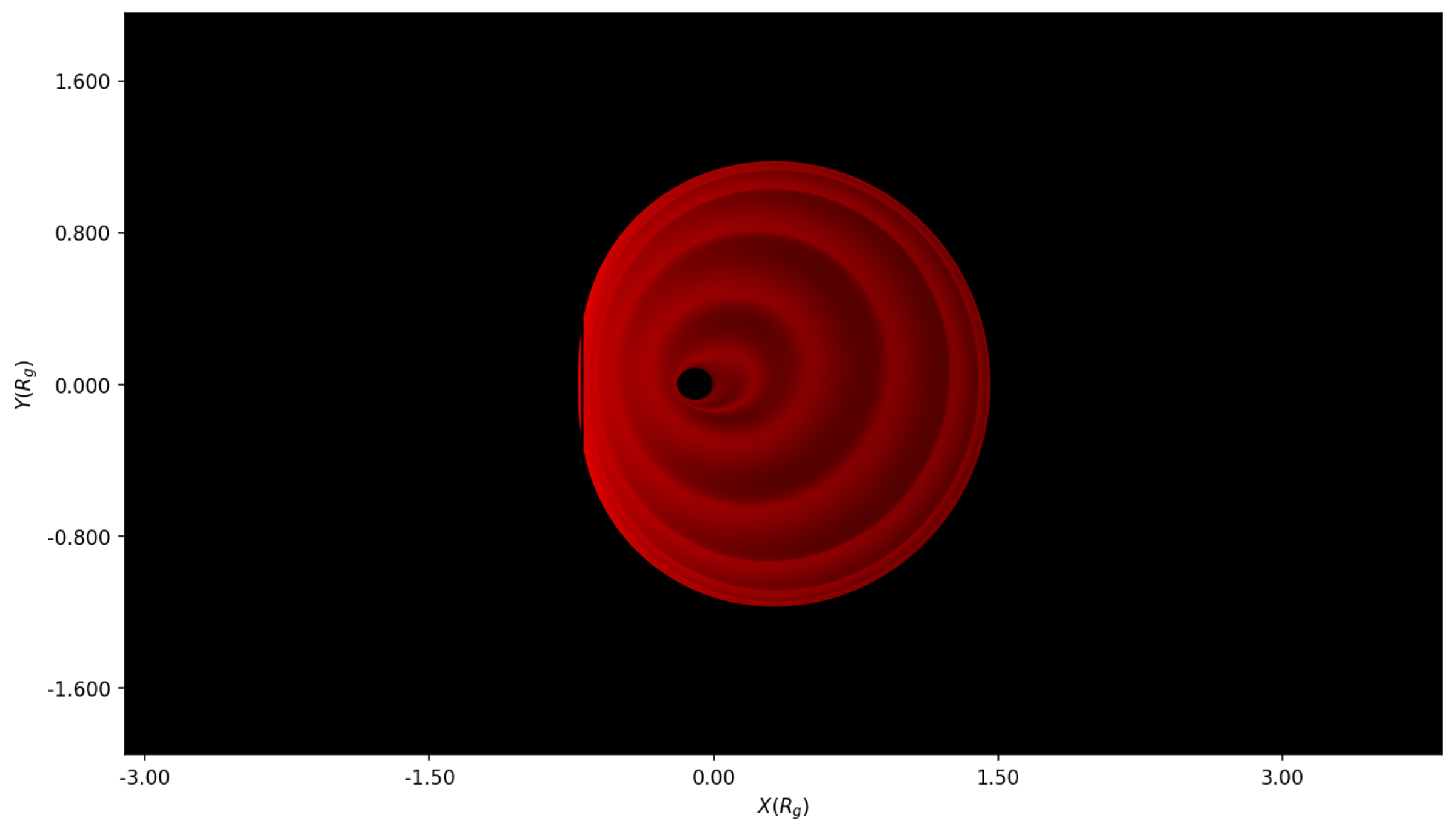}\\
\includegraphics[width=0.15\textwidth]{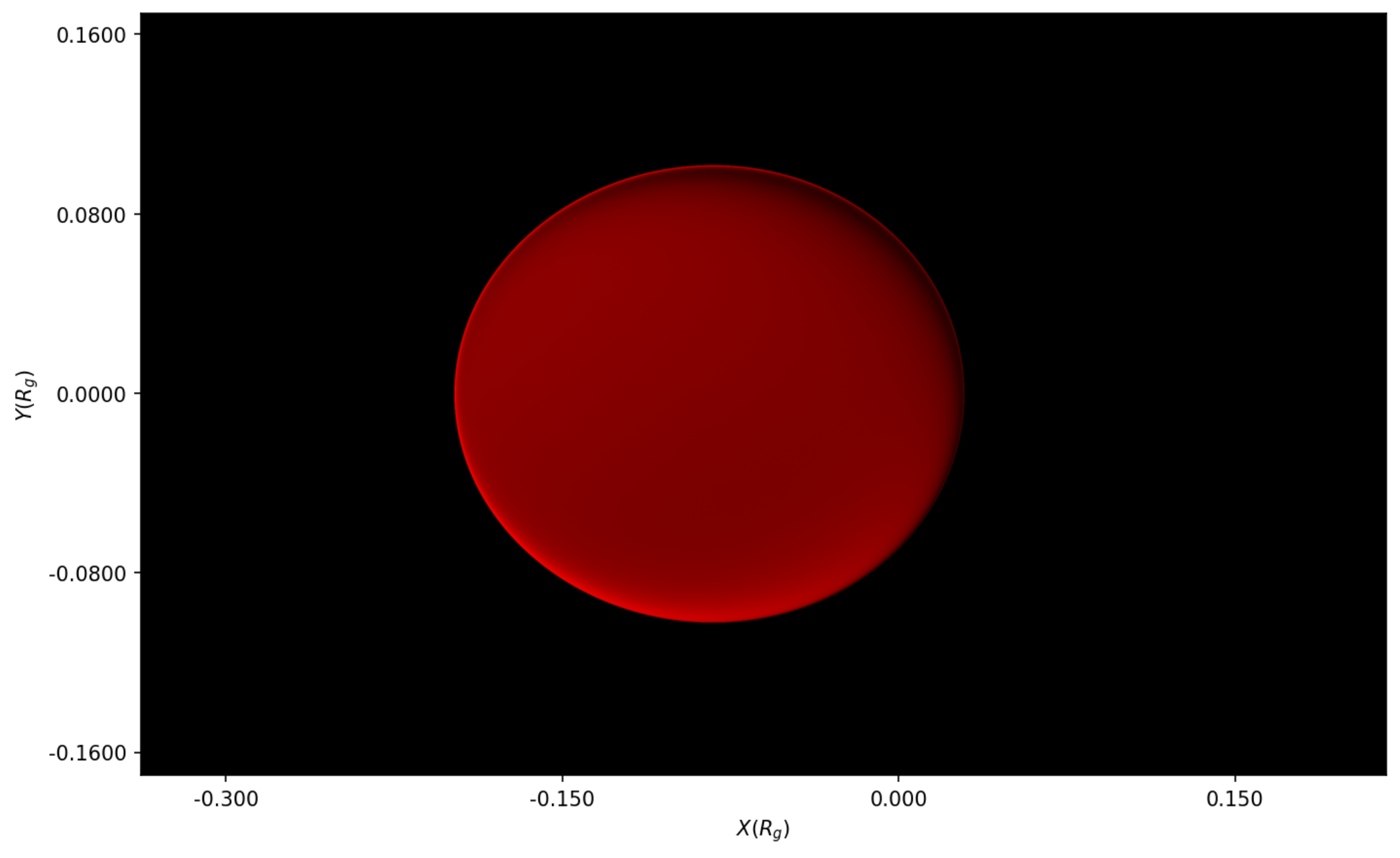}&
\includegraphics[width=0.15\textwidth]{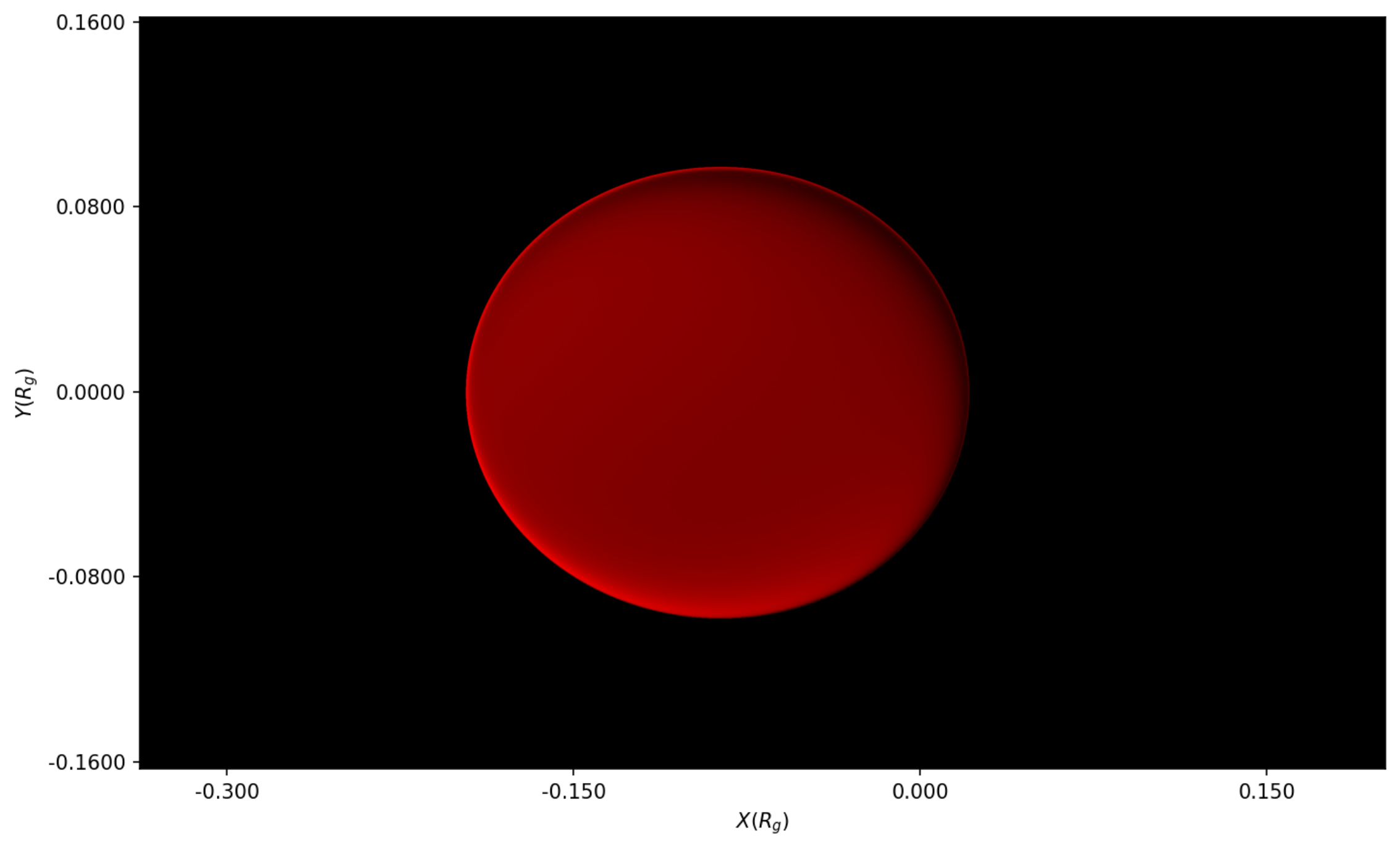}&
\includegraphics[width=0.15\textwidth]{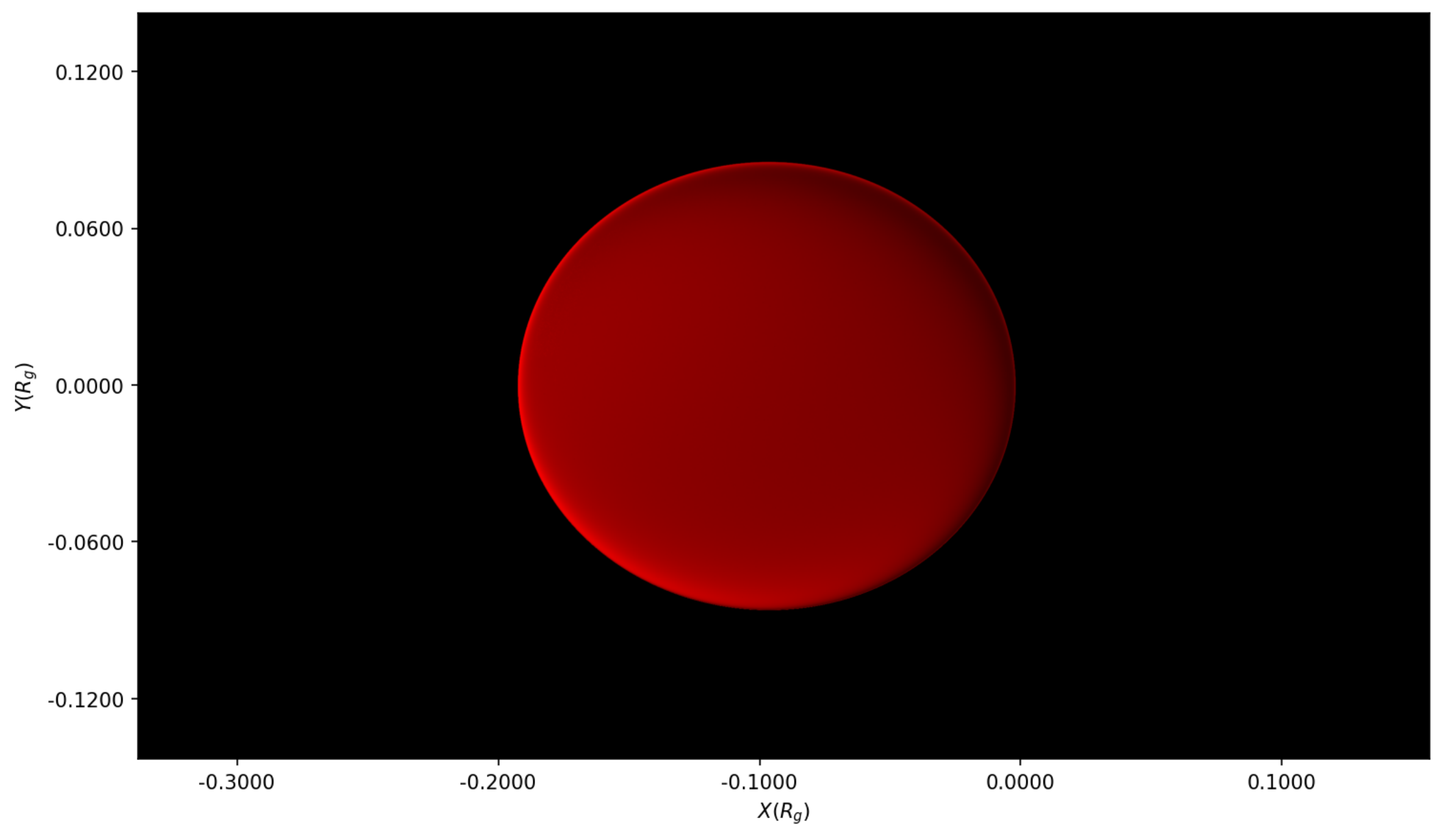}
\end{tabular}
    \caption{Dependence of black hole/white hole images on electric charge. Left column $Q=0.1$, middle column $Q=0.3$, right column $Q=0.6$. This figure corresponds to Fig.~\ref{fig:BH_vs_WH} but with an electric charge.}
    \label{fig:charge_effect}
\end{figure}

In addition to the spin parameter $a$, the Kerr-Newman (KN) spacetime metric may contain a charge parameter $Q$. To examine the influence of charge on nested ring features of white hole images we compare image behaviors for different charge parameter systems in Fig.~\ref{fig:charge_effect}. Under identical mass and spin conditions, the presence of charge causes the gravitational potential well and photon orbits to contract inward overall. Reflected in the images, the apparent angular sizes of the black hole shadow region and the white hole internal intensity nested ring structures decrease to varying degrees. Accordingly the nested ring feature remains well.

\begin{figure}
\begin{tabular}{cc}
\includegraphics[width=0.24\textwidth]{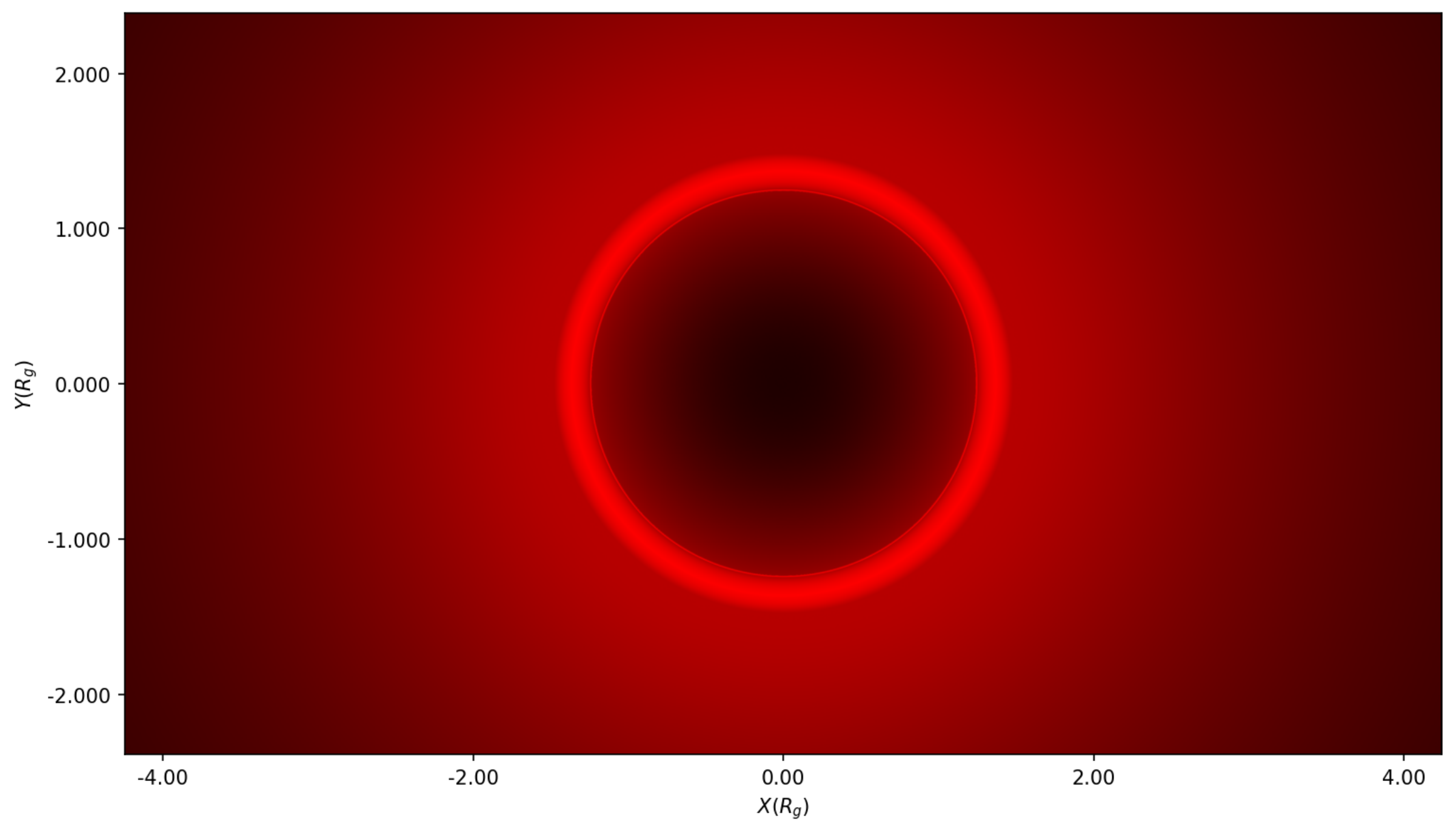}&
\includegraphics[width=0.24\textwidth]{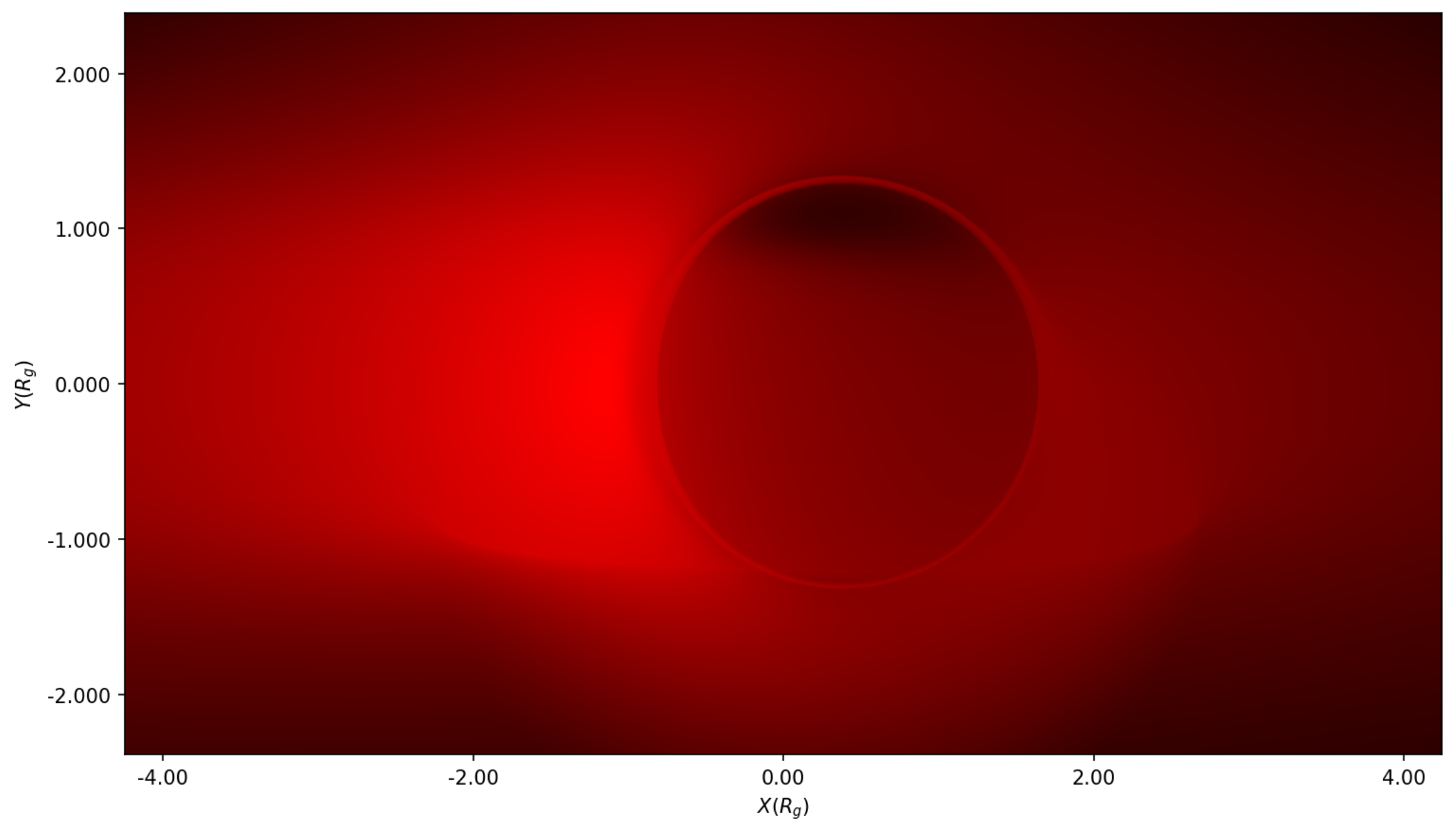}\\
\includegraphics[width=0.24\textwidth]{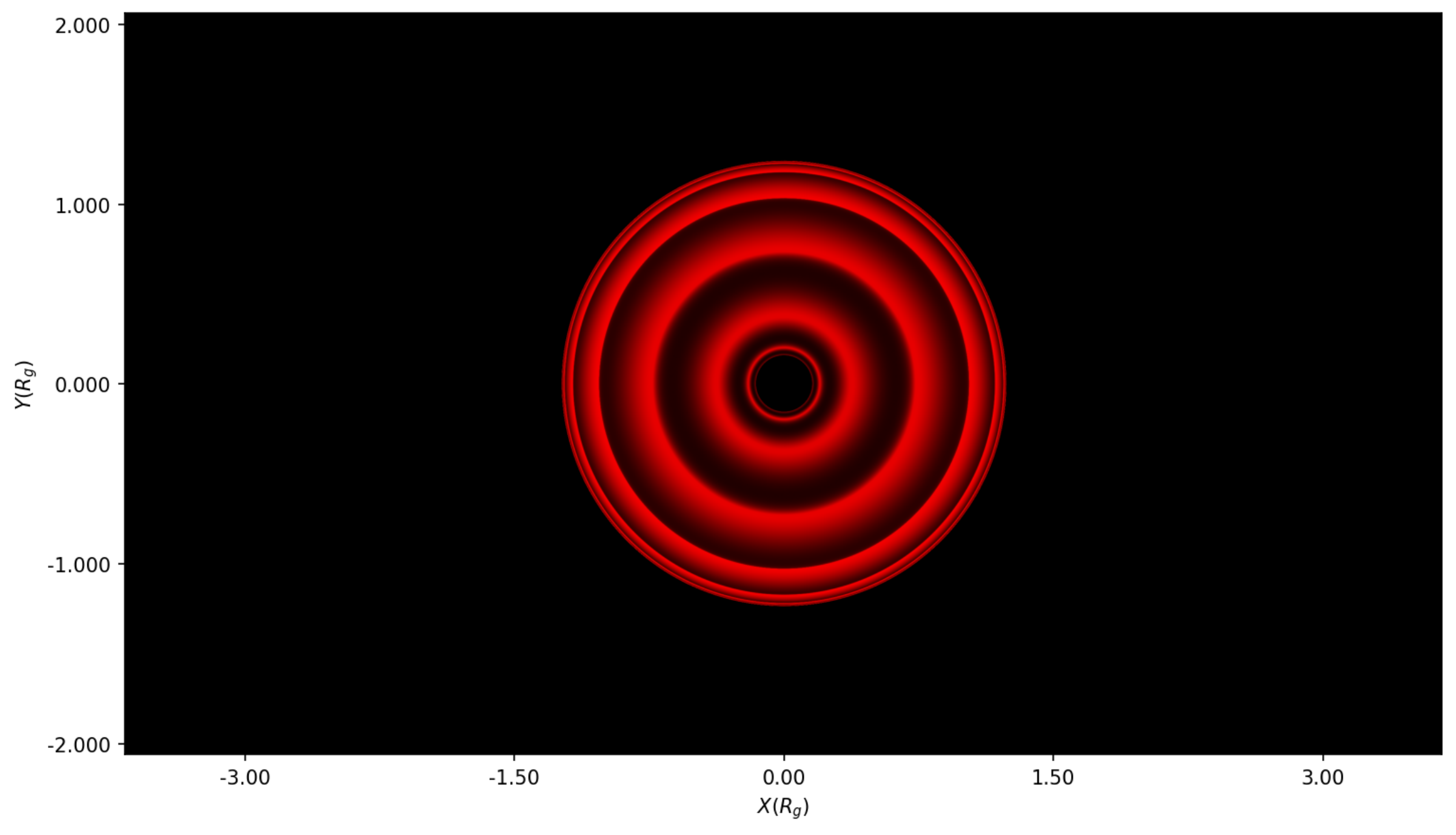}&
\includegraphics[width=0.24\textwidth]{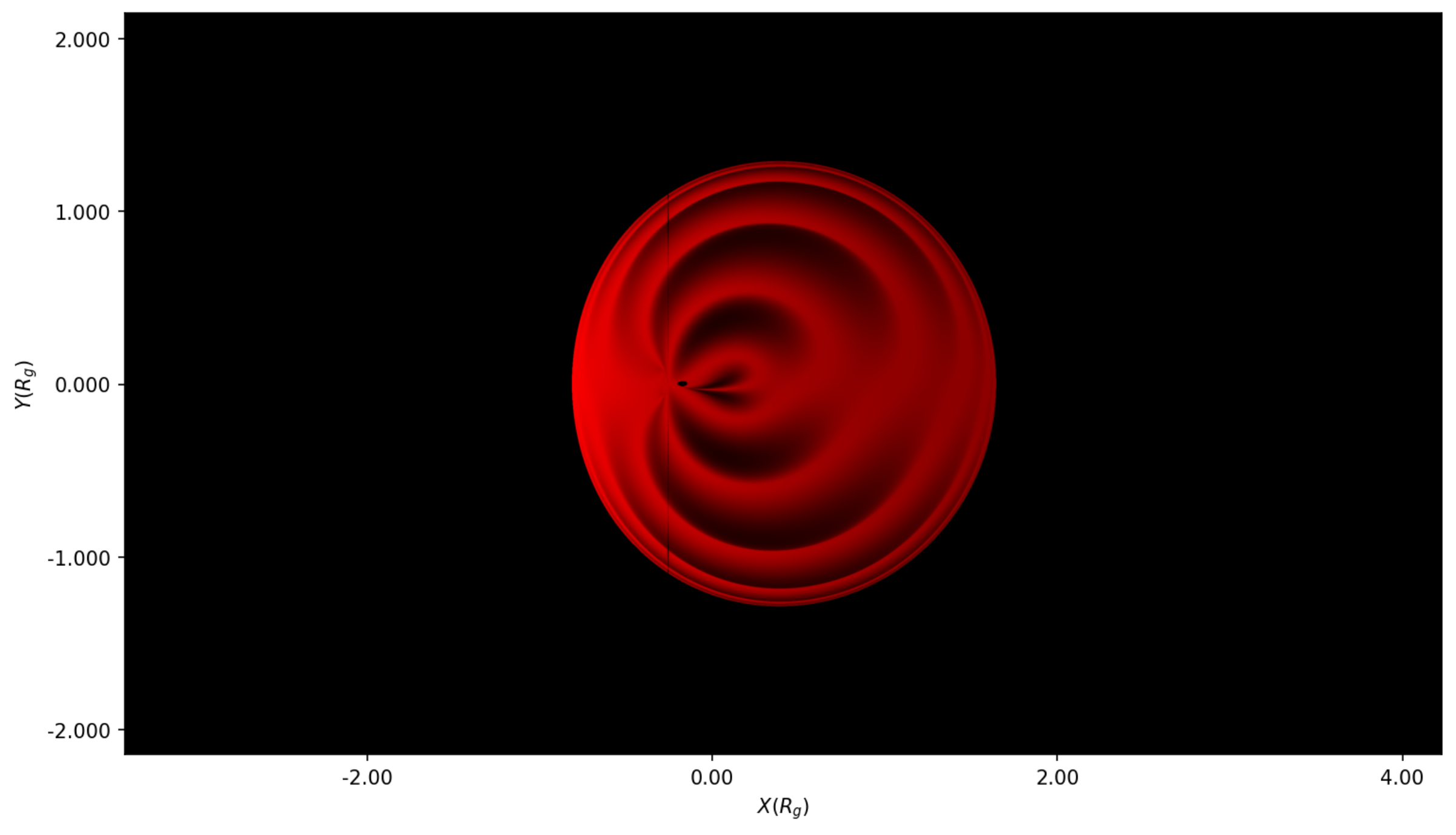}\\
\includegraphics[width=0.24\textwidth]{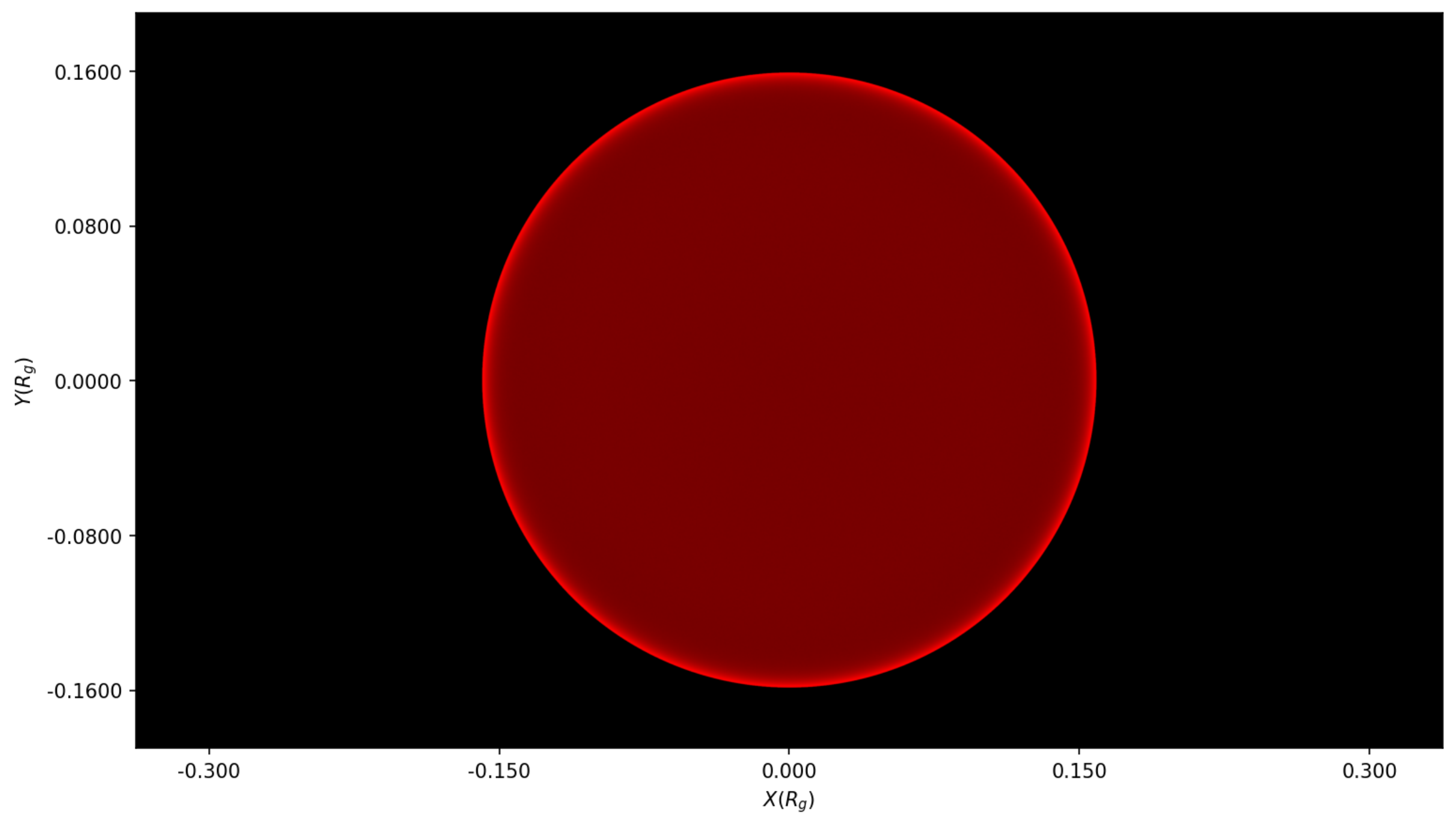}&
\includegraphics[width=0.24\textwidth]{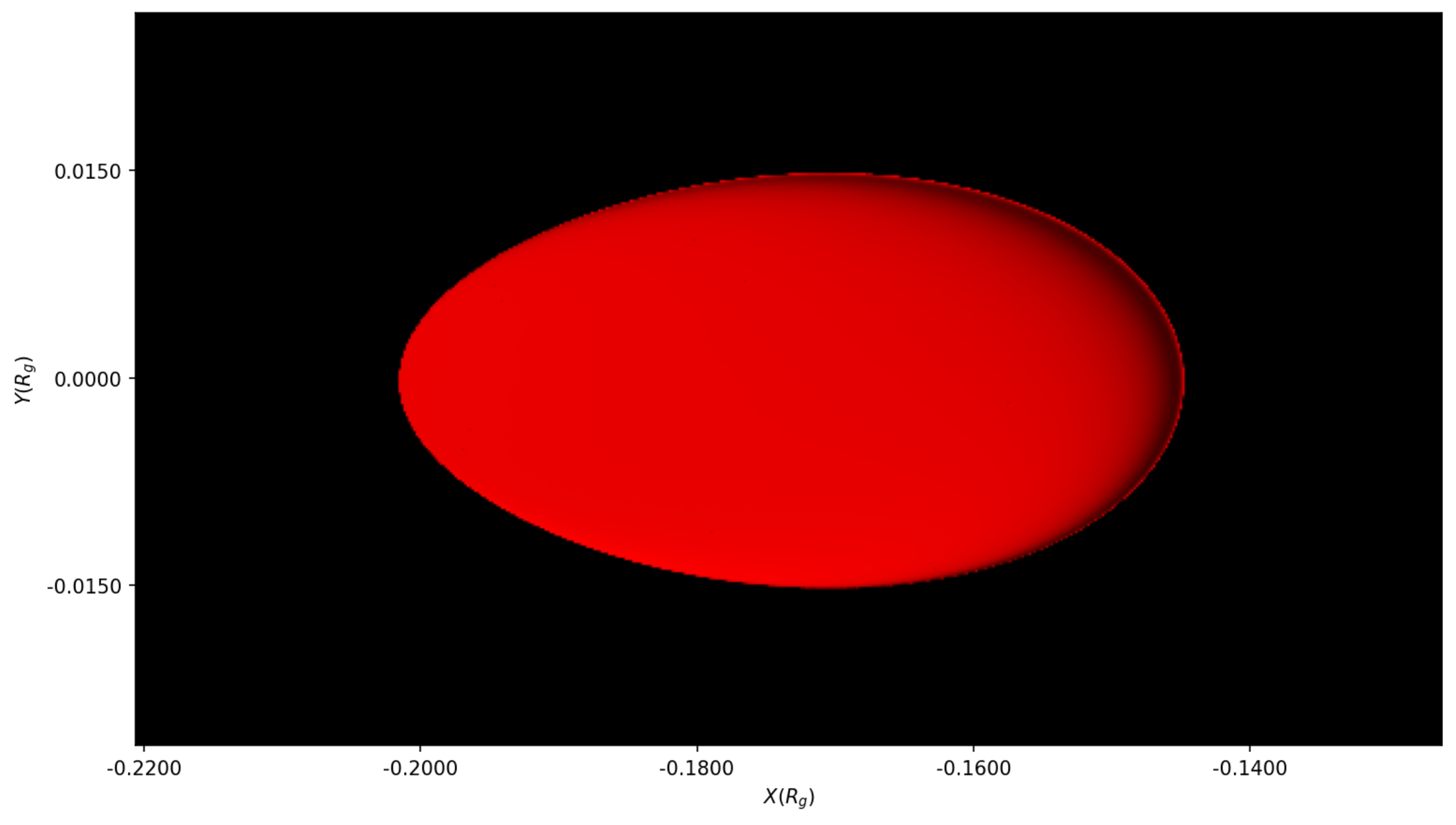}
\end{tabular}
    \caption{Dependence of black hole/white hole images on observation inclination. Left images are for polar observation $\Phi=0^\circ$, and right images are for large inclination observation with $\Phi=80^\circ$. Other setting is the same to Fig.~\ref{fig:BH_vs_WH}.}
    \label{fig:phi_effect}
\end{figure}

We check the effect of the observation inclination on the nested ring feature in Fig.~\ref{fig:phi_effect}. For comparison we fix spin parameter $a=0.8$. When observed in the polar axis direction corresponding to $\Phi=0^\circ$, the line of sight is collinear with the spin axis. Both black hole and white hole images maintain rotational symmetry, and the horizontal and vertical intensity profiles coincide. When the observation angle shifts close to the equatorial plane with $\Phi=80^\circ$, affected by Doppler beaming and gravitational lensing, the longitudinal cross-section still roughly exhibits a symmetric state, while the transverse cross-section shows significant asymmetry.

\begin{figure}
    \centering
\begin{tabular}{cc}
\includegraphics[width=0.24\textwidth]{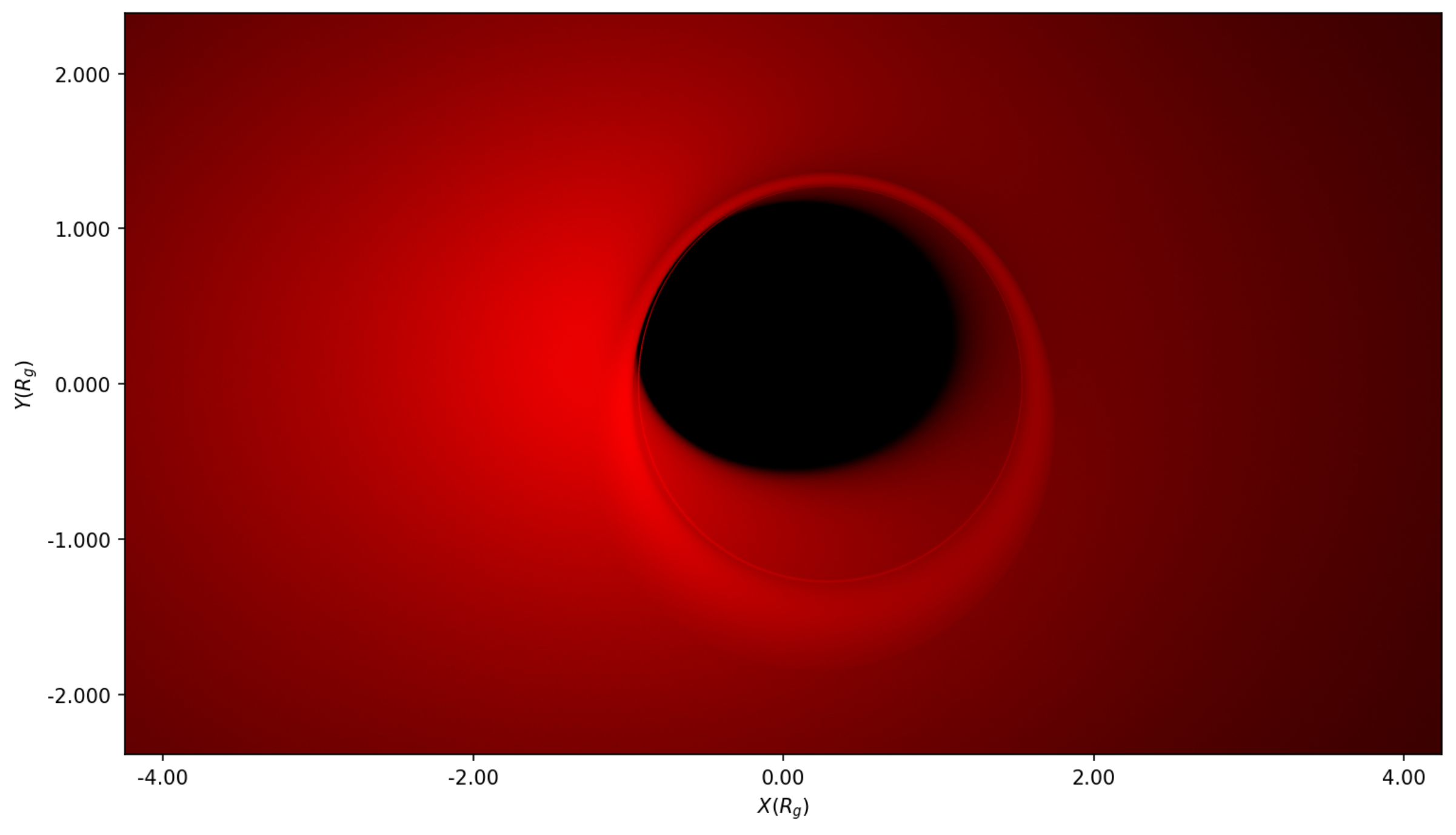}&
\includegraphics[width=0.24\textwidth]{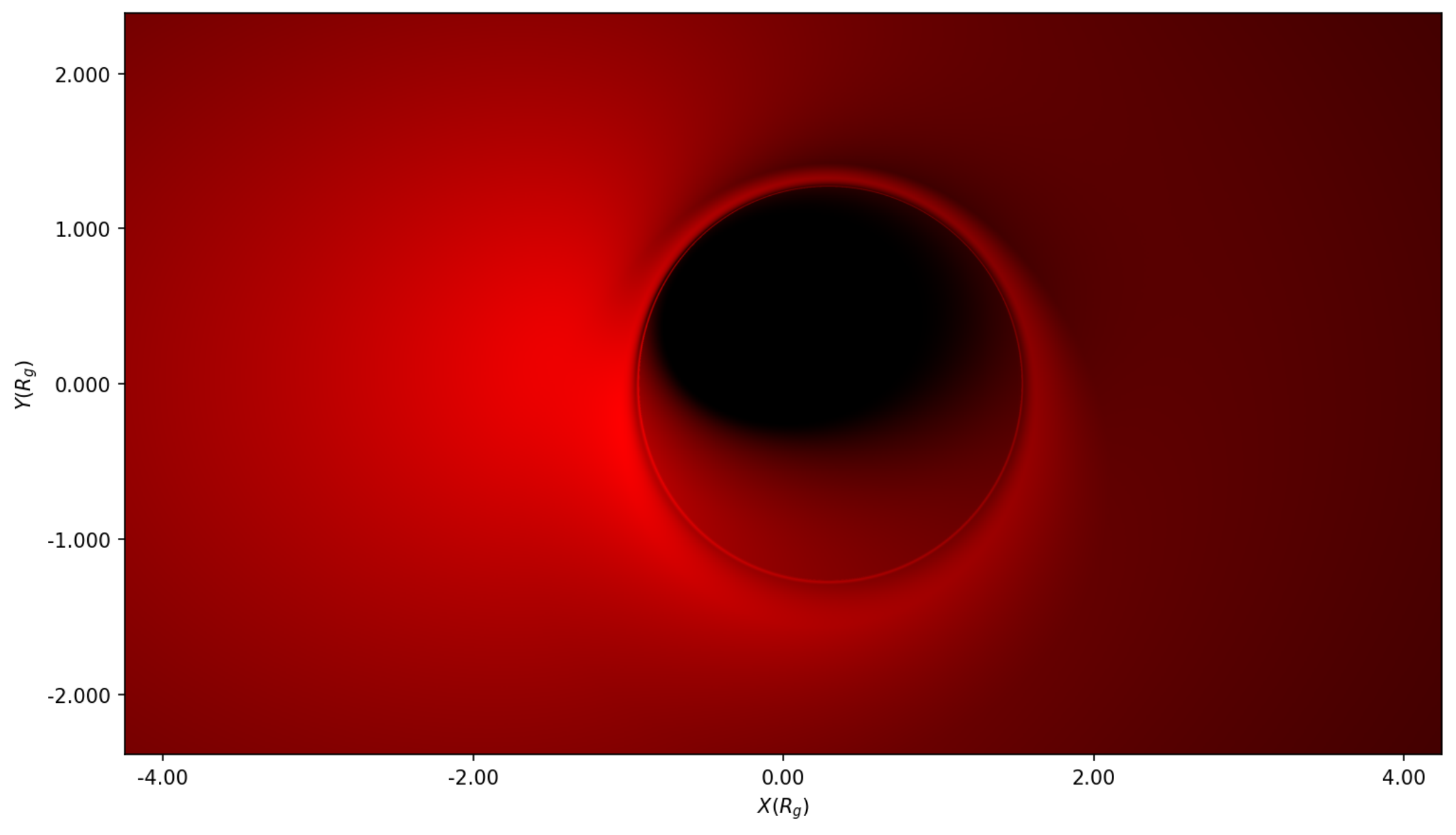}\\
\includegraphics[width=0.24\textwidth]{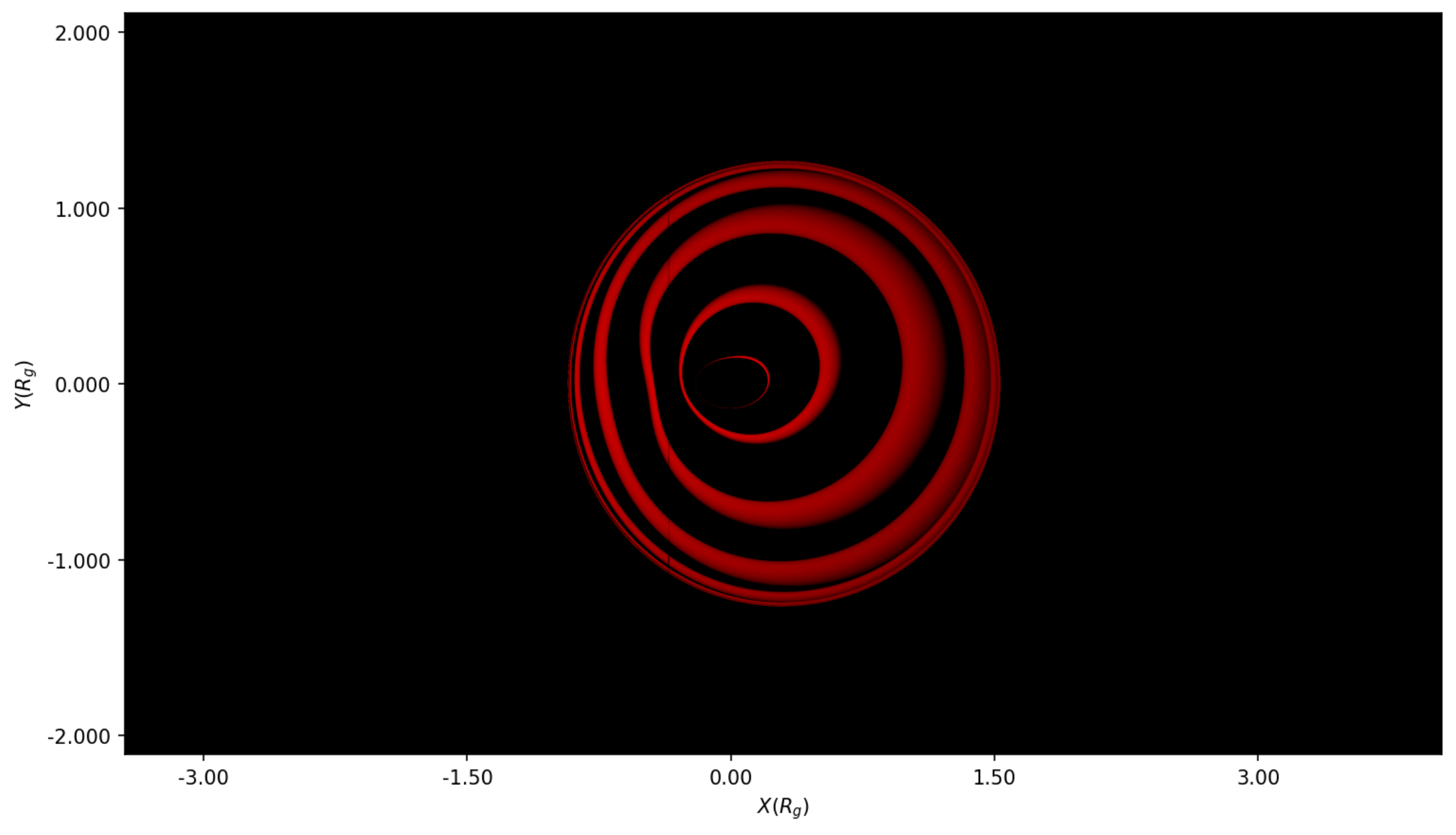}&
\includegraphics[width=0.24\textwidth]{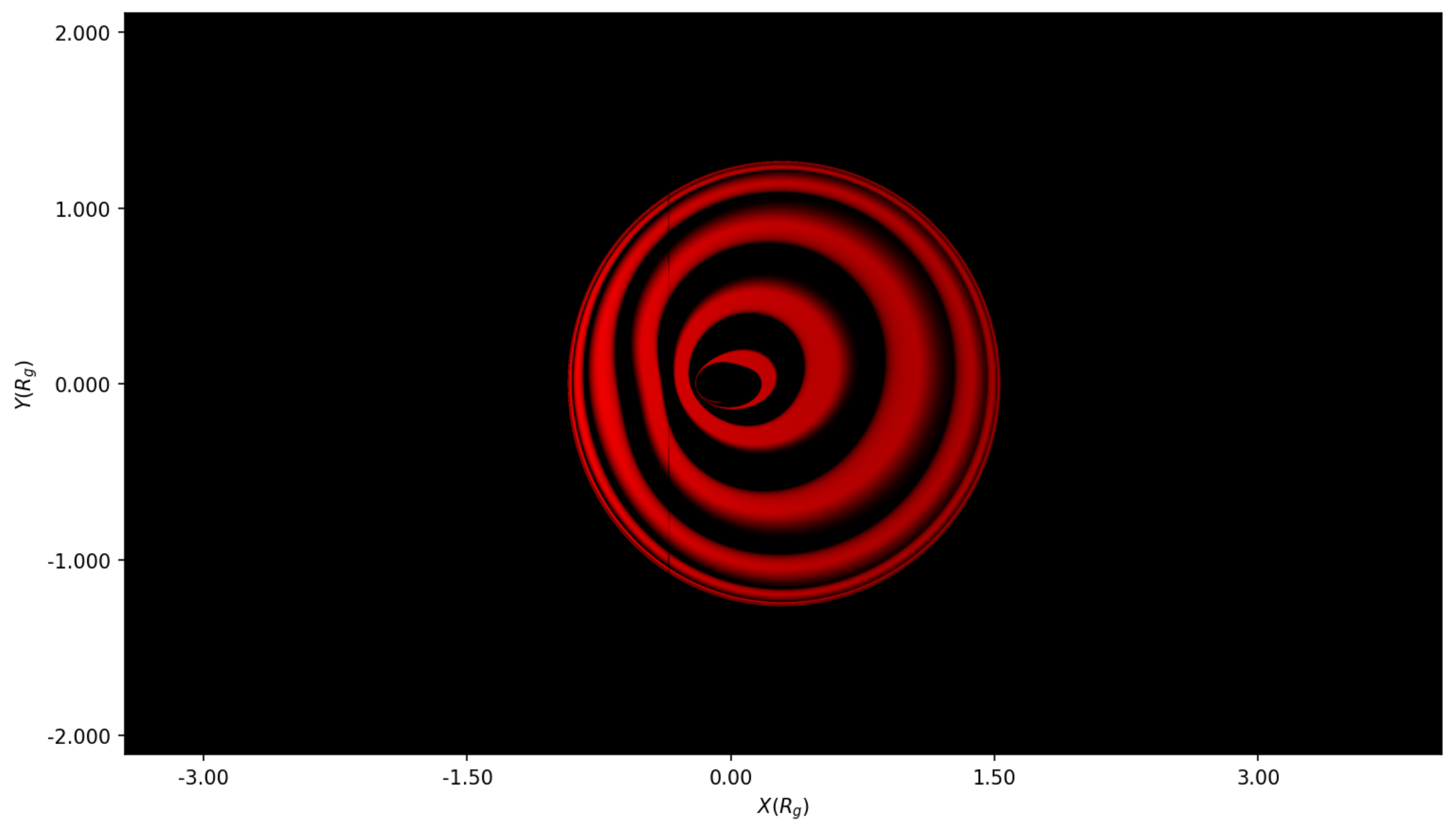}\\
\includegraphics[width=0.24\textwidth]{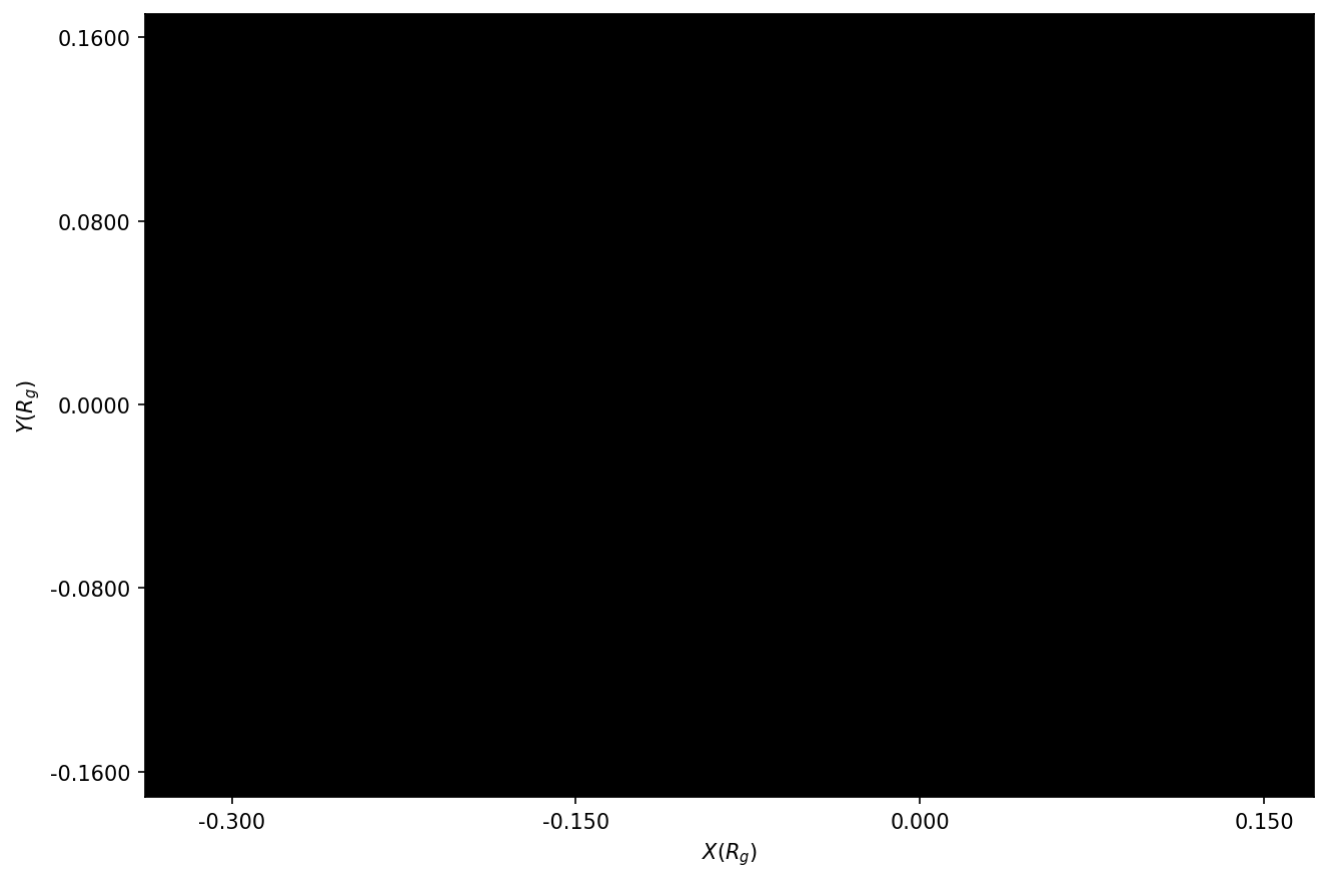}&
\includegraphics[width=0.24\textwidth]{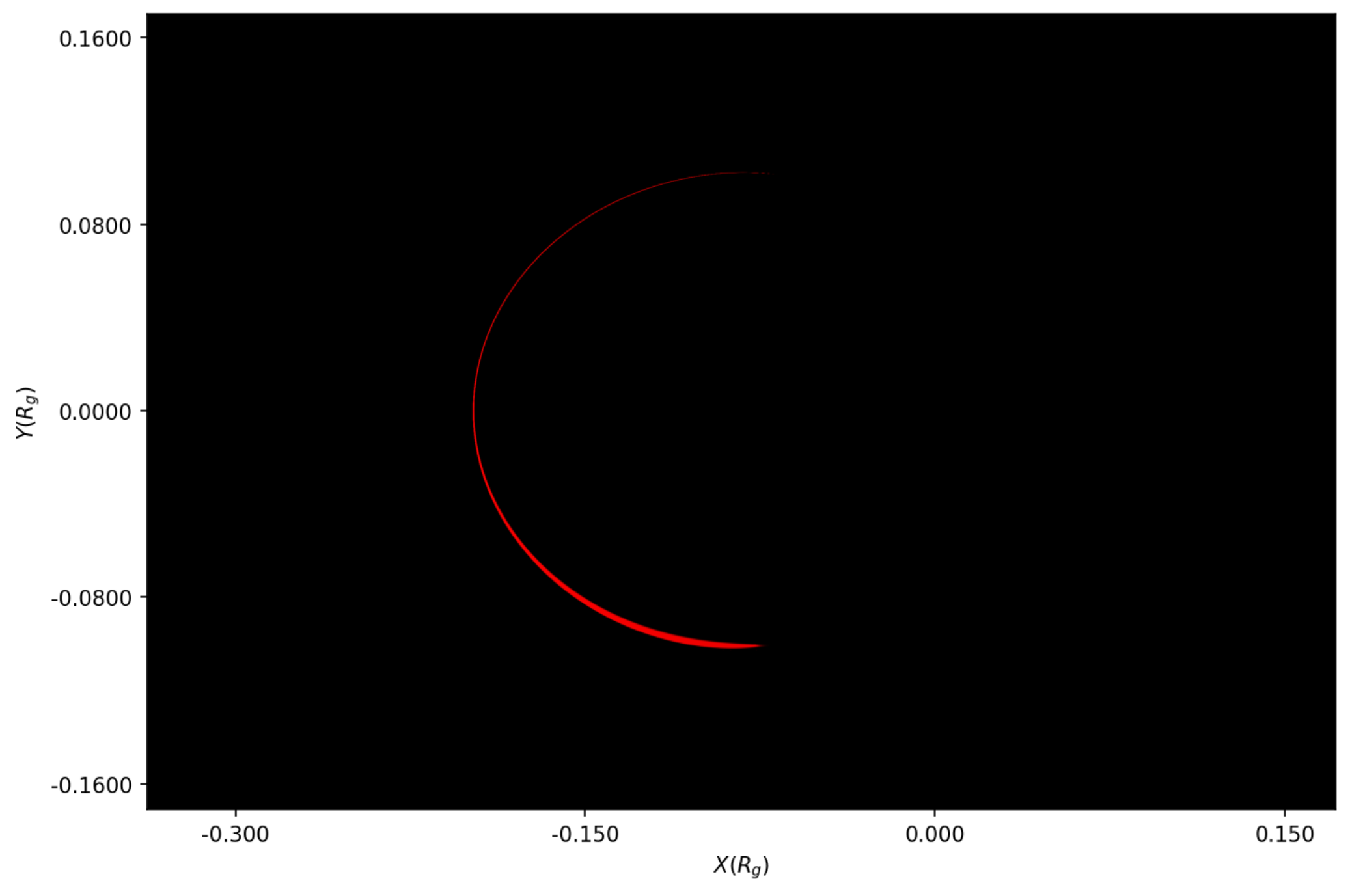}
\end{tabular}
    \caption{Influence of accretion disk configurations on Kerr black hole/white hole imaging. The setting is the same as Fig.~\ref{fig:BH_vs_WH} but for different disk models. Left column is for the thin disk model and the right column is for the hopper disk model.}
    \label{fig:disk_geometry_effect}
\end{figure}

We check more the effect of the geometric distribution of the accretion disk fluid on the nested ring features. Specifically we compare three different disk models in Fig.~\ref{fig:disk_geometry_effect}. The observation inclination and the spin are respectively set as $\Phi=45^\circ$ and $a=0.8$ in this figure.

Under these three accretion disk models, the horizontal profile curves of the white hole all maintain an asymmetric distribution state with higher intensity on the prograde side and multiple discrete oscillations. The geometric thickness of the accretion disk alters the image background intensity. In the thin disk model, the radiation source is concentrated in the equatorial plane, the section intensity profile has a low background, and the peaks of each order of emission rings separate. In the thick disk and hopper disk models, the distribution of radiation sources in the polar angle direction widens. The radiation accumulation amount of transmitted photons on the propagation path increases, filling the low-intensity regions between the photon rings. The profile curves show an overall background elevation and partial fusion of adjacent oscillation peaks.

These results indicate that the radiation source geometric parameters mainly affect the specific numerical values and background intensity of the intensity. The position features of the intensity nested rings distinguishing black holes and white holes are dominated by spacetime parameters. In the future if we observe such images, we can claim the existence of white holes.

\begin{figure}
    \centering
\begin{tabular}{c}
\includegraphics[width=0.24\textwidth]{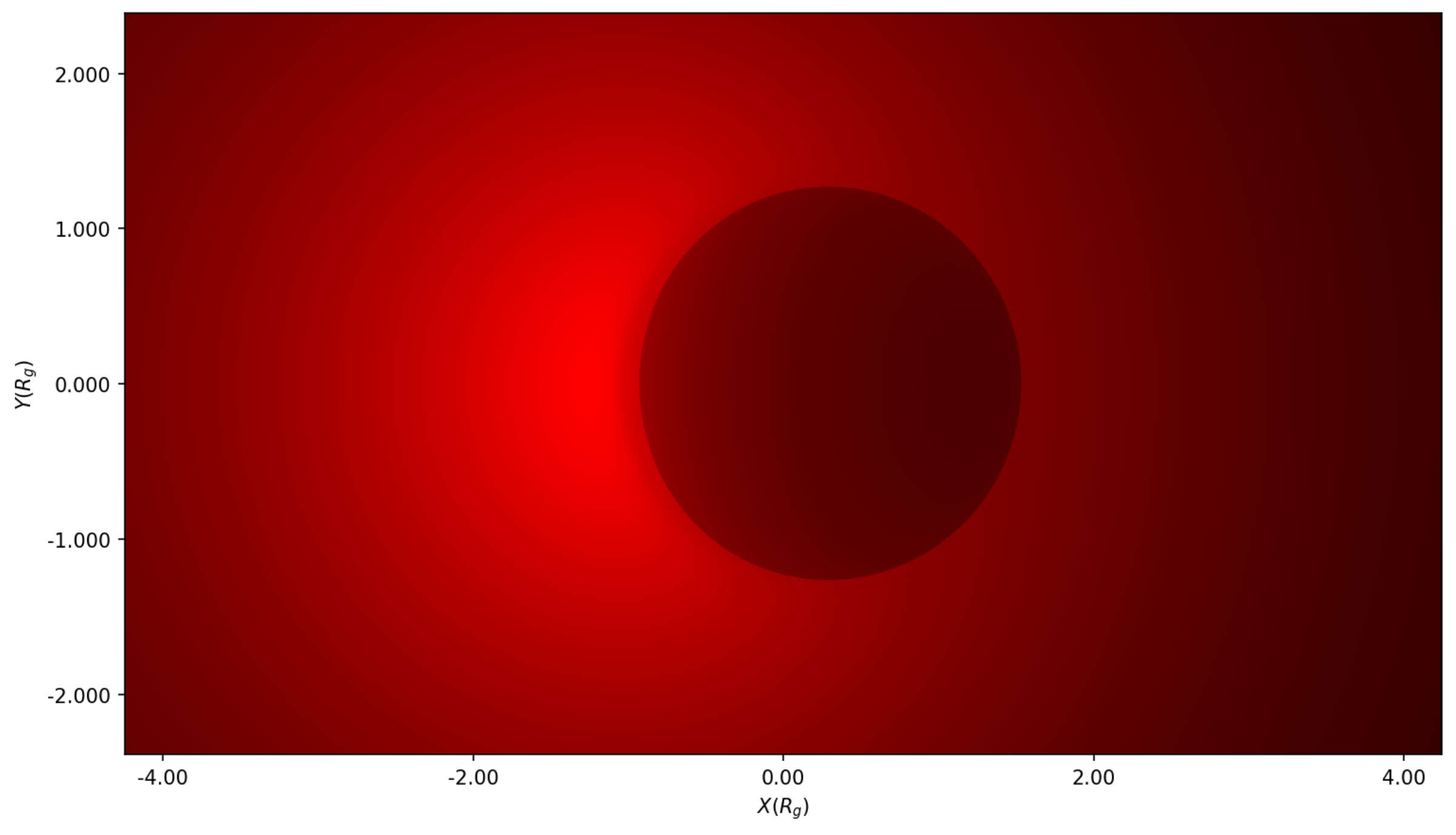}\\
\includegraphics[width=0.24\textwidth]{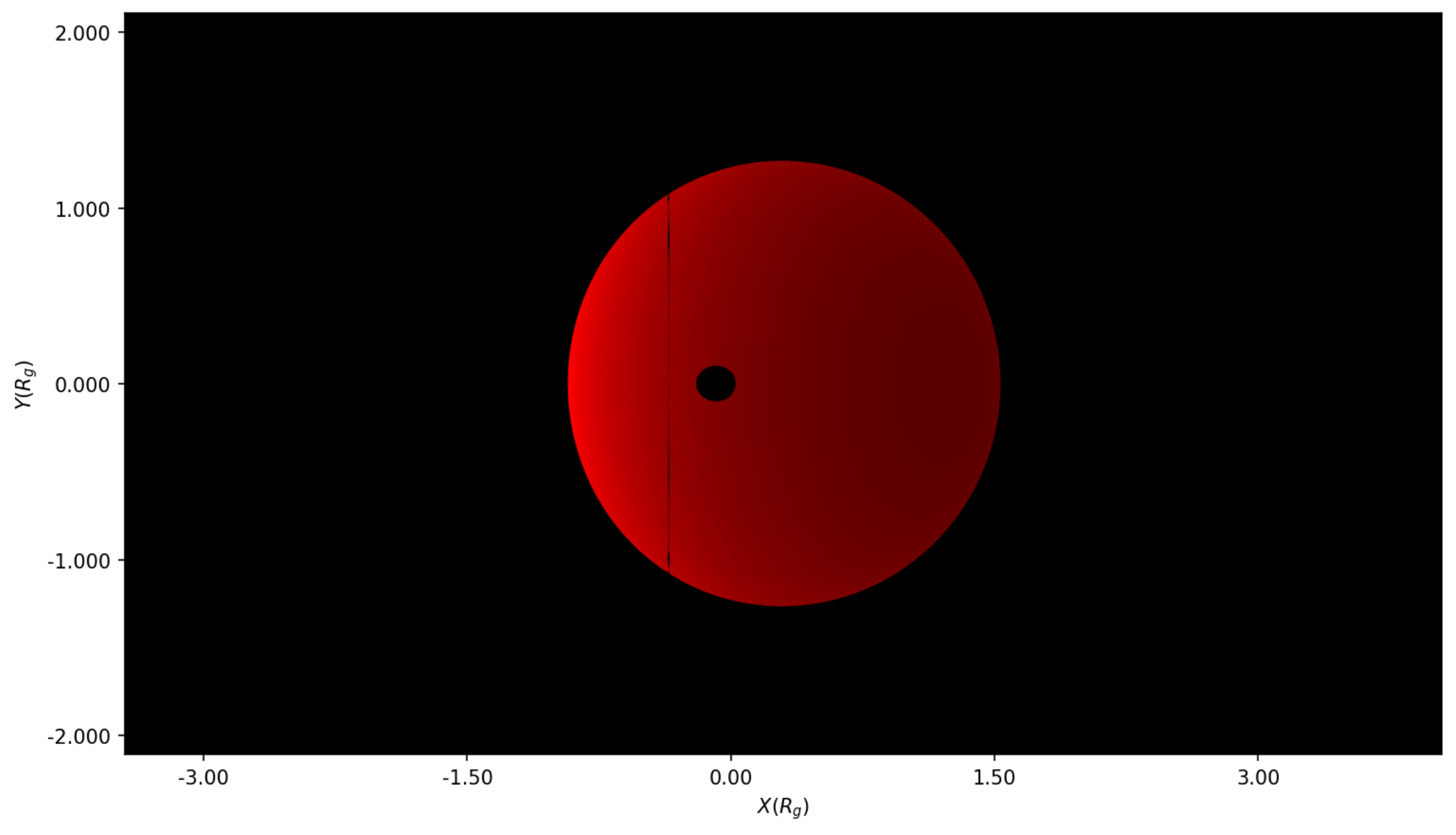}\\
\includegraphics[width=0.24\textwidth]{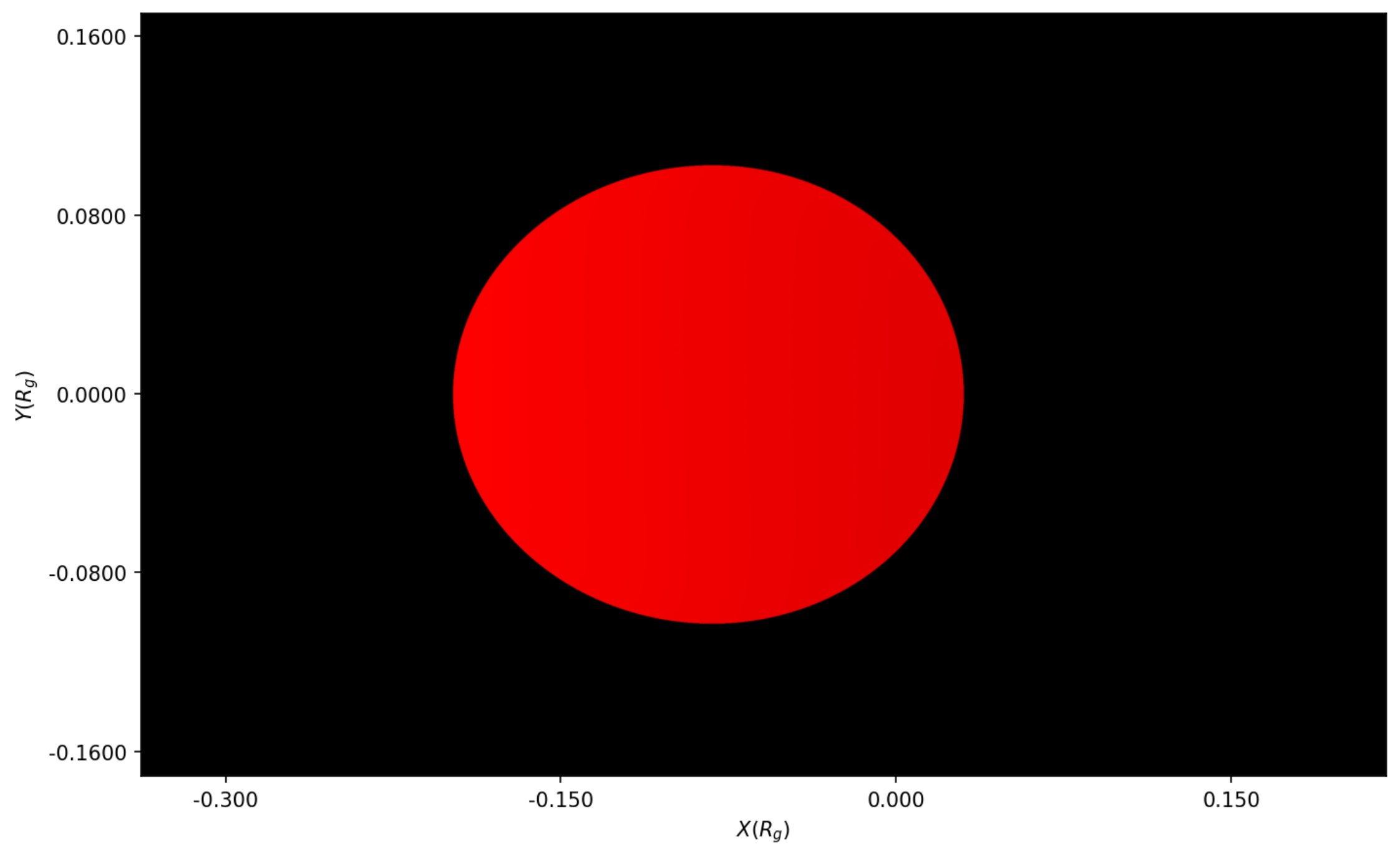}
\end{tabular}
    \caption{Intensity imaging comparison for black holes and white holes in a spherical accretion flow environment. Top row is for the black hole image, middle row is for the white hole image of observer1, and bottom row is for the white hole image of observer2. Other settings are the same to Fig.~\ref{fig:BH_vs_WH}.}
    \label{fig:thick_disk_degeneracy1}
\end{figure}

If the accretion flow is spherical and geometrically thick, morphological degeneracy in intensity observations is more pronounced. As shown in Fig.~\ref{fig:thick_disk_degeneracy1}, in a spherical accretion flow environment, uniformly accumulated radiation along the line of sight covers the intensity nested rings inside the white hole in the total intensity image. This presents as a uniform intensity peak. In this case, distinguishing black holes and white holes directly through intensity morphology is difficult.

\section{Polarization feature of white hole images}\label{sec4}
In this section, we use general relativistic polarized ray-tracing programs to calculate the polarization behaviors for images of black holes and white holes. We find a unique polarization inter-ring discontinuity behavior of white holes.

\begin{figure}[htbp]
    \centering
    \includegraphics[width=0.35 \textwidth]{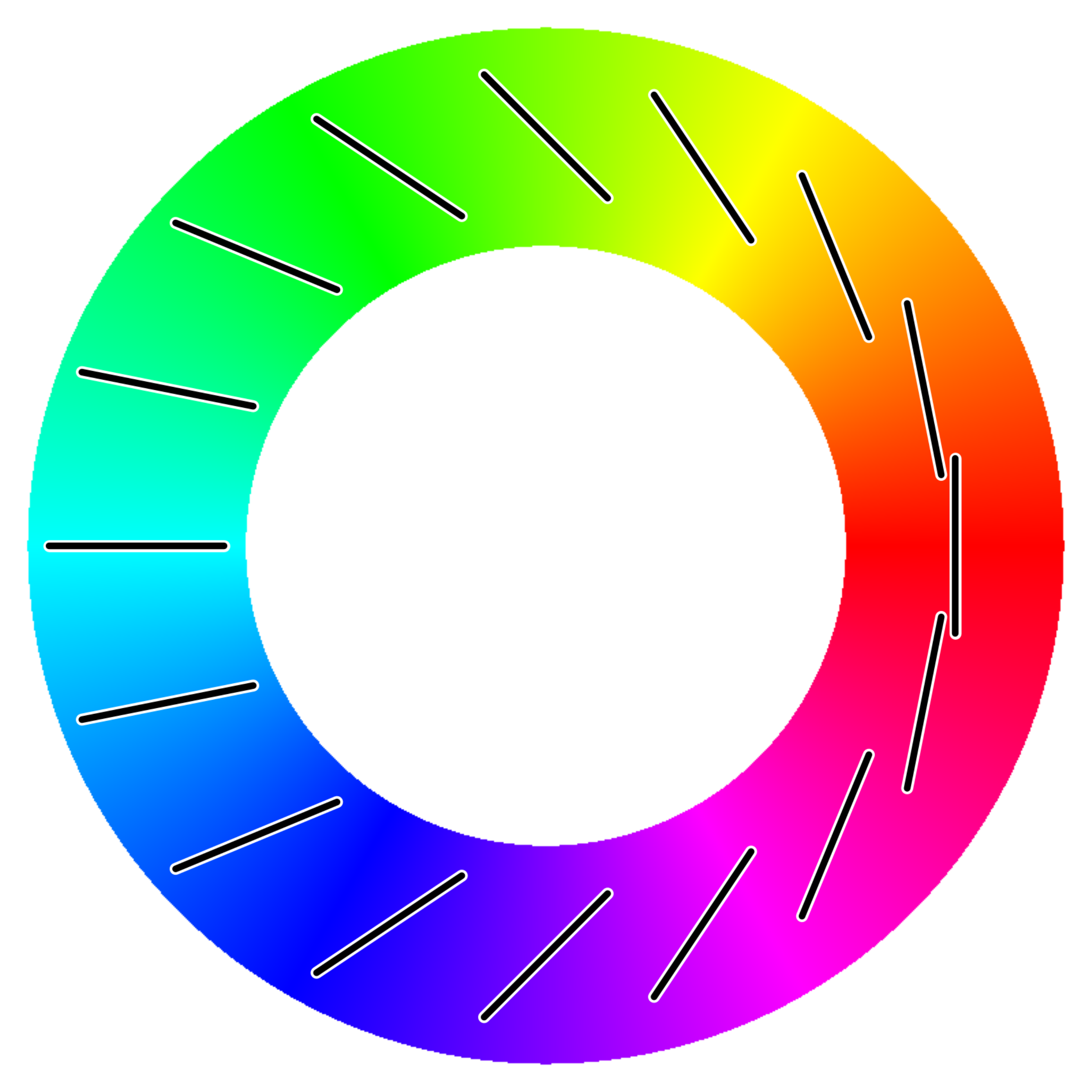}
    \caption{Linear mapping relationship between EVPA and Hue.}
    \label{fig:pol_colormap}
\end{figure}

We map the radiation EVPA $\chi = \frac{1}{2} \arctan(U/Q)$ to the pixel Hue. We linearly correspond $\chi$ in the range $[0, \pi]$ to $[0^\circ, 360^\circ]$ on the colormap as illustrated in Fig.~\ref{fig:pol_colormap}. The image luminance is modulated by radiation intensity, directly presenting the spatial evolution of the polarization field through color gradient continuity or abrupt changes.

High-order photon orbits in a gravitational field cause nonlinear twisting of the polarization plane. As the imaging order increases, each time the photon orbit winds around the central object by an extra azimuth angle of approximately $\pi$, its projected polarization vector on the observation plane undergoes one reversal. On the polarization image, this manifests as the polarization direction of adjacent higher-order photon rings being opposite, presenting an inter-ring polarization discontinuity feature.

\begin{figure}[htbp]
    \centering
    \includegraphics[width=0.48\textwidth]{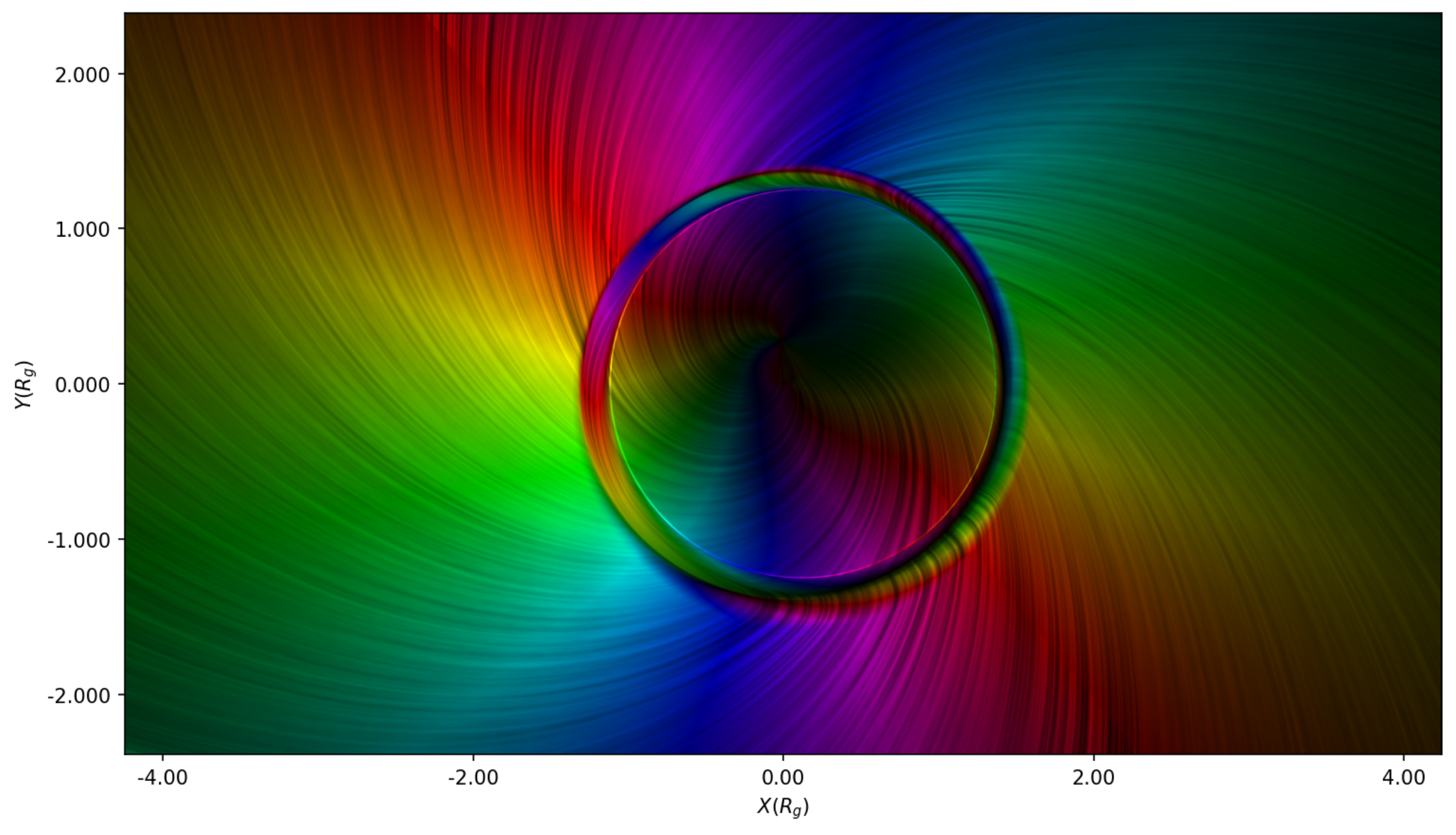} \hfill
    \includegraphics[width=0.48\textwidth]{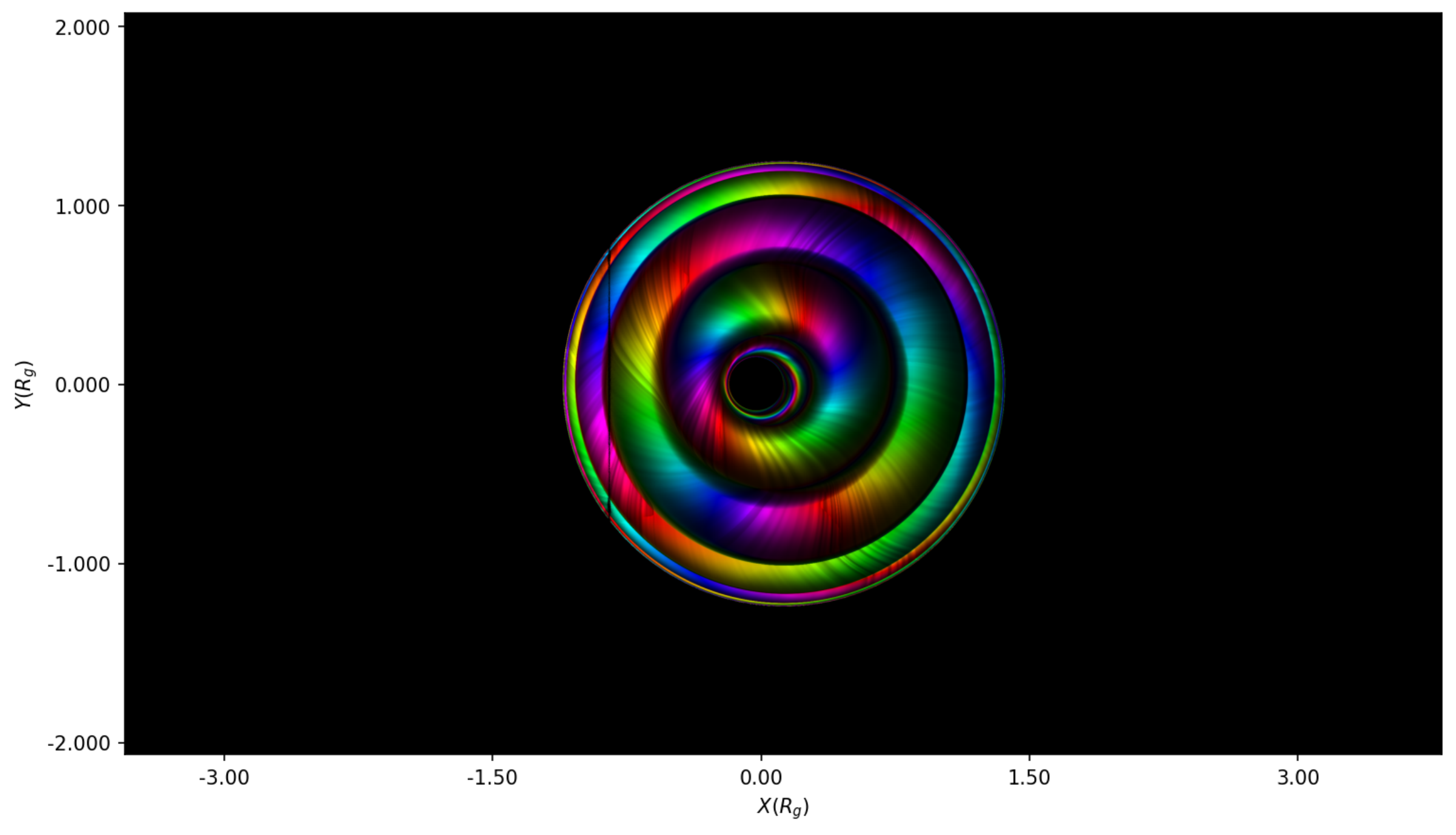} \hfill
    \includegraphics[width=0.48\textwidth]{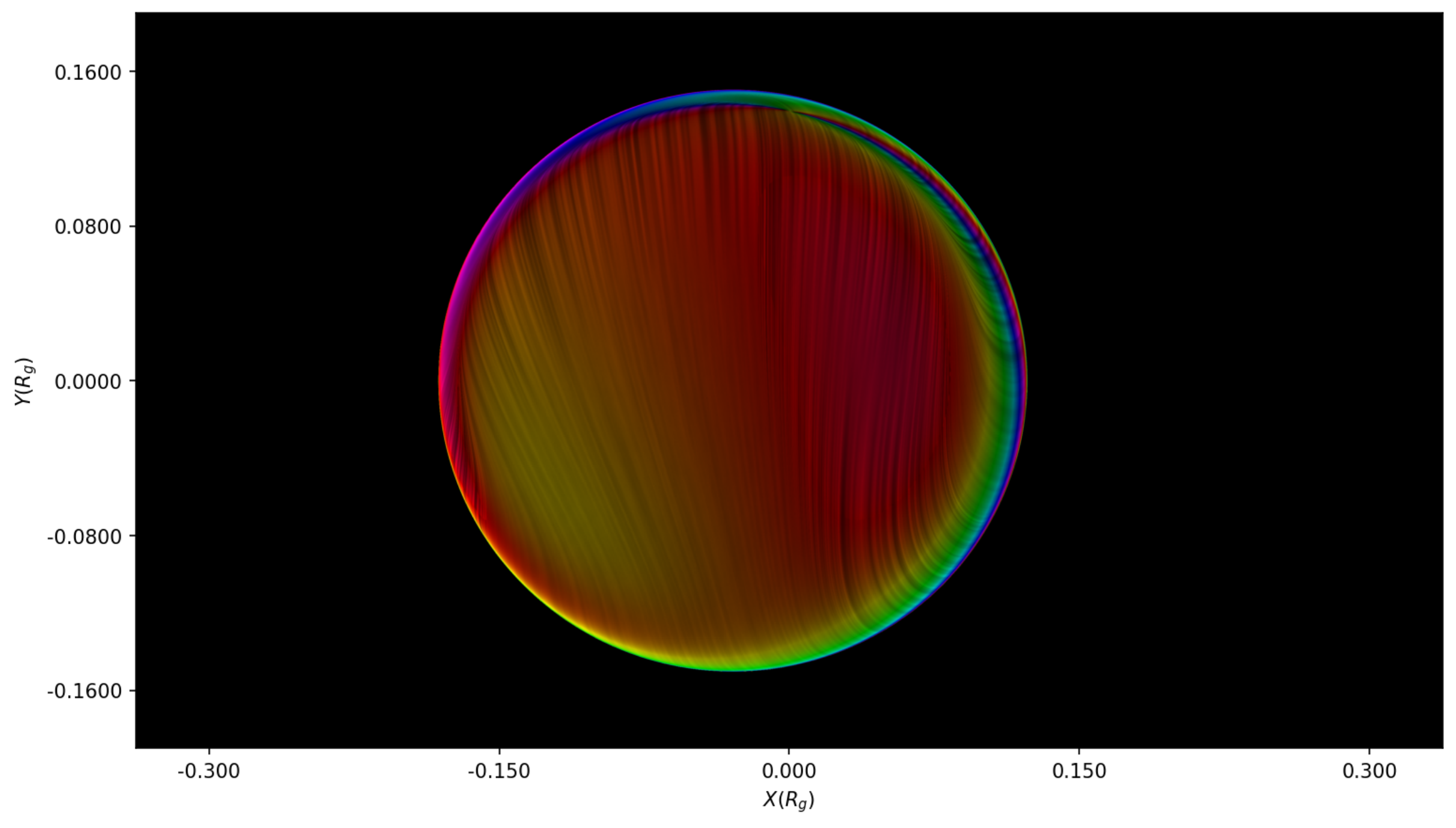}
    \caption{Global polarization of images for black holes (top) and white holes (middle and bottom). Here we have used spin parameter $a=0.8$ and inclination $\Phi=17^\circ$. Thick disk model is used in this figure.}
    \label{fig:pol_global_compare}
\end{figure}

As shown in the top subplot of Fig.~\ref{fig:pol_global_compare}, high-order photon rings of black holes also experience polarization reversal phenomena. Because black hole photon rings are highly compressed at the edge of the horizon, the physical spacing of each order of photon rings is usually smaller than the resolution of existing and near-future VLBI observations. Their reversed polarization signals cancel out in observations due to beam smearing.

\begin{figure}[htbp]
    \centering
    \includegraphics[width=0.48\textwidth]{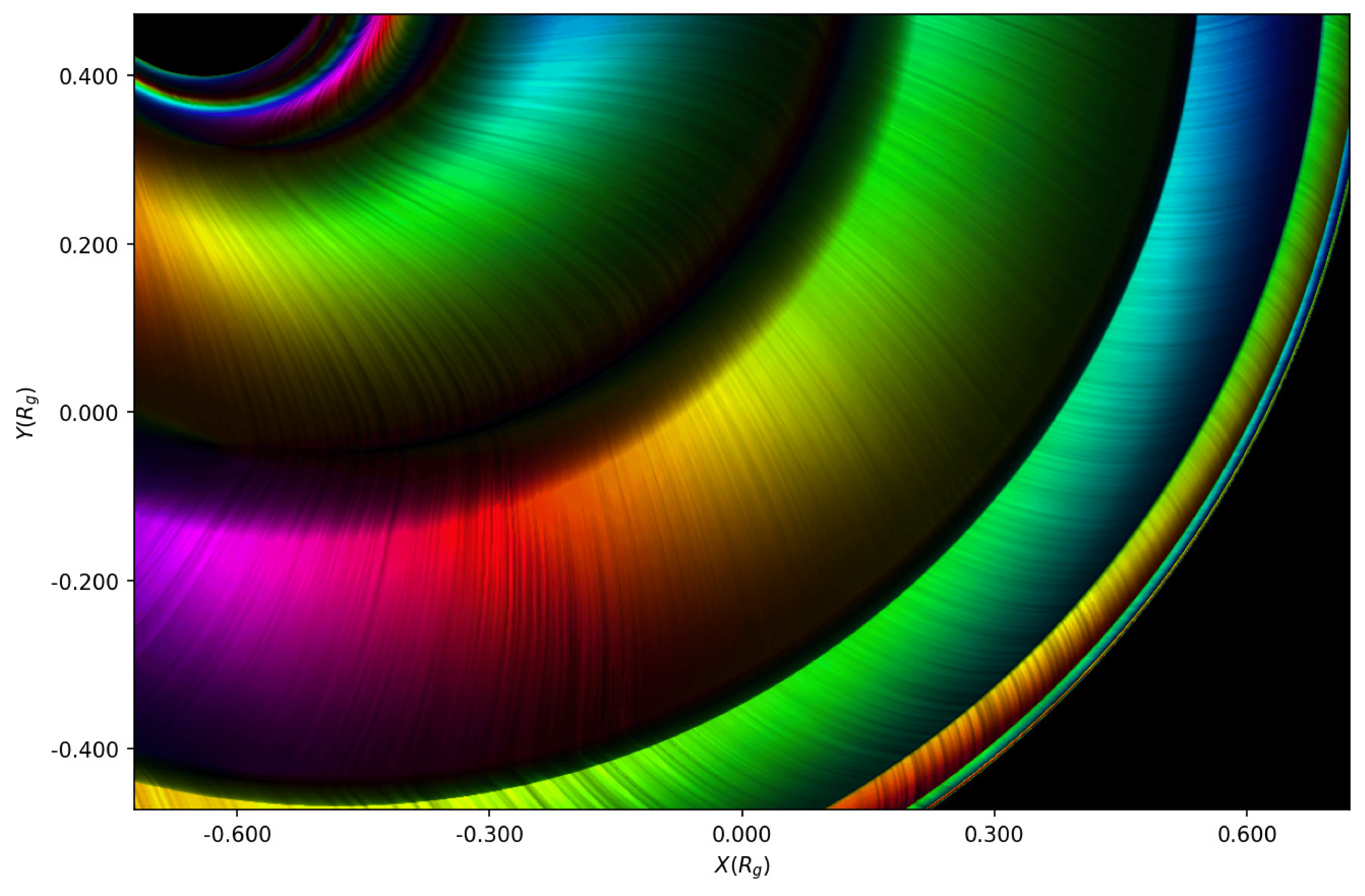}
    \caption{Local details of the white hole polarization image corresponding to the middle subplot of Fig.~\ref{fig:pol_global_compare}.}
    \label{fig:pol_detail_arrows}
\end{figure}

Unlike black holes, transmitted nested rings originating from the previous universe in white hole images as shown in the middle and bottom subplot of Fig.~\ref{fig:pol_global_compare} cover a wider field of view area. In order to highlight the feature, we zoom in the detail of the middle subplot of Fig.~\ref{fig:pol_global_compare} in Fig.~\ref{fig:pol_detail_arrows}. Due to the large projection area, polarization features carried by photon trajectories of different orders are easy to resolve spatially. This manifests as distinct polarization reversals and abrupt color changes between adjacent concentric rings. This polarization inter-ring discontinuity feature constitutes another independent observational parameter distinguishing white holes from black holes.

\begin{figure}
    \centering
\begin{tabular}{c}
\includegraphics[width=0.48\textwidth]{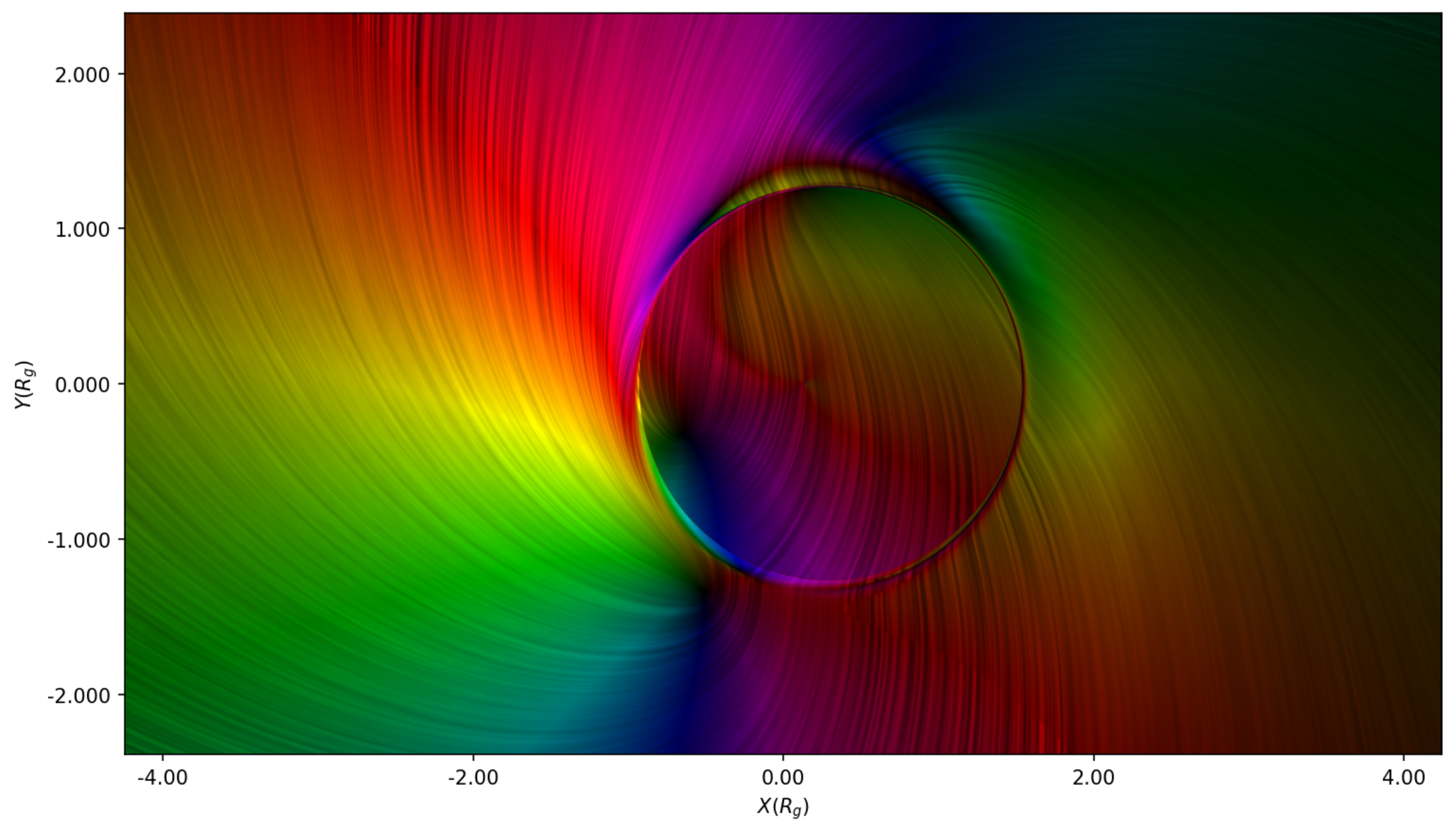}\\
\includegraphics[width=0.48\textwidth]{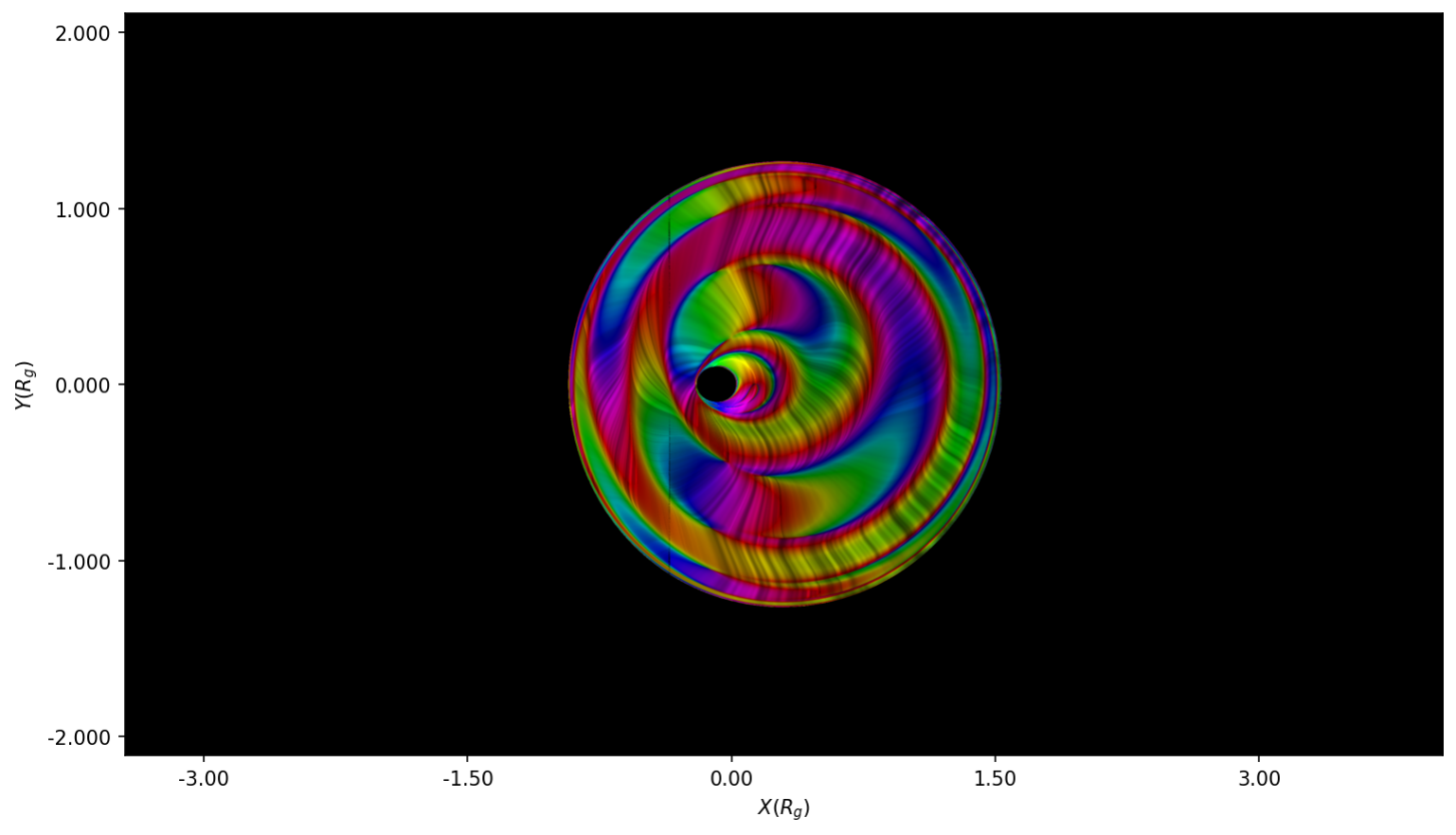}\\
\includegraphics[width=0.48\textwidth]{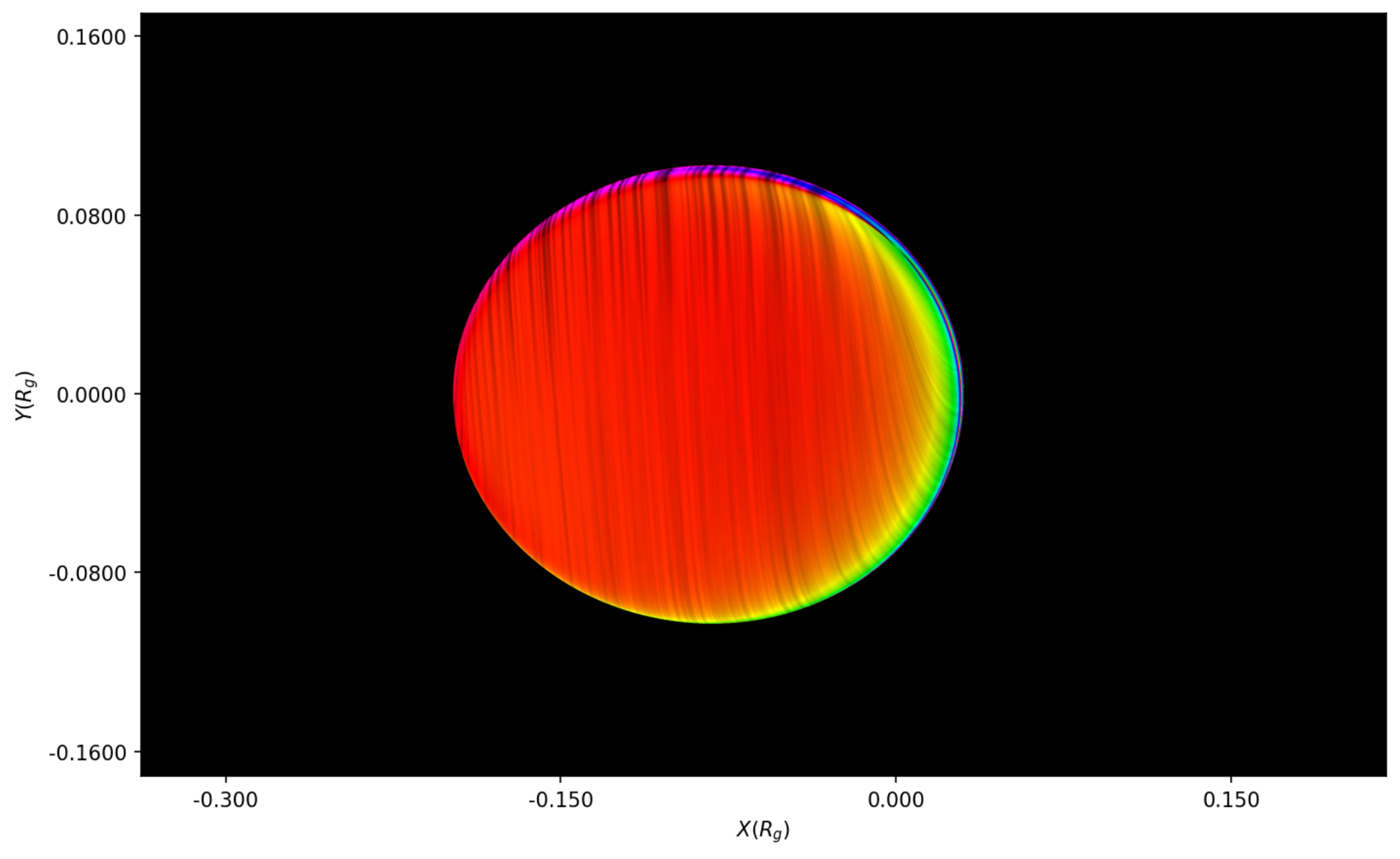}
\end{tabular}
    \caption{Polarization imaging comparison for black holes and white holes in a spherical accretion flow environment. This figure corresponds to the setting of Fig.~\ref{fig:thick_disk_degeneracy1}.}
    \label{fig:thick_disk_degeneracy2}
\end{figure}

In Fig.~\ref{fig:thick_disk_degeneracy1}, we have seen that when the accretion flow is spherical and geometrically thick, the nested ring feature of white hole images may be hidden.
However, polarization images experience less masking effects from background intensity as shown in Fig.~\ref{fig:thick_disk_degeneracy2}. Although the background fluid increases scalar intensity, the polarization phase shift is determined by the geometric path of photons penetrating the gravitational field. In the white hole polarization phase angle image, the inter-ring discontinuity structures hidden in the total intensity image persist. Results show that polarization detection provides spacetime topological information unconstrained by intensity distribution. This assists in comparative identification of compact object categories.

\section{Discussion and conclusion}

\begin{figure}[htbp]
    \centering
    \includegraphics[width=0.24\textwidth]{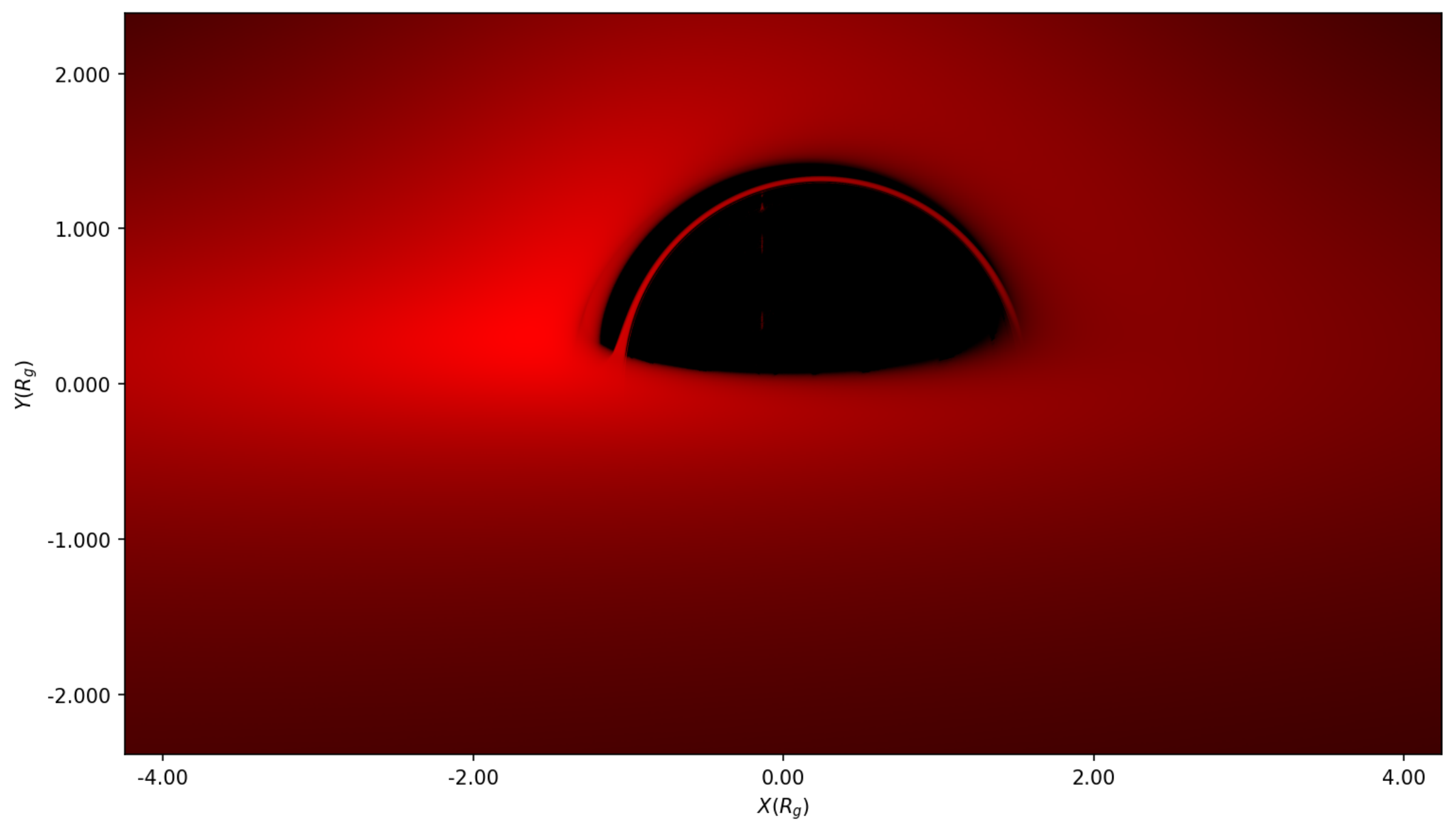}\hfill
    \includegraphics[width=0.24\textwidth]{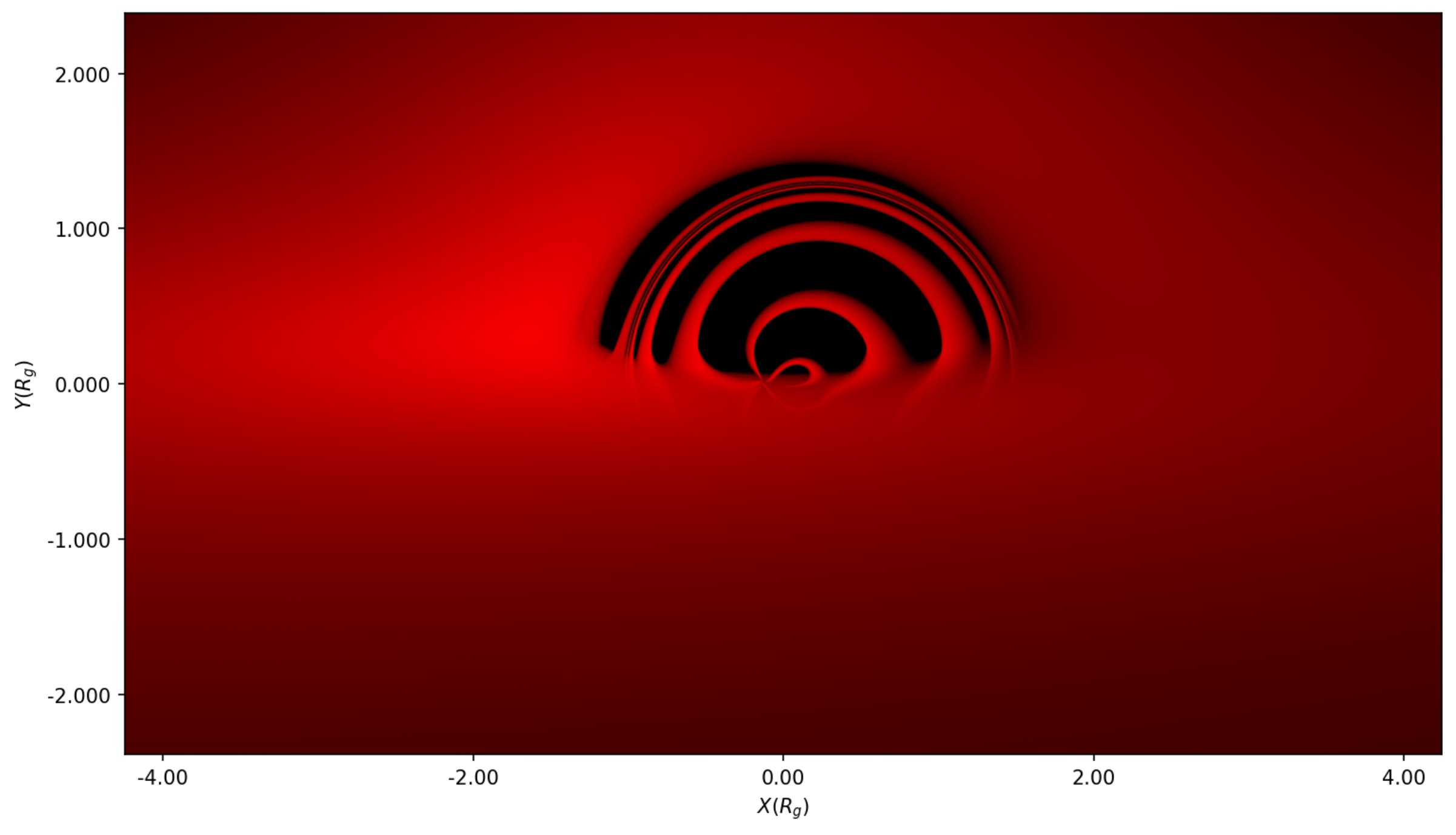}
    \caption{Comparison of geometrically thin but optically thick foreground disks at large inclination ($\Phi = 80^\circ$). Concentric ring signals from the white hole leak through the ``shadow lobe" window on one side.}
    \label{fig:leakage_thin}
\end{figure}

Regarding the observer1, realistic compact objects usually have foreground accretion disks sitting in universe1 around them. These not only produce a radiation background but also cause physical obscuration to the central field of view. When the disk is optically thin or moderately optically thick ($\tau \lesssim 1$), the transmitted intensity nested ring structures from universe0 in the center of the white hole field of view remain clear. Because their spatial stripe features differ from diffuse foreground emission, they possess certain observational discernibility. When the foreground fluid is in a dense state that is both optically thick and geometrically thick ($\tau \gg 1$), internal radiation is masked by the physical barrier. At this time, black holes and white holes tend to be identical in apparent morphology and intensity profiles. This causes single-band morphological identification to fail.

Although a thick disk environment blocks the line of sight, specific combinations of geometric configurations and observation inclinations still provide transmission windows. Taking a geometrically thin but optically thick accretion disk distributed on the equatorial plane as an example, when observed at a large inclination such as $\Phi = 80^\circ$, the disk body only blocks lines of sight near the equatorial plane. As shown in Fig.~\ref{fig:leakage_thin}, the black hole presents a crescent shadow lobe on one side of the disk. In the white hole model, truncated universe0 intensity nested ring arcs reveal themselves within these shadow lobe regions. This indicates that even with foreground physical obscuration, transmitted ring signals across evolutionary stages inside white holes maintain detectability under specific conditions.

This study uses general relativistic ray-tracing techniques to compare the imaging differences between Kerr black holes and their bounce-stage white holes under identical parameters. In a spacetime background without considering foreground interference, the center of a black hole manifests as a horizon shadow. The corresponding field of view for a white hole is filled by penetrating radiation from the accretion disk in the previous universe. Influenced by gravitational lensing and frame-dragging effects, the interior of the rotating white hole image presents a series of asymmetric intensity nested ring structures, which contrasts with regular black holes. Under high spin and large observation inclination conditions, the intensity profile in the horizontal direction shows dense and intensity peak features on the prograde side and sparse and low-intensity features on the retrograde side.

After introducing a foreground accretion disk model in the current universe, the image differences between black holes and white holes are regulated by physical obscuration. For Kerr black holes, the central region unmasked by the foreground accretion flow maintains the horizon shadow feature. The corresponding dark region of a white hole transmits truncated asymmetric nested emission rings.

The observation window to distinguish these two types of compact objects depends on the thickness and observation angle of the foreground disk. When observed in the polar axis direction or when the accretion environment is dominated by a thin disk, the unique intensity nested ring structures of white holes are preserved. Conversely, if an accretion disk with large geometric and optical thickness exists in the current universe, opacity masking causes the apparent morphology of white holes to become degenerate with regular black holes. To this end, this paper calculates the polarization evolution process of electromagnetic waves originating from the previous evolutionary stage under a relativistic framework. Polarization image analysis shows that after experiencing frame-dragging and propagation across the horizon, adjacent high-order photon rings transmitted by white holes present clear polarization reversals. This forms an inter-ring polarization discontinuity. Such polarization features determined by spacetime geometric topology are less susceptible to masking by local perturbations compared to background scalar intensity. The analysis concerning intensity nested ring structures, physical obscuration effects, and polarization inter-ring discontinuity features provides a theoretical reference for comparing and identifying related compact objects using radio observation equipment such as VLBI in the future.

\acknowledgments
We thank Xiaokai He and Yan Liu for useful discussion. This work is supported by the National Key Research and Development Program of China (Nos. 2021YFC2203001), the National Natural Science Foundation of China (No.~12475049).

\end{document}